\documentclass[%
 book,
 aip,
 jmp,%
 amsmath,amssymb,
 reprint,%
]{revtex4-2}

\usepackage{graphicx}
\usepackage{dcolumn}
\usepackage{bm}
\newcommand*{\plimsoll}{{\ensuremath{-\kern-4pt{\ominus}\kern-4pt-}}}
\usepackage{color}

\usepackage{graphicx}
\usepackage{dcolumn}
\usepackage{bm}
\usepackage{cancel}

\begin{document}

\preprint{AIP/123-QED}

\begin{titlepage} 

	\centering 
	
	\scshape 
	
	\vspace*{\baselineskip} 
	
	
	\rule{\textwidth}{1.6pt}\vspace*{-\baselineskip}\vspace*{2pt} 
	\rule{\textwidth}{0.4pt} 
	
	\vspace{0.75\baselineskip} 
	
	{\LARGE The long and short of templated copying\\} 
	
	\vspace{0.75\baselineskip} 
	
	\rule{\textwidth}{0.4pt}\vspace*{-\baselineskip}\vspace{3.2pt} 
	\rule{\textwidth}{1.6pt} 
	
	\vspace{2\baselineskip} 
	
	
	Jenny Poulton\\
	Imperial College London\\
	Department of Bioengineering\\
	PhD Thesis
	
	\vspace*{3\baselineskip} 
	
\end{titlepage}

\title[The long and short of templated copying]{The long and short of templated copying}

\author{Jenny Poulton, Imperial College London}

\date{\today}

\begin{abstract}
Templated copying is the central operation by which biology produces complex molecules. Cells copy sequence information from DNA to RNA and on into proteins, which are the molecules responsible for the function and regulation of cellular systems.  In the templated copying process the template catalyses the formation of a second molecule carrying the same sequence. Traditionally, people have ignored the separation of the template and copy at the end of the process, but separation is necessary and fundamentally changes the thermodynamics of the process.

In general, creating an accurate polymer costs free energy. Omitting separation, this cost can be compensated for by the extra free energy released by “correct” copy/template bonds. Separation requires these bonds be broken, so true copying requires an input of free energy. Equally the fact that copy/template bonds are temporary means there is no thermodynamic bias towards accuracy, instead copying relies only on kinetic effects to promote accuracy. In general, transducing energy can only happen reversibly when the transduction process is quasistatic and time varying; something that cannot be true when you are relying on kinetic discrimination.  Copying is a far from equilibrium process. This thesis explores the consequences of this observation. We start in the limit of infinite length copies where the costs of accuracy represent hard thermodynamic bounds and then moves to the finite length limit where these same limits can be understood as kinetic barriers. We then discuss copying systems as non-equilibrium steady states, which can be analysed as information engines moving free energy between out-of-equilibrium baths.
\end{abstract}

\maketitle

\newpage

\section*{Copyright and originality statement}
This work is the work of Jenny Poulton, and any parts which are adaptations of other work are clearly referenced.

The copyright of this thesis rests with the author. Unless otherwise indicated,
its contents are licensed under a Creative Commons Attribution-Non
Commercial 4.0 International Licence (CC BY-NC).
Under this licence, you may copy and redistribute the material in any medium
or format. You may also create and distribute modified versions of the work.
This is on the condition that: you credit the author and do not use it, or any
derivative works, for a commercial purpose.
When reusing or sharing this work, ensure you make the licence terms clear to
others by naming the licence and linking to the licence text. Where a work has
been adapted, you should indicate that the work has been changed and
describe those changes.
Please seek permission from the copyright holder for uses of this work that are
not included in this licence or permitted under UK Copyright Law.

\tableofcontents
\listoffigures
\newpage

\setcounter{equation}{0}
\setcounter{figure}{0}
\setcounter{table}{0}
\setcounter{section}{0}

\section*{Introduction}

\setcounter{equation}{0}
\setcounter{figure}{0}
\setcounter{table}{0}
\setcounter{section}{0}

\section{Self assembly}

\begin{figure}
    \centering
    \includegraphics{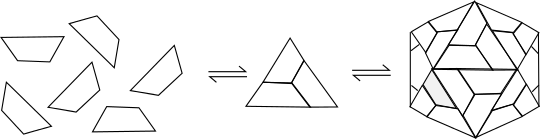}
    \caption{Image adapted from\cite{matsuurua2014rational}, here trapezium shaped capsomers first form triangular units, which then form into a complex spherical covering. The spherical covering is highly symmetric and identical capsomers can occupy multiple locations in the structure without changing its final shape. This allows the programming of the final shape of the spherical covering to be programmed into a very small number of subunits (here one).}
    \label{Capsid}
\end{figure}

Biological systems rely on complex processes requiring complex molecules, which must be assembled from available materials without external manipulation\cite{murugan2015undesired}. An example of a complex biological molecule is a virus capsid; a protective protein covering which surrounds viral genetic material. As illustrated schematically in fig. \ref{Capsid}, this protective covering is highly symmetric and is made up of a large number of similar subunits called capsomers. Due to local interactions between capsomers the capsid is able to self assemble. This means that the capsid structure is the ground state of the system, and favourable bonds between capsomers causes them to assemble themselves into a specific structure. This process is completely autonomous; it requires no external manipulation and is encoded into the capsomers themselves\cite{johnston2010modelling}.

One reason that capsids are well suited to self assembly is that they are highly symmetric, and identical capsomers can occupy multiple different locations within the capsid structure without changing the structure of the capsid. These properties are not ubiquitous to complex biological molecules.

Another group of complex molecules is proteins, which have diverse roles in the body including acting as enzymes, antibodies, message carriers, providing structure, transport and storage. Capsomers are themselves complex proteins. Proteins are made up of subunits called amino-acids, which come in a library of 20 distinct types\cite{Alberts2002}. Proteins have very complex and precisely defined folded structures, lacking symmetry, which allow the protein to perform its complex functions. As an example, many enzymes form shapes which have spaces which closely fit a specific substrate. These active sites can be used to bring reactant molecules close to each other and encourage their transformation into useful product molecules\cite{Alberts2002}. Even very small differences in the composition of a protein can cause it to misfold, and this can have serious consequences for organisms. In the human body misfolding is responsible for diseases such as Alzheimer's, Parkinson's and type 2 diabetes\cite{dobson2002protein}.

\begin{figure}
    \centering
    \includegraphics[scale=0.25]{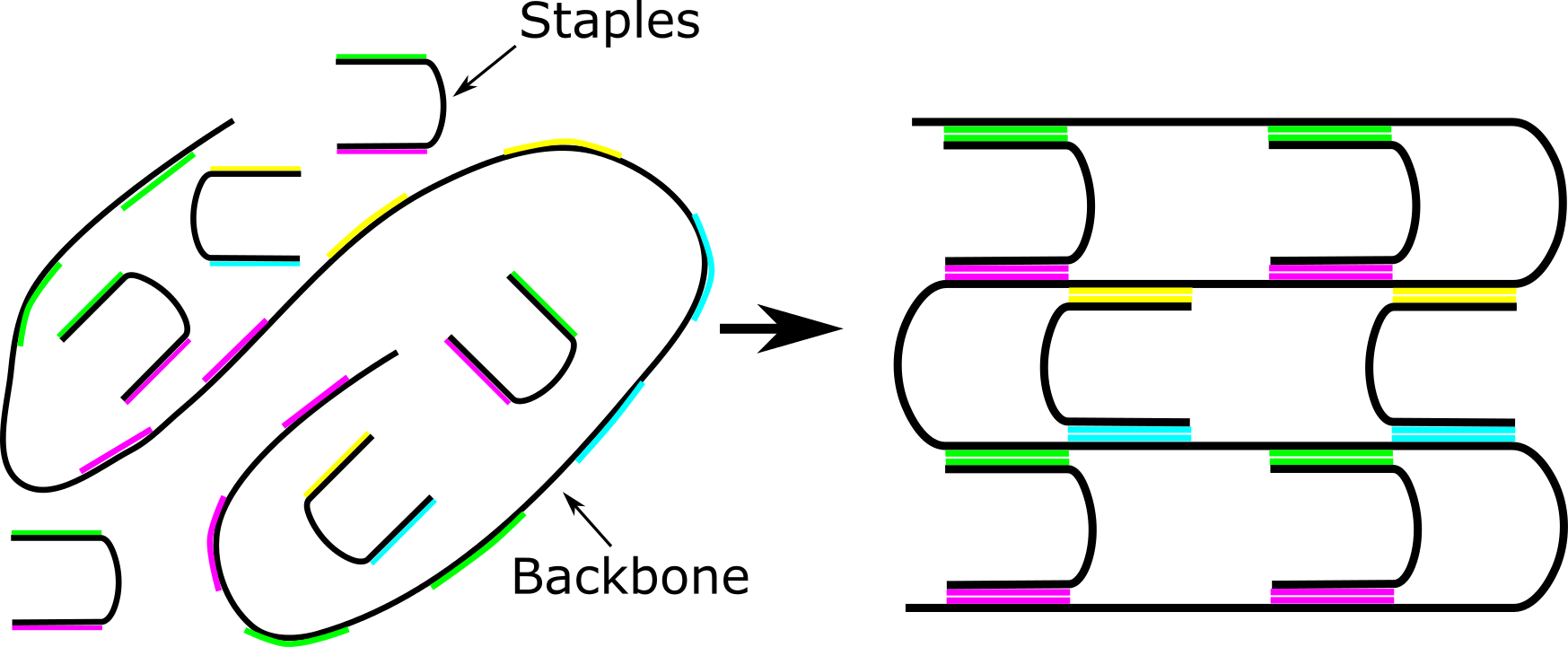}
    \caption{DNA origami is often created from a single long scaffold which is pinned into place with staple molecules which are complementary to sections of the scaffold (matching colour ends). Notice that in this case, while we have folded the scaffold into this configuration using the available staples, there are many redundant configurations that this mixture of scaffold and staple could fold into. Imagine, for example if the yellow ends of the two middle staples were to swap, the scaffold would fold into an entirely different configurations. To design out redundancy, we would have to have many more than two unique staples. This is the core of the problem of targeting specific complex molecules with self assembly.}
    \label{Origami}
\end{figure}

In order to understand the difficulty in self assembling complex molecules without symmetry we will first consider synthetic self assembly mechanisms. It is possible to design synthetic self assembly mechanisms to target complex structures. In these systems, a pool  of subunits is allowed to combine and in doing so relax into a ground state which has been pre-designed through specific interactions between the subunits\cite{he2008hierarchical,chhabra2010dna,li2009nanofabrication}. An example of this is DNA origami\cite{wang2017beauty,sacca2012dna,castro2011primer}, in which a long single strand of DNA is folded into a specific topology by specially designed ``staples" which have favourable bonds with two different regions of the long strand as shown as  fig. \ref{Origami}. It should be noted here that in order to assemble one specific structure accurately, a large library of subunits is required. For a structure with no symmetry, then you need \(n\) unique molecules for a \(n\)-particle assembly. This is in order to prevent competing ground states--and therefore configurations--the system might relax into different from the one being targeted\cite{murugan2015multifarious,murugan2015undesired}. In structures that have high symmetry, like the virus capsid, it is possible to reduce the number of subunits, but it is not the case for most biomolecules. To illustrate the issue of competing ground states; the process showed in fig. \ref{Origami} has multiple ground states, and so would not reliably form the molecule shown. Thus, every target molecule would require a different and distinct set of pre-designed subunits which themselves would have to be created. Given there are around 50,000 distinct functional proteins, often made up of hundreds of amino-acids\cite{dobson2002protein}, predesigning interactions between subunits is clearly not an optimal strategy for living organisms in many situations.

\section{Templating}

Biology manages to solve the problem of targeting these complex structures using a library of only 20 amino-acids. The assembly method now relies on interactions between amino-acids and a template, here mRNA. The mRNA sequence encodes the order in which amino-acids will be arranged in the final protein chain. Favourable interactions between amino-acids and RNA codons allow a specific polymer to be formed. The sequence of the polymer is therefore determined via the favourable interaction with the template, and the amino-acids themselves are only necessary to encode the folding of the polymer into its functional form given the sequence. This requires fewer components than encoding both the sequence and the folding. Given there are \(20^{100}\approx1\times10^{130}\) ways to combine a library of 20 amino-acids into a chain of length 100, the use of templates to target one of only 50,000 functional protein sequences, which then fold into complex molecules, is clearly vital to the functioning of the body\cite{dobson2002protein}.

We therefore argue that templated copying is central to the development of biological complexity; there is no other way for an organism to realistically manufacture the variety of complex molecules required for it to function. Unsurprisingly, therefore, organisms have developed sophisticated machinery to perform templated copying. 

\begin{figure}
    \centering
    \includegraphics[scale=0.5]{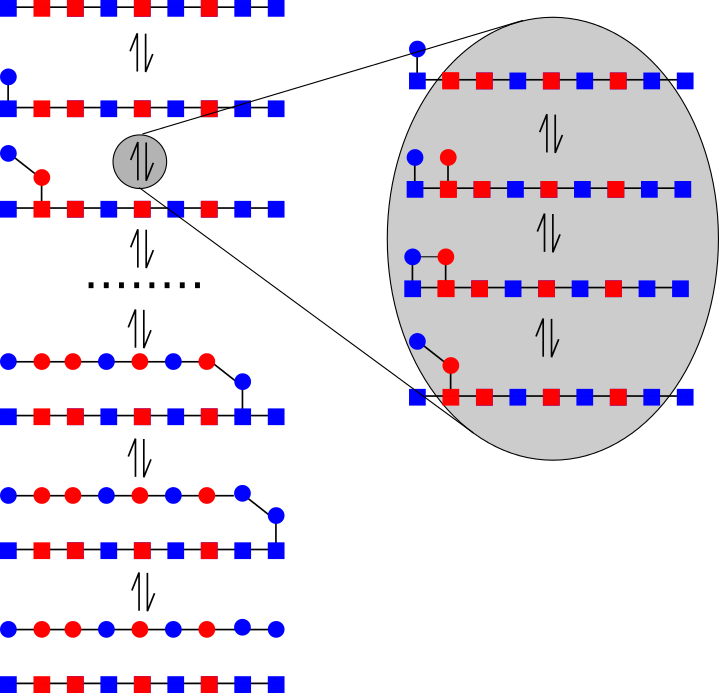}
    \caption{A generic copying process. Coarse grained steps on left show a monomer coming out of solution and attaching to the left-most free site on the template. The copy then grows on the template, site by site, with the tail detaching sequentially, before the final copy polymer detaches from the rightmost site on the template. Detailed sub-steps on right in which a monomer comes out of solution, is polymerised into the chain, and then the previous monomer detaches from the template. In this generic procedure we are enforcing an specific ordering of steps, which is a simple analogy for the process of translation and transcription. The mechanisms by which this state order is enforced is beyond the scope of the main part of this work, but will be discussed further in the conclusion.
}
    \label{generic}
\end{figure}

To aid understanding we outline a generic template copying scheme as shown in fig. \ref{generic}, which is a rough analogy to the processes of RNA transcription and translation. On the left side we see a monomer come out of solution and attach to the first site of a template. The copy then grows on the template, site by site, with the tail detaching sequentially before the final copy polymer detaches at the final step. On the right we zoom into the details of a single step, in which a monomer comes out of solution, is polymerised into the chain, and then the previous monomer detaches from the template. In this generic procedure we are enforcing an specific ordering of steps, which is a simple analogy for the process of RNA translation and transcription. The mechanisms by which this state order is enforced is beyond the scope of the main part of this work, but will be discussed further in the conclusion.

The central dogma of molecular biology describes the way information is transferred via templating from DNA via RNA to proteins. DNA is able to self replicate by using itself as a template. DNA forms a double stranded structure, with two long strands made of DNA nucleotides spiralling around each other in a shape known as the double helix. There are four DNA nucleotides; adenine, cytosine, guanine and thymine. These nucleotides have complementary interactions: adenine and thymine form one attractive base pair, and cytosine and guanine form the other. The two strands of DNA are therefore complementary to each other, but encoding the same information\cite{Alberts2002}. 

These complementary interactions are central to DNA's ability to be used as a template. The first step of the process is to separate the two strands of DNA which is accomplished by a helicase enzyme. DNA nucleotides then come out of solution and bind with the exposed strand which acts as a template. The stability of that interaction dictates whether the nucleotide is incorporated permanently into the new strand via the formation of a backbone bond, or whether the nucleotide falls back off the exposed strand into solution first\cite{Hopfield}. Thus the more stable complementary nucleotide is more likely to be incorporated than a mismatch. This, in essence, is the source of all accuracy in templated copying.

RNA transcription is a very similar mechanism to DNA self replication except that RNA is a single stranded molecule. RNA also has four bases, but here uracil takes the place of thymine. Here, once again, the two strands of DNA must be separated, this time by a polymerase enzyme. RNA nucleotides then come out of solution, bind to the exposed DNA strand, are polymerised into the chain and then separate from the DNA strand. Only active regions of the DNA are copied into RNA, and active regions code for many types of RNA. A proportion of RNA strands are functional complex molecules. These include the ribosome, and other structures which support RNA translation, as well as catalytic RNA which has specific enzymatic roles in an organism\cite{Alberts2002}.

RNA translation is the process by which the sequence of messenger RNA is then copied into proteins via the ribosome in a process called translation. Thus the structure of proteins is directly coded for by short sections of the DNA sequence.

\begin{figure}
    \centering
    \includegraphics[scale=0.25]{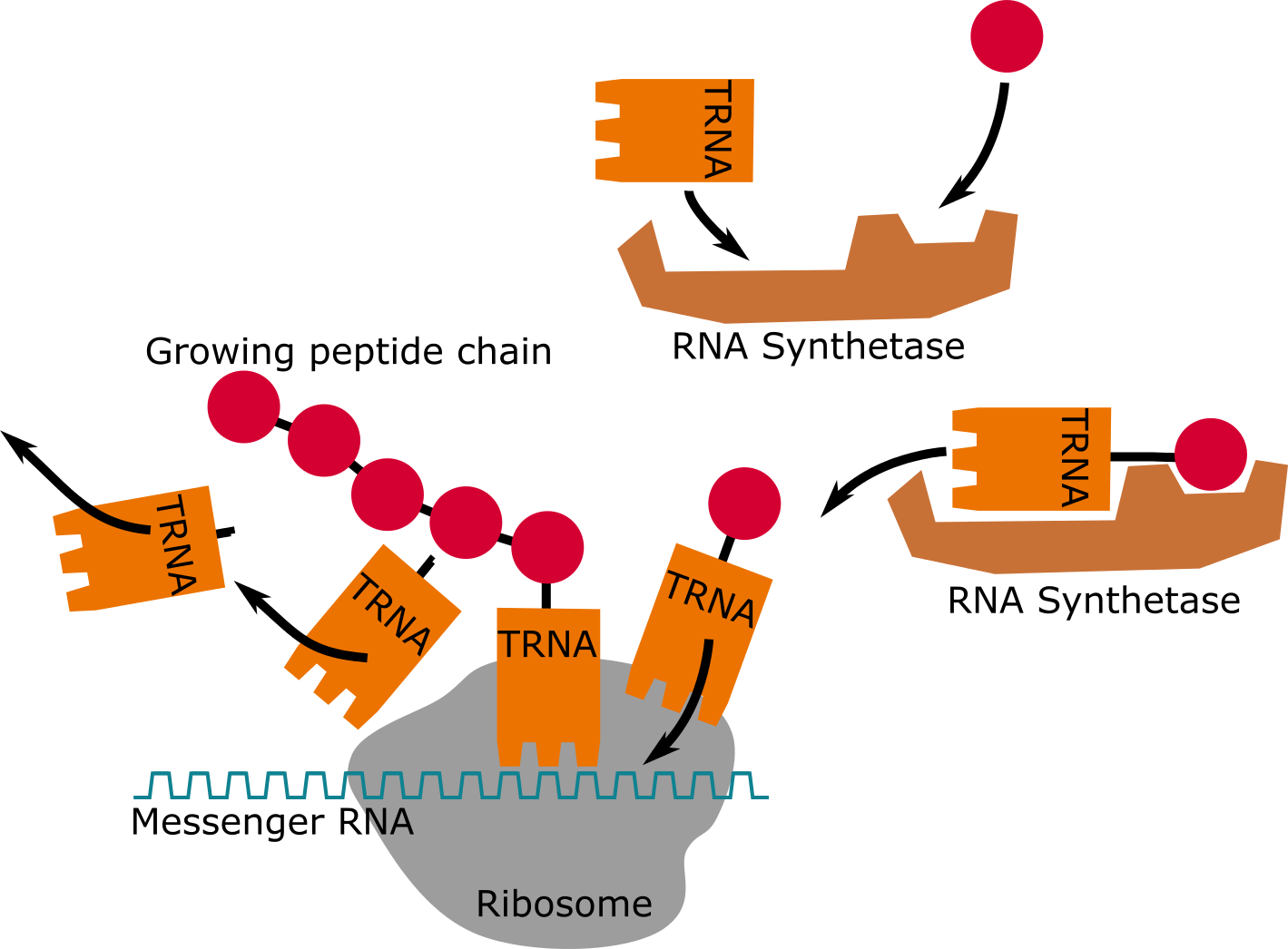}
    \caption{The process of RNA translation is two-fold. First, enzymes called RNA synthetases catalyse the binding of tRNA codons of length three to their matching amino acid. RNA synthetases are specific, they are complementary to a specific pairing of tRNA and amino-acids\cite{gomez2020aminoacyl}. This creates a pool of dimers of one codon and one amino acid. The tRNA/amino acid dimer then binds with the exposed mRNA and the amino acid is bound into the growing protein chain, catalysed nonspecifically by the ribosome. Thus the process of copying RNA into proteins involves both the copying of the information in a synthetase into a dimer, and the information in a long strand of mRNA into a protein.}
    \label{Ribosome}
\end{figure}

Translation is slightly more complex as it is necessary to go from an alphabet of four RNA nucleotides, to one of 20 amino-acids. Instead of a one-to-one mapping as in DNA self-replication and RNA transcription, it takes an RNA codon of length three to encode for a single amino acid. As shown in fig. \ref{Ribosome}, molecules of tRNA which are complementary to the three bases of the exposed mRNA codon, act as intermediaries, binding first to the amino-acid corresponding to the exposed mRNA codon using a synthetase enzyme\cite{gomez2020aminoacyl}, and then to the exposed mRNA codon. This brings the amino-acid alongside the growing protein chain, where it is polymerised into the chain.

It should be noted that the creation of a hybrid tRNA/amino acid is itself a form of templating. The synthetase enzyme is specific and corresponds to  a specific pairing of amino acid and tRNA.

While stabilising interactions between copy and template are central to the generation of accuracy in templated copying, it is important to note that templated copying is only an effective solution to the issue of creating complex molecules if the template can be recycled\cite{o2017evolution}. Without this, we would be faced with the prospect of predesigning the template for every assembly, a challenge no easier to solve than designing the target molecule directly. The DNA sequence acts as a stable store of information, which can be readily accessed by the cell, and which does not have to be created from scratch for every process. Indeed the evolution of DNA genes through successive generations of an organism is itself a form of templating, albeit one with extra steps which cause variation. 

Beyond this, in the vast majority of cases the purpose of templated copying is to create a second functional molecule which can act separately to its template. In the case of DNA self replication, the two DNA strands end up in different cells, and mRNA and proteins are required to be able to function separately from their template within and without the cell. 

This separation of copy and template can happen as part of the process copy creation in the case of RNA transcription and translation, leaving the copy and template separated as a by-product of creating the copy. Alternatively, in the case of DNA, the copying process involves separating the two strands of DNA in the helix (which represent copy and template from the previous round of copying) which are then used as templates to create another double stranded copy/template double helix system. Separation is therefore vital and the very interactions between complementary molecules which promote accuracy, now make it difficult for copy and template to separate\cite{Ouldridge}. Understanding this is key to understanding templated copying.

\section{Synthetic Copying Systems}

The ability to design a synthetic templated copying system, without leveraging existing biological machinery, is an essential step in humanity's understanding of the development of biological complexity. However, success in building synthetic self replicators has been very limited. 

The earliest work focused on the accumulation of monomers on a template and subsequent polymerisation, but did not demonstrate subsequent separation of copy and template\cite{Tjivikua, Feng}. While this does not fulfil our requirements for a templated copying system, instead achieving templated self-assembly, it does have some interesting uses. Templated self-assembly has been explored as a computational paradigm\cite{Leupold}, with information being propagated between layers of a polymer in predetermined ways to generate patterns or perform functions. Erik Winfree is perhaps the best-known name in this field, demonstrating the production of self-assembled and algorithmically patterned two-dimensional lattices of DNA in his PhD thesis\cite{Winfree}, and expanding into other complex algorithms using the same technique\cite{Barish}.

A system by which monomers settle on a template, polymerise and then dissociate from the template completely autonomously, i.e. without external manipulation, has been demonstrated by G. von Kiedrowski's group in their 1994 paper ``Self-replication of complementary nucleotide-based oligomers"\cite{Sievers}. In this system, monomers comprising of three nucleotides bind on a six-nucleotide dimer template, polymerise into a six-nucleotide dimer, and then dissociate. They demonstrate accumulation of product above that caused by the spontaneous binding of the two three-nucleotide monomers into a six-nucleotide dimer in the absence of a template. They acknowledge that product inhibition due to complementarity between copy and template allows at best parabolic growth of dimer concentration. Product inhibition due to cooperativity grows with length\cite{Vidonne, orgel, colomb2015} because the entropy of a single long monomer in solution is much lower than multiple short molecules in solution. It is therefore telling that this paper only succeeds in the accumulation of dimers, and above this, only in a system where the monomers are very short strands of nucleotides.

In order to achieve accumulation of products other than dimers, people have resorted to various novel methods to enforce separation. One method has been used by the group of S. Otto in his paper ``Exponential self-replication enabled through a fibre elongation/breakage mechanism"\cite{colomb2015}. Here the replicating unit is a hexamer, which accumulates into long strands, with the end of the strand acting as a template for the next hexamer. Through constant agitation of the mixture, these strands break, causing an exponential growth of free ends and therefore templates. A similar method has been used by R. Schulman's group\cite{Schulman} in which water currents are use to break apart crystals which build up information in layers.

Other systems require several different reaction environments in order to perform one copying iteration of template interaction, polymerisation and separation. Various groups have achieved the copying of dimers and trimers\cite{Wu}, heptamers\cite{Wang}, and even 24mers\cite{Li}. In ``Chemical self-replication of
palindromic duplex DNA"\cite{Li}, Li and Nicolaou use a double strand of DNA as a template, with DNA fragments as the copy monomers. At low pH these DNA fragments settle onto the copy, a chemical reagent is added in order to bring about the formation of backbone bond, and then the pH is raised in order to force the single strand of DNA off the double strand template. The pH is then lowered and the single strand copy accumulates complementary molecules.  The chemical reagent is added for a second time to encourage backbone bond formation and the system now has two double stranded templates. This process requires three different environments (two of which are used twice) arranged into a sequence of five steps, with external manipulation required to move between each environment. 

A cellular system does not have the luxury of moving between many different environments in order to enforce separation. Therefore, if the aim is to understand the process which underlies the development of biological complexity, some form of autonomy in the process seems essential. After all, as previously stated, biological systems rely on complex processes requiring complex molecules {\it which must self-assemble in an autonomous manner}. In order to understand the development of the complex machinery required for templated copying in a biological setting, it seems essential to probe the design of autonomous synthetic copying mechanisms.

One particularly compelling piece of work is that by Dieter Braun's group\cite{braun2004}. Here they use a spatially non-uniform environment, namely a capillary which has a temperature gradient over it. This sets up a convection current, where monomers settle on a polymer in the cool region, polymerise and then separate in the hot region. They have successfully demonstrated the accumulation of polymers of length 143\cite{Mast2010}, an extremely impressive achievement. Despite still requiring two different environments (hot and cold), because a thermal vent is a plausible candidate for an origin of life scenario\cite{orgel}, this work is of interest to those who wish to probe early life-like systems. It should however be noted that the accumulated polymers are random, and do not reflect the sequence of the seeded polymer.

It is therefore clear, given our failure to achieve synthetic copiers, that we do not fully understand the process of templated copying. This thesis is inspired by an observation that none of the theoretical literature explicitly considers the separation of copy and template, in many cases actively omitting it\cite{Bennett,Cady,Andrieux,Sartori1,Sartori2,esposito2010,EHRENBERG1980333,Johansson}. Given the issue of product inhibition is well-known by those performing experiments in the field, it seems critical that this aspect of copying is included in theoretical models of copying. Previous theoretical works omitting separation have successfully highlighted a range of phenomena, such as entropy-driven growth \cite{Bennett, esposito2010} and the possibility of using kinetic proofreading \cite{Hopfield,ninio} to enhance accuracy \cite{Bennett, Sartori2}. However the failure to consider separation as part of the copying process has led to the erroneous suggestion that there is a so-called ``energetic limit"\cite{Sartori1} in which accuracy can be generated due to permanent thermodynamic stabilisation of complementary monomers in the copy and template. As soon as separation is explicitly considered, it is clear that these stabilising forces must be temporary. My thesis will explore this in more detail.

\section{Theoretical Modelling of Chemical Systems}

\begin{figure}
    \centering
    \includegraphics[scale=0.25]{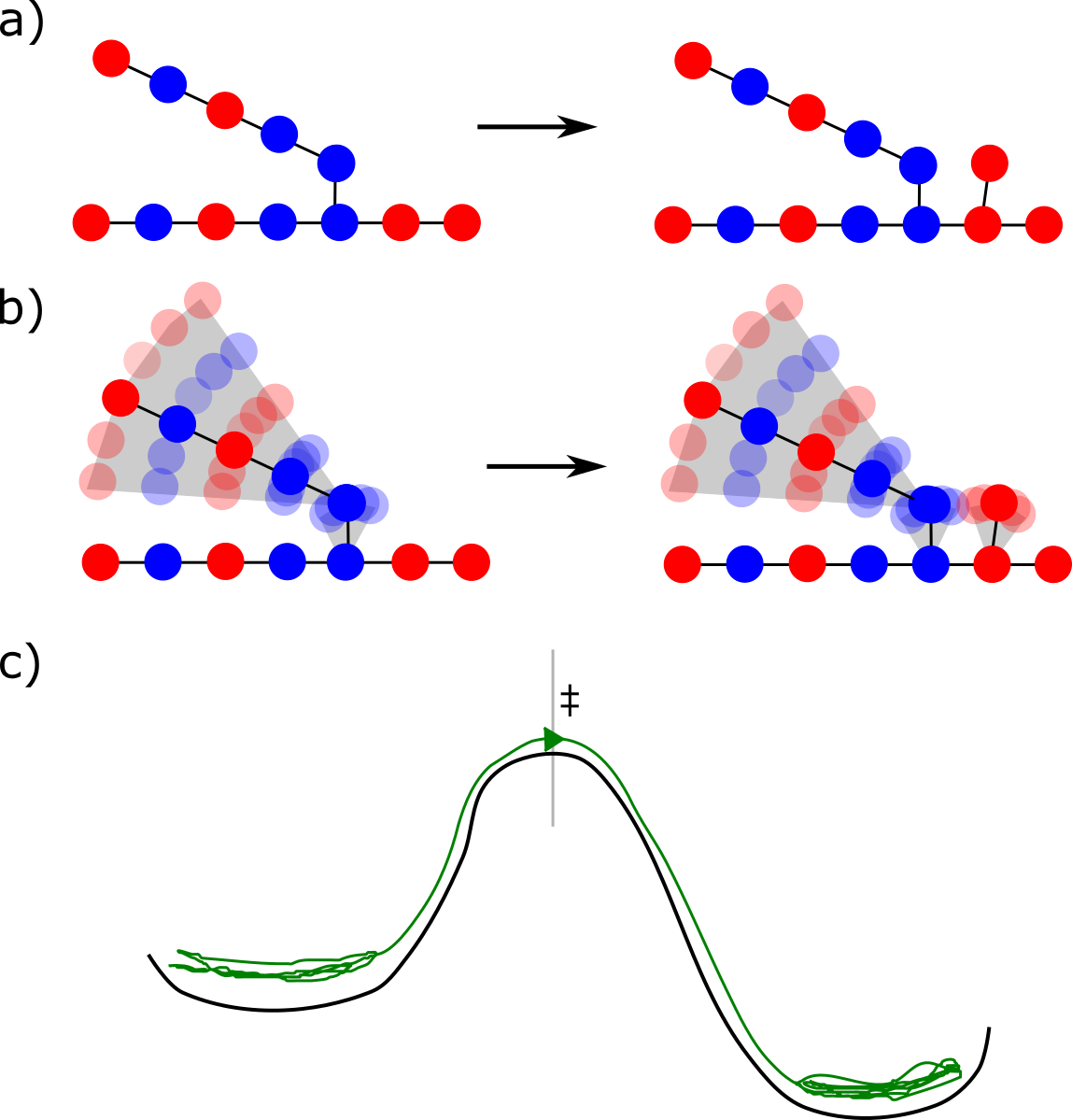}
    \caption{a) A representative configuration of a given pair of neighbouring macrostates, the first a growing copy attached by its final monomer to the template, and the second, in which a monomer has come out of solution and bound to the next empty site on the template. b) Each of these macrostates contains many microstates, for example changes in the physical configuration of the trailing end. c) A hypothetical energy landscape for the transition, with a free-energy barrier between the two states, topped by the ``activated complex" referred to in transition rate theory. A plausible trajectory through the system is shown; the system initially quickly moves between microstates within the left hand basin, before it gains enough energy due to the fluctuating environment to overcome the barrier, releasing the energy as it travels into the right hand stable basin, where it remains.}
    \label{Baths}
\end{figure}

Biological copying systems, such as polymerases interacting with a DNA template during RNA transcription, are inherently unpredictable, or stochastic. In order for the reaction to progress, resources must be acquired from the surrounding environment via collision and it is not possible to predict the exact moment that two molecules collide, even in an environment of relatively consistent resource density. For RNA transcription these resources include RNA nucleotides, the monomer required to build the copy, and ATP, which is a fuel source for the process. Even the process of the polymerase first interacting with the DNA strand must occur stochastically. 

We can imagine a hypothetical energy landscape for the general process of making a templated copy, which starts with a single monomer about to bind to an empty template and finishes with a full-length copy polymer having just separated from the template. Every iteration of the process will move through the energy landscape differently but on average they will behave in predictable ways. The field of stochastic thermodynamics is designed to analyse such systems.

 The free energy landscape for a single copy being built on a single template can be accurately modelled as a series of meta-stable macrostates separated by high free-energy barriers. For example, as shown in fig. \ref{Baths}a, a growing polymer on a template, attached by its final monomer, would be a meta-stable macrostate, as would the same system with a monomer having come out of solution and bound to the template at the next available site. While there are in fact lots of different microstates within each of these meta-stable macrostates e.g. multiple different physical arrangements of the trailing tail (fig. \ref{Baths}b), transitions between these microstates within the meta-stable macrostates are fast, while transitions between meta-stable macrostates are slow (fig. \ref{Baths}c)\cite{peters2017reaction,seifert2011stochastic}. Given this state space, it is reasonable to consider that we are working in a discrete state setting, with our trajectories spending most of their times in the stable basins, and only occasionally moving between neighbouring macrostates. 

Stochastic thermodynamics gives us tools to evaluate discrete state systems which include many biological systems such as molecular motors\cite{sakaguchi2006efficiency,seifert2005fluctuation}, membrane transport\cite{berezhkovskii2008counting}, population dynamics\cite{naasell2001extinction, mangel1993dynamics} and disease spreading models\cite{gomez2010discrete}. There are two broad classes of models with discrete state space; the first is for settings with both discrete state space and discrete time, and the second is for settings with discrete state space but continuous time. While in order to understand the full dynamics of a templated copying system it is necessary to work in the discrete state space and continuous time limit, there are specific cases where it is sufficient to work in the discrete state space and discrete time limit. I will therefore give an overview of methods of solving both classes of system.

In the discrete state space and discrete time case, systems can be analysed by discrete Markov state models. Given that systems spend a long time in meta-stable free-energy basins, with fast transitions between microstates within the meta-stable basin, the system quickly forgets how any particular trajectory entered the meta-stable state. Therefore the next step {\it between} meta-stable basins is dependent only on the current basin, and not on any previous steps taken. This is the condition for a system being Markovian and is true in both the discrete and continuous time cases. If a system has both temporal homogeneity and is ergodic then there will be a unique steady state for a given system, giving the probabilities of being in each of the macrostates of the system in the long time limit\cite{stopnitzky2019physical}. Temporal homogeneity is the property that the probability of a given transition \(i\) to \(j\), \(P_{i\rightarrow j}\) is dependent only on relative time, not on absolute time. This means that the probability that a transition happen in one second is not dependent on whether that is between second 2 and 3 of a process or between second 12 and 13. Ergodicity is the property that it is possible to reach every possible other state from any given state in a finite number of steps, so that there are no regions of the state space that are isolated from others.

In order to find the unique steady state of a system, it is necessary to quantify the probabilities of each possible transition between two meta-stable basins in a certain unit of time. In the discrete time case, we model time as a series of discrete ``ticks" of length \(\Delta t\), after which exactly one transition has happened (including the possibility of a transition to the current state). An example of a truly discrete time system is a turn-based board game; after each round, one tick has happened and each player (each representing a single trajectory) has taken one turn's worth of action with a given probability. Many systems are not truly discrete in time, but can be discretised by integrating the continuous probability of making any given move from any given basin (or failing to make a move) in the time \(\Delta t\). We can use these probabilities in our discrete time Markov state model.

\section{Solving Markov models}

\begin{figure}
    \centering
    \includegraphics[scale=0.4]{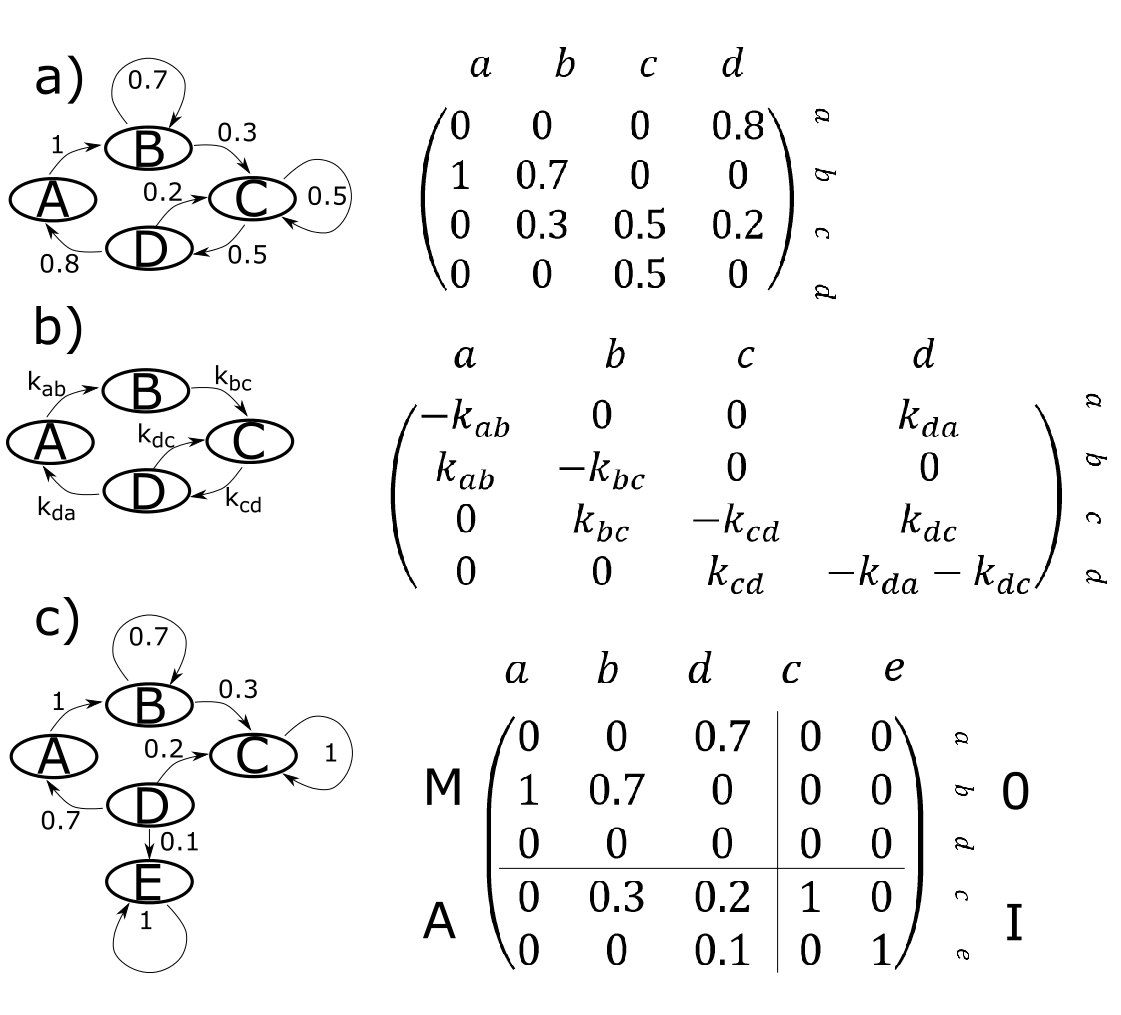}
    \caption{a) The transition matrix for the discrete state-space, discrete time network shown. Note that all quantities are probabilities, and all columns sum to one. The long time limit is the zeroth eigenvector of the matrix. b) The rate matrix for the discrete state-space, continuous time network shown. Note that all quantities are rates, and all columns sum to zero. The long time limit is again the zeroth eigenvector of the matrix. c) The absorbing state matrix for the discrete state-space, discrete time network shown. States C and E are absorbing states, and A, B and E are transient states. Thus \({\bf M}\) is \(3\times 3\) and \({\bf A}\) is \(2\times 3\).}
    \label{Matrix}
\end{figure}

\subsection{Transition matrices}

One common method of solving discrete time Markov state models is by using a transition matrix\cite{peters2017reaction,maruyama2009colloquium}. For a  network which has well defined probabilities for moving between discrete states in a discrete unit of time, each location in the transmission matrix \({\bf T}\) represents the probability of moving from the state represented by the row, to the state represented by the column in that given time, ie \(T_{ij}\) is the probability of moving from state \(j\) to state \(i\). An example transition matrix for a system is shown in fig. \ref{Matrix}a. If we represent a given initial state by a probability vector \({\bf P}(0)\) where the \(i\)th position in \({\bf P}(0)\) represents the probability of starting in state \(i\) then the probability distribution after one tick is just \({\bf P}(\Delta t)={\bf T}{\bf P}(0)\). For a given number of ticks \(n\) the probability distribution is \({\bf P}(n\Delta t)={\bf T}^{n}{\bf P}(0)\) and the long time steady state is just the zeroth eigenvector of \({\bf T}\).

Return to our hypothetical energy landscape for a single copy growing on a single template, which starts with a single monomer about to bind to an empty template and finish with a full length copy polymer having just separated from the template. It is clear than in this case, what we are most interested in is not the probability density of being in any given intermediate state at any given time, but instead the probability of us ending up with one of a number of empty template end states, with either the first monomer dissociating from the template or a full length copy dissociating from the template. Methods such as these also gives us tools to calculate the probability of ending up any given final state, given a starting state (such as the first monomer bound to the otherwise empty template).

\subsection{Absorbing state matrices}

In many systems, what we are most interested in is not the probability density of being in any given intermediate state at any given time, but instead the probability of us ending up with one of a number of designated ``endpoint" states. Stochastic thermodynamics also gives us tools to calculate the probability of ending up any given final state, given a starting state.

To do this we use an absorbing state matrix similar to a transition matrix, in which the probability of leaving the absorbing state is zero\cite{peters2017reaction}. These matrices can be split into four parts as shown in fig. \ref{Matrix}c, known as \({\bf M}\), \({\bf A}\), \({\bf I}\) and \({\bf 0}\). Here, \({\bf M}\) is a \(t\times t\) matrix where \(t\) is the number of transient states, \({\bf T}\) is a \(r\times t\) matrix where \(r\) is the number of absorbing states and \({\bf I}\) and \({\bf 0}\) are identity and zero matrices respectively. It is straightforward to calculate the fundamental matrix \({\bf F}=({\bf I}-{\bf M})^{-1}\) which gives the expected number of times a system passes through state \(i\) given it started in state \(j\). This fundamental matrix \({\bf F}\) can be used to calculate a whole range of properties of the system, including the matrix of probabilities that the system ends in absorbing state \(j\), given it started in transient state \(i\) which is given by \({\bf Z}={\bf F}.{\bf A}\). This is key to our analysis of the creation of short oligomer copies being created with respect to short oligomer templates.

\subsection{Rate matrices: for the continuous time case}

In the continuous time limit we use not a probability based transition matrix, but a rate matrix which defines the master equation for each state\cite{peters2017reaction}. A master equation describes a system in terms of flows in and out of a state. The probability of being in a single state \(i\) at a given time \(t\) is denoted \(P_{i}(t)\). The master equation for the rate of change of the probability of being in state \(i\) is then,
\begin{align}
    \frac{dp_{i}(t)}{dt}=\sum_{j\neq i}W_{j\leftarrow i}P_{j}(t)-\sum_{j\neq i}W_{i\leftarrow j}P_{i}(t).
\end{align}
where the transition frequency from state \(i\) to state \(j\) is given as \(W_{j\leftarrow i}\)\cite{seifert2005entropy}. We can roll the entire set of master equations for a system into a rate matrix \({\bf W}\) (fig. \ref{Matrix}, b), where \(W_{ij}=W_{i\leftarrow j}\) is the flow from state \(j\) to state \(i\) for \(i\neq j\) and where the diagonals \(W_{ii}\) are the sum over the negative of the other values in the column\cite{stopnitzky2019physical}. The solution to the master equation \(\frac{d{\bf P}}{dt}=-{\bf W}{\bf P}\) is just \({\bf P}(t)=e^{{\bf W}t}{\bf P}(0)\), where \({\bf P}\) is the vector of probabilities, so by diagonalising \({\bf W}\) we can straightforwardly find the steady state of the system\cite{stopnitzky2019physical}.

\subsection{Embedding}

It is possible to embed a continuous time process into a discrete time process, which will have the same probability of visiting a given sequence of states, but may not hold any information about the timing of the process. This most commonly done in one of two ways, one in which the properties of the continuous time process is broken down into a series of discrete ``ticks", and one in which normalised rates are used as probabilities of taking a step. 

In the first of these two methods, the continuous system is integrated over a defined time period in order to calculate the probabilities of any given transition, or failure to transition. These probabilities are then used as the probabilities in the discrete time transition matrix. This retains some information about timings, albeit limited to the resolution of the length of a single ``tick".

The second method is arguably simpler than the first. Here for a given state, the rates of transitioning out of that state are normalised and used as probabilities in the discrete time transition matrix. This method can give the same probability for a given sequence of transitions and states as the continuous time case, but all information about timings is lost\cite{peters2017reaction}.

\subsection{Monte Carlo simulations}

It can often be the case that systems are not analytically tractable using the various matrix methods above. Systems with more than a small number of intermediate states quickly become ungainly to work with analytically. In these cases we can use Monte-Carlo simulations of a process to generate many individual stochastic trajectories, the average properties of which should converge on our analytical results for a sufficiently large amount of trajectories\cite{peters2017reaction}.

The basic mechanism of a Monte-Carlo simulation is straightforward. Like the transition matrices above, we start by defining the probabilities of where the system will move to at the next step, given your current position. We then draw a random number, and use this to choose which of the possible steps it might take. For example if from your current state you can go to one of 2 states with equal probability, a random number of less than or equal to \(\frac{1}{2}\) would take you to the first state, and of greater than \(\frac{1}{2}\) would take you to the second state. You then update the time, by adding an exponentially distributed random number, distributed around the inverse of the sum of the rates from the current state to all adjoining states \(\frac{1}{\sum_{j} k_{ij}}\) where \(i\) is the current state. If a system is initialised in a given state, it is possible to observe the trajectory it traces through the different states. Multiple iterations of this allows you to calculate the average properties of such a system.

\subsection{Using methods in combination}
It is often the case in copying systems that several of these methods will be used in combination. As an example an absorbing state matrix being used to calculate the final state probabilities of an underlying network which represents a small part of a more complex systems, which are the used in a further analytical calculation or simulation of a higher level system. This is especially useful for systems with a high level of modularity.

Over the next sections I will outline some important concepts in stochastic thermodynamics and information theory, which give us constraints on our models, and allow us to interpret them as physically consistent systems.

\section{Stochastic Thermodynamics}

\subsection{Thermodynamics of trajectories}

Consider a specific trajectory, which encompasses several steps, in a system which is in a stationary free energy landscape. Throughout this subsection we are not considering systems undergoing time dependent protocols unless explicitly stated otherwise, because we are interested in autonomous systems typical of biochemistry. Here the system starts in state \(0\), and waits until time \(t_1\) until transitioning to state \(1\), and then waits until time \(t_2\) until transitioning to state two, all the way up to waiting time \(t_{n}\) to transition from state \(n-1\) to n. There exists a time reversal of this trajectory in which the system starts in state \(n\), and waits time \(t_{n}\) to transition to state \(n-1\) and so on, all the way back to waiting time \(t_1\) to transition from state one to zero\cite{parrondo2015thermodynamics,jarzynski2011equalities}.

In the case that this trajectory is markovian, as is the case throughout this thesis where the system undergoes many fast transitions between microstates within a meta-stable macrostate, and only rarely transitions between meta-stable macrostates, then the probability of the forward and backward trajectories, given their starting states are,
\begin{equation}
    P_{0\rightarrow n|0}(t)=P_{0}({\rm wait}=t_{0})P_{{0}\rightarrow {1}}\times P_{1}({\rm wait}=t_{1})P_{{1}\rightarrow {2}}\times ...\times P_{n-1}({\rm wait}=t_{n-1})P_{{n-1}\rightarrow {n}},\\
\end{equation}
\begin{equation}
    P_{0\rightarrow n|0}(t)=\prod_{i=1}^{n}P_{i-1}({\rm wait}=t_{i-1})P_{{i-1}\rightarrow {i}},\\
\end{equation}
for the forward trajectory and
\begin{equation}
   P_{n\rightarrow 0|n}(t)=P_{n}({\rm wait}=t_{n})P_{{n}\rightarrow {n-1}}\times P_{n-1}({\rm wait}=t_{n-1})P_{{n-1}\rightarrow {n-2}}\times ...\times P_{1}({\rm wait}=t_{1})P_{{1}\rightarrow {0}},\\
\end{equation}
\begin{equation}
  P_{n\rightarrow 0|n}(t)=\prod_{i=1}^{n}P_{i}({\rm wait}=t_{i})P_{t_{i}\rightarrow t_{i-1}},\\
\end{equation}
for the reverse trajectory. Here \(P_{n}({\rm wait}=t_{n})\) is the probability of the system spending time \(t_n\) in state \(n\) before transitioning and \(P_{n\rightarrow n+1}\) is the instantaneous probability of transitioning from state \(n\) to \(n+1\), which is identical to the rate of transition \(k_{n\rightarrow n+1}\). Each individual transition in the path corresponds to an amount of entropy released into the environment. We denote the individual contributions to the ``environmental entropy" of the path \(\Delta s^{i+}_{\rm env}=\ln{\frac{P_{n\rightarrow n+1}}{P_{n+1\rightarrow n}}}\) for the individual reactions in the forward trajectory or \(\Delta s^{i-}_{\rm env}=-\Delta s^{(i+1)+}_{\rm env}\) for the reverse trajectory. Thus for the forward trajectory the total entropy released into the environment \(\Delta S_{\rm env}=\sum_{i=0}^{n}\Delta s^{i+}_{\rm env}\). This entropy change accounts for the entropy due to the transitions between states, but it is also necessary to consider the individual entropies of the initial and final states. Because the system has been coarse grained, the initial and final states are macrostates which themselves contain multiple microstates. There is thus an internal entropy of the initial and final states, and we denote the difference between them \(\Delta S_{\rm mac}\)\cite{van2015ensemble}. 

This leads us neatly to fundamental assumption of stochastic thermodynamics; that the relative probabilities of a pair of forward and backward trajectories are given by
\begin{equation}
    k_{B}\ln{\frac{ P_{0\rightarrow n|0}(t)}{P_{n\rightarrow 0|n}(t)}}=\Delta S_{\rm env + mac}=\Delta S_{\rm env}+\Delta S_{\rm mac}.
\end{equation}
Here \(\Delta S_{\rm env}\) is the entropy change of the environment due to the forward trajectory and  \(\Delta S_{\rm mac}=S(z(t))-S(z(0))\) is the difference between the macrostate entropy of the macrostate in which the trajectory begins and the entropy of the macrostate in which the trajectory ends\cite{crooks1999entropy,seifert2005entropy,Esposito,ouldridge2018importance,seifert2011stochastic,parrondo2015thermodynamics,jarzynski2011equalities}. In a system without coarse graining, \(\Delta S_{\rm mac}\) would be zero because a microstate has no internal entropy.  Observe that in the ratio \(\frac{ P_{0\rightarrow n}(t)}{P_{n\rightarrow 0}(t)}\) the waiting times cancel, leaving just the ratio of the rates. This equation is basis of a large part of stochastic thermodynamics, from which you can derive fluctuation relations, uncertainty relations and the second law\cite{ouldridge2018importance}.

As a result the rates can be given by
\begin{align}
    \frac{k_{\rm forward}}{k_{\rm backward}}=e^{\Delta S_{\rm env + mac}},
    \label{reverse}
\end{align}
where \(k_{\rm forward}=\prod_{i=1}^{n}k_{i-1\rightarrow i}\) and \(k_{\rm forward}=\prod_{i=1}^{n}k_{i\rightarrow i-1}\)\cite{seifert2011stochastic}. Given that a simple molecular system such as a templated copying system can only exchange energy in the form of heat with its environment, then the entropy change of the environment takes the form of heat dissipated into the environment (the canonical ensemble), ie \(T\Delta S_{\rm env}(z(t))=\Delta U_{\rm env}=-\Delta U_{\rm mac}\)\cite{ouldridge2018importance}, where \(\Delta U_{\rm mac}\) is the heat dissipated free energy difference between the initial and final macrostate. This allows us to write the entropy change of the environment and the macrostate as
\begin{align}
    \Delta S_{\rm env + mac}=T\Delta S_{\rm env}+T\Delta S_{\rm mac},\\
     \Delta S_{\rm env + mac}=-\Delta U_{\rm mac}+T\Delta S_{\rm mac},\\
     \Delta S_{\rm env + mac}=T\Delta S_{\rm env}+T\Delta S_{\rm mac}\\
     \Delta S_{\rm env + mac}=-\Delta G_{\rm mac},
\end{align}
where \(\Delta G_{\rm mac}\) is the change in free energy of the system due to changing from one macrostate to another\cite{ouldridge2018importance}. This tells us that for any trajectory, including individual transitions,
\begin{align}
    \frac{k_{\rm forward}}{k_{\rm backward}}=e^{-\Delta G_{\rm mac}}.
\end{align}
This result will underly the design of the models throughout this thesis.

\subsection{Thermodynamic reversibility}

\begin{figure}
    \centering
    \includegraphics[scale=0.3]{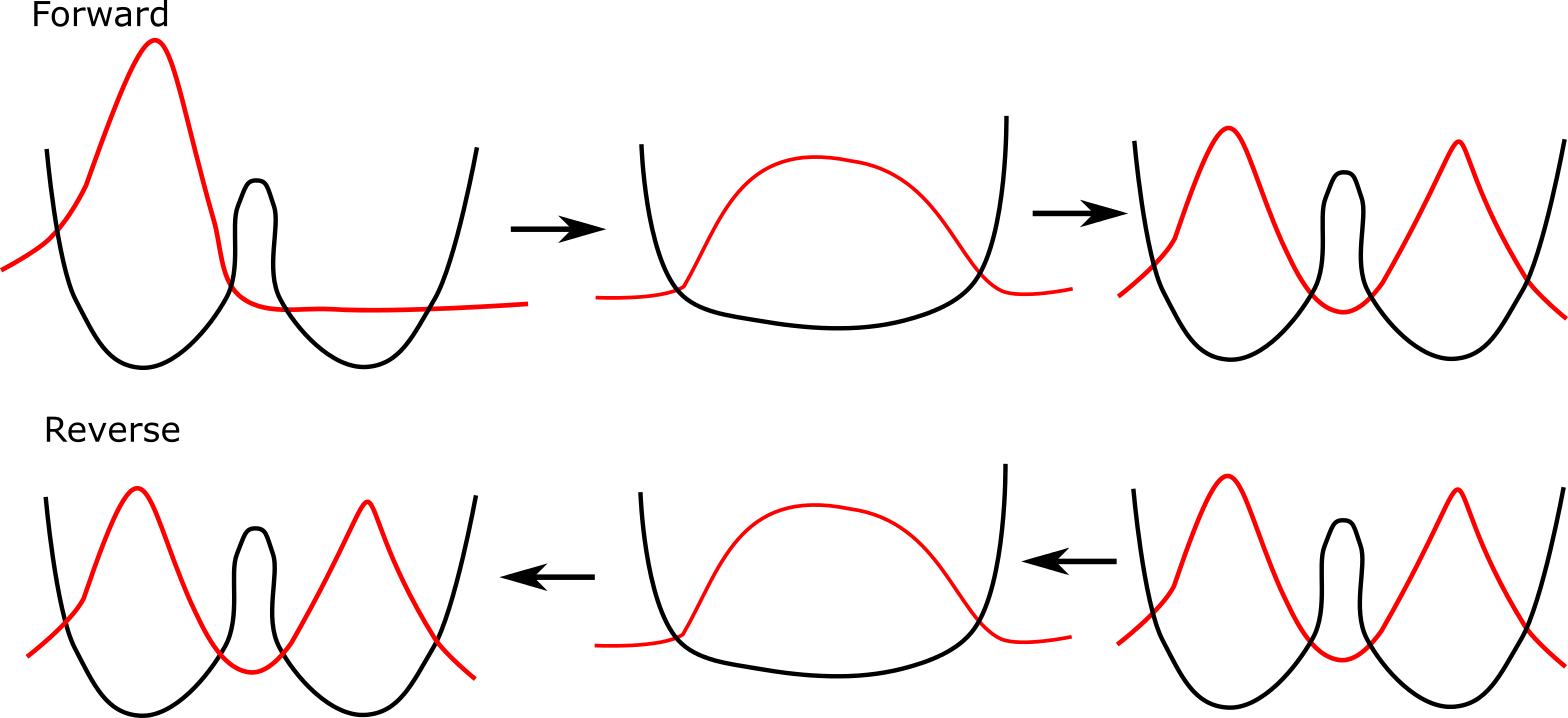}
    \caption{The probability distributions (red) for system with a time varying free energy landscape (black). The system shown undergoes the following protocol; the system is switched to a single monostable well by lowering the free energy barrier, allowing the system to equilibrate, and then the barrier is raised. If this system is initialised with all trajectories in the left hand well, then after this protocol half will be in the right hand well and half will be in the left hand well. If we reverse this protocol starting with the trajectories in the mixed final state, we will not return to all trajectories being in the left hand well, but instead the trajectories would remain with half in the left and half in the right.}
    \label{reversible}
\end{figure}
It is worth taking a moment to talk about different types of reversibility. {\em Microscopic reversibility} is related to eq. \ref{reverse}. It requires that if there is a trajectory that moves through a system via a specific path, that there is an equivalent trajectory which can move through the system along the same path, but in the opposite direction. The probability of the two trajectories is related by eq. \ref{reverse}. This concept is closely related to that of detailed balance, which gives the probability of changing states \textit{without reference to a specific pair of trajectories}. It is characterised by the relation that for states \(a\) and \(b\) that the instantaneous probabilities of transitioning between the states are related by \(P(a\rightarrow b)=P(b\rightarrow a)e^{\beta\Delta G}\), where \(\Delta G\) is the difference in free energy between states \(a\) and \(b\)\cite{crooks1999entropy,Esposito,jarzynski2011equalities}. 

{\em Thermodynamic reversibility} is a property of an ensemble of transitions, which are moving through a system as it undergoes a protocol. An example of a protocol illustrated in fig. \ref{reversible} could be on a double well potential. The protocol would be the switching of the system to a single monostable well by lowering the free energy barrier, allowing the system to equilibrate, and then raising the barrier again. If this system is initialised with all trajectories in the left hand well, then after this protocol half will be in the right hand well and half will be in the left hand well. If we reverse this protocol starting with the trajectories in the mixed final state, we will not return to all trajectories being in the left hand well, but instead the trajectories would remain with half in the left and half in the right. Thus the process is thermodynamically irreversible This is despite the fact that some individual trajectories will return exactly to the state they came from (microscopic reversibility). Thermodynamically reversible process do not increase the entropy of the universe, but processes which are irreversible do\cite{ouldridge2018importance,seifert2011stochastic,landauer1961irreversibility,jarzynski2011equalities}.

In the case of a copying system our protocol is very simple; merely ``do nothing". This is because our trajectories are moving through an unchanging free energy landscape. In these cases, the system is inevitably thermodynamically irreversible if a non-zero net evolution occurs\cite{ouldridge2018importance}. 

Returning to 
\begin{equation}
    k_{B}\ln{\frac{ P_{0\rightarrow n|0}(t)}{P_{n\rightarrow 0|n}(t)}}=\Delta S_{\rm env + mac},
\end{equation}
and noting that \(P_{0\rightarrow n|0}(t)\) is conditional on the system starting in state zero. If we add \(k_{B}\ln{\frac{P_{0}(0)}{P_{n}(t)}}\), where \(P_{n}(t)\) is the probability of being in state \(n\) at time \(t\), to both sides of the equation we get
\begin{equation}
    k_{B}\ln{\frac{ P_{0\rightarrow n}(t)}{P_{n\rightarrow 0}(t)}}=\Delta S_{\rm env + mac}+ k_{B}\ln{\frac{P_{0}(0)}{P_{n}(t)}},
\end{equation}
where \(P_{0\rightarrow n|0}(t)P_{0}(0)=P_{0\rightarrow n}(t)\) in general\cite{seifert2005entropy,seifert2011stochastic}. 
If we now average over all possible trajectories \(0\rightarrow n\) then we get,
\begin{equation}
    k_{B}\sum_{0,n}P_{0\rightarrow n}(t)\ln{\frac{ P_{0\rightarrow n}(t)}{P_{n\rightarrow 0}(t)}}=\sum_{0,n}P_{0\rightarrow n}(t)\Delta S_{\rm env + mac}+ k_{B}\sum_{0,n}P_{0\rightarrow n}(t)\ln{\frac{P_{0}(0)}{P_{n}(t)}},
\end{equation}
\begin{equation}
    k_{B}\sum_{0,n}P_{0\rightarrow n}(t)\ln{\frac{ P_{0\rightarrow n}(t)}{P_{n\rightarrow 0}(t)}}=\Delta S_{\rm env + mac}+ k_{B}P_{0}(0)\ln{P_{0}(0)}-k_{B}P_{n}(t)\ln{P_{n}(t)},
\end{equation}
The term on the left hand side is the Kullback Leibler divergence between the distribution of forward and backward trajectories. This quantity is discussed in more detail below, but essentially is a non-negative quantity which quantifies how different two distributions are. The first term on the right hand side is the entropy change of the environment and the macrostates and the rightmost two terms are the differences between the Shannon entropy of the initial and final states. Thus the right hand side is the total entropy change of the system during the trajectory. Shannon entropy is also discussed in more detail below but is functionally identical to thermodynamic entropy.

Thus the right hand side is exactly the total entropy production for a trajectory. Due to Jenson's inequality\cite{cover1999elements}, the left hand side is non-negative, and only reaches zero when the process is thermodynamically reversible, ie forwards and backwards transitions have the same probability.\cite{parrondo2015thermodynamics}. Thus, a thermodynamically irreversible process generates entropy.

\subsection{Transition State Theory}

In defining the rates of reaction, it is important to understand what a transition entails. In order to leave the reactant basin, the system must gain enough free energy to reach the top of the barrier for long enough to cross the barrier, only to lose it again as it progresses to the stable product basin. It is assumed that it takes much longer for the system to fluctuate in such a way that it gains enough free energy to leave the reactant basin than it does to relax to the product basin. This is called the separation of timescales.

Transition state theory\cite{peters2017reaction} can be used to estimate the value of the rates of transition \(k_{a\rightarrow b}\) or \(k_{b \rightarrow a}\) between states \(a\) and \(b\). It considers a hypothetical dividing surface, which separates states \(a\) and \(b\). In the common case where the top of the energy barrier is a saddle point this surface is represented by a single ``activated state'' \(\ddagger\) as shown in fig. \ref{Baths}.  Transition rate theory assumes that states on the dividing surface are at equilibrium with the reactants. It further assumes that once states cross the dividing surface, they proceed to the formation of products without ``recrossing'' the dividing surface. The reaction \(a+b\rightarrow c+d\) can be written \(a+b\rightleftharpoons \ddagger \rightarrow c+d\). The rate of reaction is therefore \(k_{forward}=[\ddagger]k_{\ddagger\rightarrow {\rm prod}}\) where \(k_{\ddagger\rightarrow {\rm prod}}\) is the rate at which transition states become products and \([\ddagger]\) is the concentration of transition rates species. The rate of the reaction is
\begin{align}
    k_{\rm forward}=Ae^{-\frac{\Delta G^{\ddagger}}{k_{B}T}}
\end{align}
where A is a constant depending on a reference volume, \(V_{0}\) and \(\Delta G^{\ddagger}\) is the height of the energy barrier\cite{peters2017reaction} from the bottom of macrostate \(a\). Given the difficulty (or perhaps impossibility) of creating a reaction surface which is never recrossed, transition rate theory provides an upper bound on the rate. Transition state theory is most useful in deciding relative rates of similar processes, for which uncertainties in \(A\) cancel.

\section{Information Theory}

The key parameter in information theory is the Shannon Entropy\cite{shannon2001mathematical,cover1999elements}, which is defined as \(H=\sum_{p_{i}}p_{i}\ln{p_{i}}\) where \(p_{i}\) is the normalised probability of an outcome of a random process. It is a measure of uncertainty and is directly analogous to the thermodynamic entropy used above, different only by a factor of \(k_{B}\).

In general, when studying templated copying systems we are interested in parameters that describe the sequences of the produced polymers and how they compare to the sequences of template polymers. Both of these quantities are random variables, because they are variables where the value depends on the outcome of a process which itself contains randomness or stochasticity. We use specific quantities which describe how the two sequences relate to each other.

The error in the sequence and the mutual information between copy and template are both quantities which compare two random variables. The error \(\epsilon\) is the proportion of mismatches incorporated into a chain and the mutual information is \(I(X;Y)=\sum_{X,Y}\frac{P(x,y)}{P(x)P(y)}\) where \(X\) and \(Y\) are random variables such as the sequence of a template polymer and a copy polymer, gives the reduction in Shannon entropy about \(Y\) having measured \(X\) and vice versa\cite{mcgrath2017biochemical,sagawa2012fluctuation,horowitz2014thermodynamics}. Here \(P(x,y)\) is the joint probability of random variables \(X\) and \(Y\) and \(P(x)\) and \(P(y)\) are their marginal probabilities\cite{cover1999elements}. These quantities are related to the distribution of the quantities \(x\) and \(y\).  Because the value of \(X\) and \(Y\) are dependent on the outcome of a stochastic process, it can be useful to express these quantities as a distribution, \(P(x)\) is the probability that a randomly selected trajectory would give outcome \(X=x\)\cite{mcgrath2017biochemical}, equally the joint probability \(P(x,y)\) is the probability that a randomly selected trajectory would give outcome \(X=x\) and \(Y=y\) and the conditional probability \(P(x|y)\) is the probability that a randomly selected trajectory from the pool of trajectories that give \(Y=y\) will also give \(X=x\)\cite{jarzynski2011equalities}.

\begin{figure}
    \centering
    \includegraphics[scale=0.5]{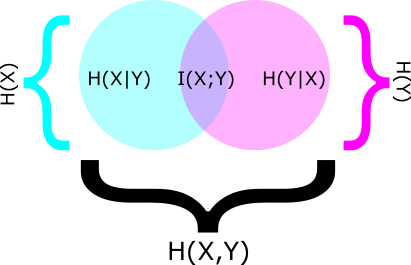}
    \caption{A venn diagram showing the additive relationship between various forms of entropy and mutual information. We can straightforwardly see from here that \(I(X;Y)=H(X)-H(X|Y)=H(Y)-H(Y|X)\) or \(I(X;Y)=H(X,Y)-H(X|Y)-H(Y|X)\) etc.}
    \label{Venn}
\end{figure}

In more detail, Shannon entropy and mutual information are related as follows; \(I(X;Y)=H(X)-H(X|Y)=H(Y)-H(Y|X)\)\cite{cover1999elements,mcgrath2017biochemical,sagawa2012fluctuation,horowitz2014thermodynamics,parrondo2015thermodynamics,jarzynski2011equalities}. Fig. \ref{Venn} shows a Venn diagram representation of entropy and mutual information. We can see that the mutual information is the intersection of the two circles and the joint entropy is the the area of the two overlapping circles. It is therefore clear that as the mutual information increases, then the joint entropy must decrease.  In our copying analogy if we define \(X\) as the monomer type at position \(n\) in the copy polymer and \(Y\) as the monomer type at position \(n\) in the template, then the more accurate the copy, the more the mutual information \(I(X;Y)\) increases and the joint entropy H(X,Y) decreases. 

The joint entropy can also be expanded to the entropy rate. The entropy rate is the limit of the average of the joint entropy over all variables in the system, ie \(H(Z)=\lim_{n\rightarrow\infty}\frac{1}{n}H(Z_{0},Z_{1},...,Z_{n})\), where \(Z_{i}\) is a random variable set by the system\cite{cover1999elements,still2012thermodynamics,jarzynski2011equalities}. As an example, for a polymer sequence, each monomer of which represents a random variable, the entropy rate is capable of taking into account correlations between monomers of any length.

Another useful quantity is Kullback-Leibler divergence \(D(P||Q)\), which quantifies the distance between two probability distributions \(P(x)\) and \(Q(x)\) of random variable \(X\)\cite{cover1999elements,jarzynski2011equalities}. For example we might quantify This for example could quantify the distance between the distribution of polymers in the baths coupled to the system and the distribution of copy polymers produced by the system.

\section{Thermodynamics of information}

Having outlined some of the information theoretic quantities used in this thesis, it is worth exploring the thermodynamics of information. Historically the thermodynamics of information has centred around the study of a paradox known as ``Maxwell's demon" and a hypothetical engine called a Szilard engine, designed to provide a framework in which to exorcise the demon.

\begin{figure}
    \centering
    \includegraphics[scale=0.2]{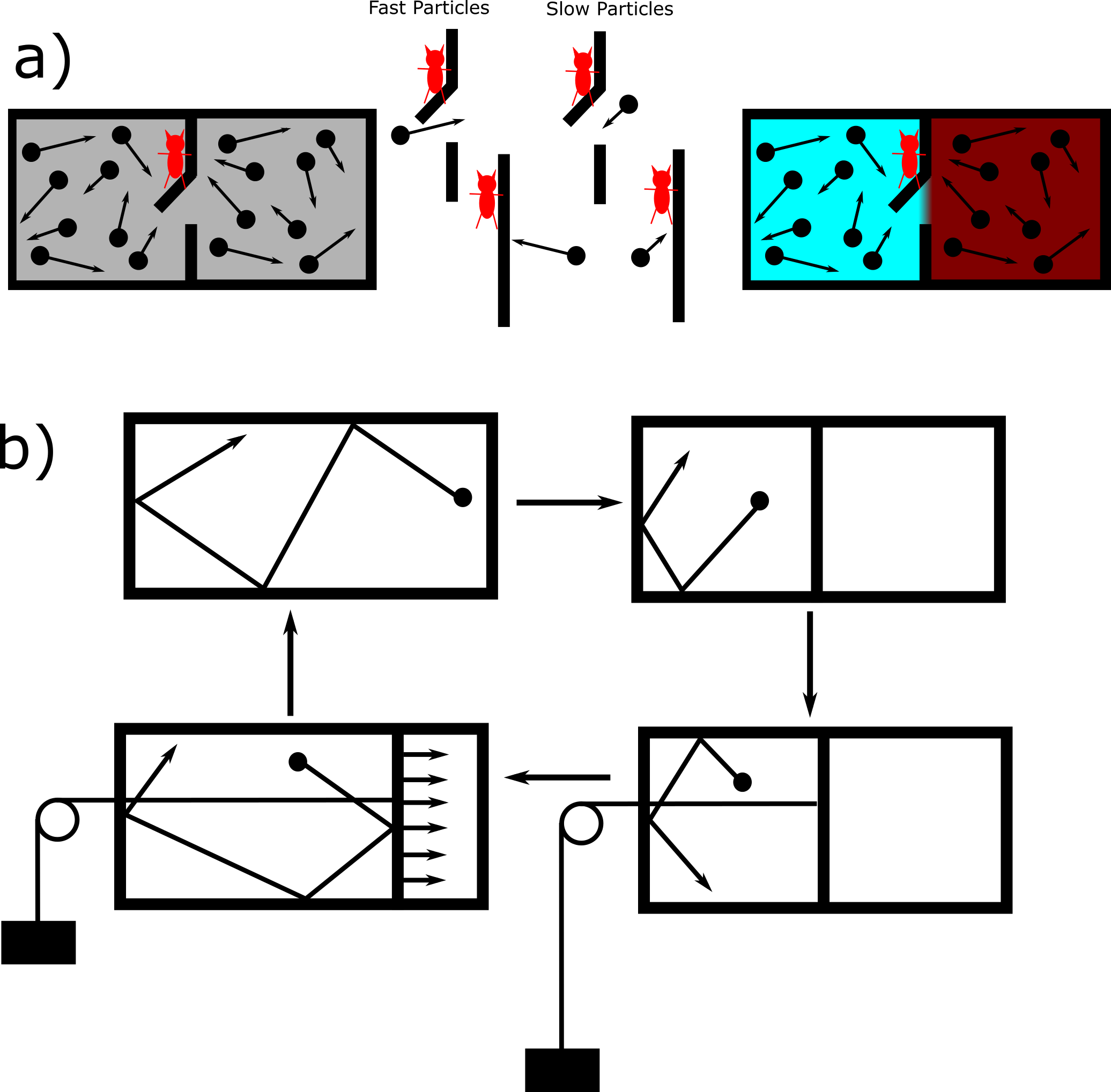}
    \caption{(a) The Maxwell's demon paradox\cite{maxwell1867letter}. A box containing a uniform gas at equilibrium is partitioned down the middle. That partition contains a door which is operated by a ``demon". Without spending any energy, the demon can selectively allow particles to cross the partition. It allows fast particles from left to right, while blocking slow particles, and allows slow particles from right to left, while blocking fast particles. This results in the system now having a cold section and a hot section, which can hypothetically be used to do work. Thus the system has gone from an equilibrium state, to an out-of-equilibrium state without work being done on it, breaking the second law. (b) The solution to the Maxwell's demon paradox is best understood by simplifying the system into a Szilard engine\cite{szilard1929reduction}. Here a box contains a single particle gas. A partition is dropped into the box, confining the particle to one side of the box. The particle then pushes the partition back to the edge of the box. If a weight is coupled to the partition, then the particle can lift the weight doing a maximum of \(k_{B}T\ln{2}\) of work. This problem is identical to that posed by Maxwell, fluctuations in an equilibrium gas are being harnessed to do work, without a corresponding input of work. The key to exorcising the demon comes from the observation that in order to extract the work, the position of the pulley must be correlated with the position of the particle. If it was always attached on e.g. the right hand side, then half the time the particle would lift the weight, but half the time it would lower it, on average doing no work. The minimum cost of correlating two bits of information is \(k_{B}T\ln{2}\), exactly the same as the maximum amount of work that can be extracted from the system. Thus the second law is restored and the demon is exorcised.}
    \label{demon}
\end{figure}

The Maxwell's demon paradox was first outlined by James Clerk Maxwell in 1867\cite{maxwell1867letter}. He suggested that the second law of thermodynamics could be broken using the following method. Illustrated in fig. \ref{demon}(a) is a box containing a uniform gas. This box is partitioned down the middle, and the partition contains a door operated by a demon. Although the temperature of the box is initially the same on both sides of the partition, it is in theory possible for the demon to open the door and allow through any fast moving particle to the left box, and to allow slow particle through to the right box, and block all others. If we assume that the demon can act for free; which is reasonable because it does no work against a force, then this would cause the left box to heat relative to the right box without an input of free energy. Thus work can be extracted from the box which was previously at equilibrium. This breaks the second law of thermodynamics.

In 1929, Leo Szilard came up with a model system which illustrates this issue without the invocation of a difficult to quantify ``demon". The Szilard engine\cite{szilard1929reduction}, illustrated in fig. \ref{demon}(b), works as follows. Consider a gas made up of a single particle in a box. If a partition is instantaneously dropped into the box, the single particle will now be trapped on one side of the partition. The particle will now expand against the partition, pushing it back to the edge of the box. If the partition is coupled to a pulley and weight, then this system is able to perform a maximum of \(k_{B}T\ln{2}\) of work in raising the weight. This framework allows us to identify that this system is in fact exploiting fluctuations in the particles position in order to extract work\cite{mandal2012work}. This is however absolutely forbidden by the second law.

The solution to this paradox has been much discussed\cite{parrondo2015thermodynamics,maruyama2009colloquium,leff2003maxwell,mandal2012work,stopnitzky2019physical,barato2014unifying,sagawa2012fluctuation,landauer1961irreversibility}, but the simplest explanation comes from the observation that in order to extract work from the system, it is necessary to correlate the position of the particle with the position of the pulley and weight. If the pulley and weight were always attached to the left hand side of the partition, or if the attachment was assigned randomly, then 50\(\%\) of the time the system would raise the weight and 50\(\%\) of the time the system would lower the weight; on average doing no work. Therefore the particle and pulley {\it must} be correlated in order for the engine to work as designed. Correlating two bits of information costs a minimum of \(k_{B}T\ln{2}\) of work, the same amount that can be extracted by the engine. This can be understood by considering that correlating the position of the the piston and the position of the particle is increasing the mutual information to the extent that the joint entropy becomes just the uncertainty in the position of the particle. Reducing the joint entropy of the system requires an input of free energy. Thus the second law is restored and the demon is exorcised.

It should be noted that if the pulley and the particle were correlated via a direct physical interaction, such as an attractive force, then the system would not be able to work as designed. In order for the engine to act as an  as in fig. \ref{demon} it is necessary for the pulley to remain coupled to the left hand side, even as the particle moves into the right hand side in the process of moving the partition. It is necessary to instead create a {\em temporary} coupling via a complex mechanism, such as performing a measurement. This is in neat analogy to the process of self assembly in which interaction have to be predesigned and permanent, and the process of templated copying in which interactions that generate correlations are via a template catalyst which the copy must ultimately separate from.

A class of machines known collectively as information engines have been proposed which can extract work from correlations, or spend work to generate correlations. These often take the form of a tape of bits\cite{mandal2012work,stopnitzky2019physical,barato2014unifying} interacting with a heat bath at equilibrium. Through various mechanisms which use correlations as fuel, the system is able to use this equilibrium heat bath to do work, such as lift a weight. The flaw in many of these systems however is that they leave the details of the interactions somewhat abstract, effectively mystifying correlations as a source of fuel.

The paper ``Biochemical Machines for the Interconversion of Mutual Information and Work"\cite{mcgrath2017biochemical} seeks to demystify information engines as a concept by proposing a physically realisable system based around an enzyme, which when it is in its active state, either extracts fuel by converting ATP into ADP while activating a complex \(X\rightarrow X^{*}\) or pumps fuel by converting ADP into ATP while deactivating a complex \(X^{*}\rightarrow X\). A pair of tapes \(A\) and \(B\) are simultaneously pulled past the enzyme. Tapes \(A\) and \(B\) are correlated, and tape \(B\) dictates whether or not the enzyme can interact with the tape. The system is set up so that tape \(B\) allows tape \(A\) to interact with the enzyme only when tape \(A\) contains an \(X^{*}\). The system pumps fuel by converting ADP into ATP while deactivating a complex \(X^{*}\rightarrow X\). Thus by exploiting the correlations between \(A\) and \(B\), with tape \(B\) only exposing the tape \(A\) to the enzyme when \(A\) contains an \(X^{*}\), the system can pump fuel. An explicit design for this system is then proposed using DNA origami. This very concrete example of an information engine makes explicit many things which are left abstract in more theoretical versions of information engines\cite{mandal2012work,stopnitzky2019physical,barato2014unifying}.

\section{Thermodynamics of copying}

Having defined many of the tools needed to study simple molecular systems, we now explicitly discuss the properties of templated copying systems. We start by considering an isolated bath of polymers. We make the assumption that all backbone bonds are identical. While this is not technically true for biological polymers, it is true that some backbone bonds are stronger than others and this could be leveraged if targeting a specific sequence (ie. always copying the same polymer), this is not beneficial for generating accuracy while copying a generic polymer.

This isolated bath of polymers has a macroscopic free energy, which is a property of the whole bath;
\begin{align}
    G(p({\bf s}))=U(p({\bf s}))-TS(p({\bf s})).\label{Free}
\end{align}
Here \(G\) is the free energy which is a function of the probability distribution of the \(n\) different types of polymer {\bf s}. \(U\) is the enthalpy, \(S\) is the entropy and \(T\) is a temperature. In general, for a system evolving autonomously, the rate of change in free energy \(\dot{G}<0\).

Thus far all quantities have been functions of the entire distribution of polymers, but we can expand this to be functions of the microstates of the individual polymers. For the enthalpy this is straightforwardly a weighted sum of the enthalpies of the individual polymers;
\begin{align}
    U(p({\bf s}))=\sum_{\bf s}p({\bf s})u({\bf s}).
\end{align}
For the entropy this is slightly more complex as there is a component of the individual polymer's entropies, but there is also a macroscopic entropy which is the entropy of the distribution of the polymers; ie its Shannon entropy. 
\begin{align}
    TS(p({\bf s}))=T\sum_{\bf s}p({\bf s})s({\bf s})-k_{B}T\sum_{\bf s}p({\bf s})\ln{p({\bf s})}\cite{ouldridge2018importance}.
\end{align}
Substituting these two into equation \ref{Free} and observing that the microstate free energy \(g=u-Ts\) we get
\begin{align}
    G=\sum_{\bf s}p({\bf s})g({\bf s})+k_{B}T\sum_{\bf s}p({\bf s})\ln{p({\bf s})}.
\end{align}
Thus the macrostate free energy is the sum of the microstate free energy and the negative of the Shannon entropy. Given the assumption that all backbone bonds are the same then the first term is independent of whether the polymers in the bath are randomly distributed or all identical. Thus, just considering the microstate free energy, there is no cost to creating an accurate copy, your system could produce as many perfectly accurate copies as it likes and not have to pay any extra free energy for accuracy. However it is clear that the second term is much more positive for systems with all identical copies than those with a random distribution of polymers\cite{landauer1961irreversibility}. It is the production of a biased polymer bath which costs free energy. Thus the production of an accurate copy is intrinsically the production of a high free energy state. Copying systems are not relaxing to equilibrium, they are transducing free energy from chemical free energy stored in the components and any fuel molecules such as ATP, to information free energy stored in the distribution of the polymer baths.

We therefore propose that accuracy can only be generated in an out-of-equilibrium system. Previous work has considered either self-assembly\cite{Whitelam,nguyen1,Gaspard} or templated self-assembly\cite{Bennett,Cady,Andrieux,Sartori1,Sartori2,Gaspard,esposito2010,EHRENBERG1980333,Johansson} in non-equilibrium contexts. However, in these cases, the non-equilibrium driving merely modulates a non-zero equilibrium specificity which relies on stabilising copy/template or intra-component interactions. This work is the first work which considers a fully autonomous templated copying system which includes separation of copy and template.

\section{Aims and Outline}
 
 It is clear that the full process of templated copying is insufficiently described by the existing theoretical literature. The separation of copy and template fundamentally changes the underlying physics of a copying system. This thesis attempts to fill in this gap.
 
 In chapter one I solve and analyse the first model of an autonomous templated copying process which includes the sequential separation of copy and template. Here I analyse a copy growing on, and sequentially separating from an infinitely long template, remaining attached only by its final monomer. Given the observation that the creation of an accurate polymer requires an input of free energy, I identify a general measure of the thermodynamic efficiency with which these nonequilibrium states are created and analyze the accuracy and efficiency of a family of dynamical models that produce persistent copies. For the weakest chemical driving, when polymer growth occurs in equilibrium, both the copy accuracy and, more surprisingly, the efficiency vanish. At higher driving strengths, accuracy and efficiency both increase, with efficiency showing one or more peaks at moderate driving. Correlations generated within the copy sequence, as well as between template and copy, store additional free energy in the copied polymer and limit the single-site accuracy for a given chemical work input. 
 
 In chapter two I extend this model to finite length templates, explicitly considering the final separation step in which the polymer must detach from its template and mix with surrounding baths of polymers. In this work I split the free-energy change of copy formation into informational and chemical terms.  I show that, surprisingly, copy accuracy plays no direct role in the overall thermodynamics. Instead, finite-length templates function as highly-selective engines that interconvert chemical and information-based free energy stored in the environment; it is thermodynamically costly to produce outputs that are more similar to the oligomers in the environment than sequences obtained by randomly sampling monomers.  By contrast, any excess free energy stored in correlations between copy and template sequences is lost when the copy fully detaches and mixes with the environment; these correlations therefore do not feature in the overall thermodynamics.  Previously-derived constraints on copy accuracy therefore only manifest as kinetic barriers experienced while the copy is template attached; these barriers are easily surmounted by shorter oligomers.

 In chapter three I then add in the destruction of templated oligomers, and ask whether driven cycles of creation and destruction can contribute to accuracy in the created polymers. I find regions in which a system can net create the most accurate polymer and net destroy all other polymers. This chapter seeks to question the validity of Bennett's claim that if the system creates and destroys a bit of information, that the system must disperse \(\ln{4}\) of free-energy in the form of entropy generation\cite{Bennett1982}. I calculate four constraints on a creation and destruction system which suggest that this limit can be broken. Having analysed a simple model of a system in which the template and destroyer reach steady state with the oligomer baths, we find no system which beats the limit, but do find a signature of systems which may beat the limit for longer length oligomers.

\newpage
\setcounter{equation}{0}
\setcounter{figure}{0}
\setcounter{table}{0}
\setcounter{section}{0}

\section*{Nonequilibrium correlations in minimal dynamical
models of polymer copying}

{\it This work is adapted from work published in January 2019 in Proceedings of the National Academy of Science under the same title.}

\setcounter{equation}{0}
\setcounter{figure}{0}
\setcounter{table}{0}
\setcounter{section}{0}

Living systems produce ``persistent" copies of information-carrying polymers, in which template and copy sequences remain correlated after physically decoupling. 
We identify a general measure of the thermodynamic efficiency with which these non-equilibrium states are created, and analyze the accuracy and efficiency of a family of dynamical models that produce persistent copies. For the weakest chemical driving, when polymer growth occurs in equilibrium, both the copy accuracy and, more surprisingly, the efficiency vanish. At higher driving strengths, accuracy and efficiency both increase, with efficiency showing one or more peaks at moderate driving.  Correlations generated within the copy sequence, as well as between template and copy, store additional free energy in the copied polymer and limit the single-site accuracy for a given chemical work input. Our results provide insight in the design of natural self-replicating systems and can aid the design of synthetic replicators.

\maketitle

\begin{figure}
\includegraphics[scale=0.7]{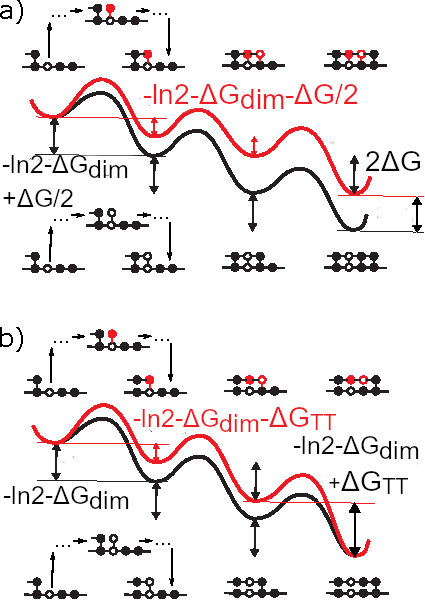}
\caption{Free-energy landscapes for simple examples of (a) templated self-assembly, in which the monomers remain bound to the template during the copy process; and (b) persistent copying, in which the monomers detach from the template after they have been incorporated into the polymer. Both diagrams show the addition of three monomers to a growing polymer, driven by a chemical free energy of backbone polymerisation \(\Delta G_{\rm pol}\). In each case, two scenarios are considered: the addition of two incorrect monomers, followed by a correct one (top), and the addition of three correct monomers (bottom). Local minima in the landscape represent macrostates following complete incorporation of monomers; intermediate configurations, illustrated schematically for the first transition, are part of the effective barriers. 
In templated self-assembly,
the chemical free-energy cost of previously incorporated mismatches is retained as the daughter grows \cite{Bennett,Cady,Andrieux,Sartori1,Sartori2,esposito2010}. Thus in (a), each mismatch in the daughter increases the chemical free energy by \(\Delta G_{\rm D}\) relative to the perfect match. 
In persistent copying (b), the chemical free-energy penalty for incorporating wrong momomers is only temporary; it arises when the incorrect monomer is added to the growing polymer, but is lost when that monomer subsequently detaches from the template. As a result, the overall chemical free-energy change of creating an incorrect polymer is the same as that for a correct one. Analyzing the consequences of this constraint, which is a generic feature of copying but does not arise in templated self assembly, is the essence of this work. The figure also shows that in our specific model, incorporating a wrong monomer after a correct one tends to reduce the chemical free-energy drop to \(\Delta G_{\rm pol}-\Delta G_{\rm TT}\), and incorporating a correct monomer after an incorrect one tends to increase it to \(\Delta G_{\rm pol}+\Delta G_{\rm TT}\); however, adding a wrong monomer to a wrong one, and adding a correct monomer to a correct one, does not change the free-energy drop \(\Delta  G_{\rm pol}\).}
\label{Phase}
\end{figure}

It is clear that a model of templated copying, which explicitly includes separation, is necessary for a fuller understanding of this crucial mechanism. Our first challenge is to analyse a family of model systems, which describe the growth and sequential separation of an infinitely long polymer growing on an infinitely long template. This allows us to consider a system which is constantly attached to the template by a single bond between the final monomer in the copy and it's corresponding template site. Here we omit the final step, where the copy polymer breaks it's final bond with the template, and mixes with surrounding baths of polymers and the first step, in which the first monomer comes out of solution and binds with the first site on the template. 

In this chapter we introduce a new metric for the thermodynamic efficiency of copying, and investigate the accuracy and efficiency of our models. We highlight the profound consequences of requiring separation, namely that correlations between copy and template can only be generated by pushing the system out of equilibrium. Previous work has considered self-assembly \cite{Whitelam,nguyen1,Gaspard} or templated self-assembly \cite{Bennett,Cady,Andrieux,Sartori1,Sartori2,Gaspard,esposito2010,EHRENBERG1980333,Johansson} in non-equilibrium contexts; in these cases, however, the non-equilibrium driving merely modulates a non-zero equilibrium specificity. Alongside the effect on copy-template interactions, we find that intra-copy-sequence correlations arise naturally. These correlations store additional free energy in the copied polymer, which do not contribute towards the accuracy of copying.

\section{Models and Methods}
\subsection{Model definition}

We consider a copy polymer \({M}=M_{1},...,M_{l}\), made up of a series of sub-units or monomers \(M_{x}\), growing with respect to a template \({N}=N_{1},...,N_{L}\) (\(l\leq L\)). Inspired by transcription and translation, we consider a copy that detaches from the template as it grows; fig. \ref{Phase}b, gives a generic coarse grained model with a strictly enforced step order. In this work we consider only the coarse grained steps in which a single monomer is added or removed, encompassing many individual chemical sub-steps \cite{Sartori2,Cady}.
After each step there is only a single inter-polymer bond at position \(l\), between \(M_l\) and \(N_l\). As a new monomer joins the copy at position \(l+1\), the bond position \(l\) is broken, contrasting with previous models of templated self assembly \cite{Bennett,Cady,Andrieux,Sartori1,Sartori2,esposito2010} (fig.~\ref{Phase}a). Importantly, as explained below, each step now depends on both of the two leading monomers, generating extra correlations within the copy sequence. It should be noted that throughout this thesis, we assume that this step order is fixed, and do not investigate the conditions needed to enforce the ordering.

\begin{figure}[t]
\centering
\includegraphics[scale=0.6]{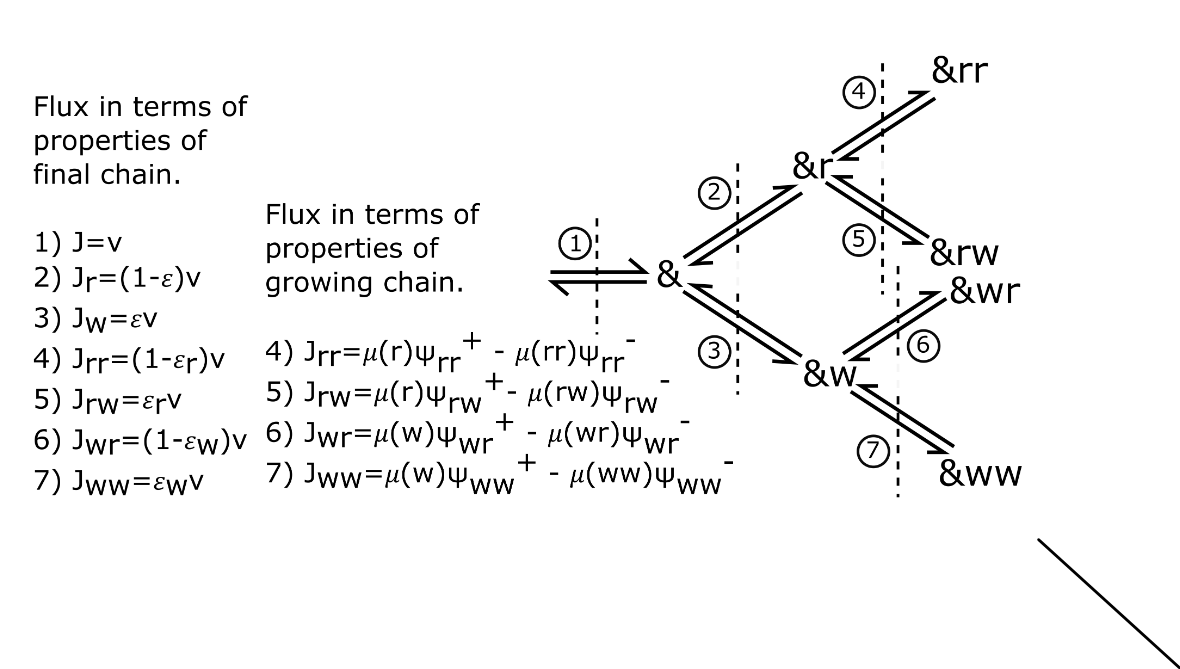}
\caption{Transitions of an arbitrary polymer \(\&\). To relate the final chain to the growing chain it is useful to consider fluxes through interfaces in this transition diagram. Using the tip and combined probabilities, along with relative propensities it is possible to describe the fluxes through interfaces 4-7 in terms of properties of the growing chain. Equally by considering errors and conditional errors and taking fractions of the overall growth velocity, it is possible to find the fluxes through interfaces 4-7 in terms of properties of the final chain and growth velocity.}
\label{Flux}
\end{figure}


Following earlier work, we assume that both polymers are copolymers, and that the two monomer types are symmetric \cite{Bennett,Cady,Andrieux,Sartori1,Sartori2,esposito2010}. Thus the relevant question is whether monomers \(M_{l}\) and \(N_{l}\) match; we therefore ignore the specific sequence of \(N\) and describe \(M_{l}\) simply as right or wrong. Thus \(M_{l}\in r,w\); with example chain \(M=rrwwrrrrrwrr\). An excess of \(r\) indicates a correlation between template and copy sequences. Breaking this symmetry would favour specific template sequences over others, disfavouring the accurate copying of other templates and compromising the generality of the process.

Given the model's state space, we now consider state free energies (which must be time-invariant for autonomy). We treat the environment as a bath of monomers at constant chemical potential \cite{Bennett,Cady,
Andrieux,Sartori1,Sartori2,esposito2010}.
%
%
By symmetry, extending the polymer while leaving the copy-template interaction unchanged involves a fixed polymerization free energy. We start by defining concentrations of monomer types with respect to a standard reference concentration. Here \([1]\) and \([2]\) are the monomer concentration of monomer types 1 and 2 respectively. The free-energy change when a matching monomer binds to the template (here both matching monomers are of type \(1\) and the template is type \(1,1\)) is \(\Delta G^{\plimsoll}_{r} -\ln[1]\) and a mismatching monomer (here type \(2\)) is \(\Delta G^{\plimsoll}_{w} -\ln[2]\), in units of \(k_{\rm B}T\). The difference between the two at the reference concentration, \(\Delta  G_{\rm TT}=\Delta G_{w}^{\plimsoll}-\Delta G^{\plimsoll}_{r}\), is the discrimination free energy, which biases the system towards accuracy. This bias can be describes as ``temporary thermodynamic" (TT) since it only lasts until that contact is broken. \(\Delta G^{\plimsoll}_{\rm dim}\) is the free-energy change of creating a backbone bond at the reference concentration and we further define \(\Delta G_{\rm pol}=\ln{[m_{l}]}-\Delta G^{\plimsoll}_{\rm dim}\) as the free energy change of a monomer coming out of solution and forming a backbone bond (neglecting copy/template interactions). Throughout this work \([m_{l}]=\frac{1}{2}\) for \([m_{l}]=r,w\) so \(\Delta G_{\rm pol}=-\ln{2}-\Delta G^{\plimsoll}_{\rm dim}\). \(\Delta G^{\plimsoll}_{\rm dim}\) incorporates the difference in entropy between a monomer in solution and as part of a polymer. If the polymer is ideal, ie does not interact with itself then it's free-energy will grow linearly with length. We omit any polymer self interactions which may cause the free energy to grow non-linearly with length. These could include attractive interactions which encourage the polymer to form a ``sticky ball", which would lead to longer polymers capable of coiling being more stable than longer polymers without this interaction, with shorter polymers not able to self stabilise.

Overall, each forward  step makes and breaks one copy-template bond. There are four possibilities:  either adding \(r\) or \(w\) at position \(l+1\) to a template with \(M_l = r\); or adding \(r\) or \(w\) in position \(l+1\) to a template with \(M_l = w\). The first and last of these options make and break the same kind of template bond, so the total free-energy change is  \(\Delta G_{\rm pol}\). For the second case there is a \(r\) bond broken and a \(w\) bond added, implying a  free-energy change of  \(\Delta G_{\rm pol}+\Delta G_{\rm{TT}}\).  Conversely, for the third case, there is a \(w\) bond broken and a \(r\) bond added, giving a free-energy change of \(\Delta G_{\rm pol}- \Delta G_{\rm{TT}}\). These constraints are shown in fig. \ref{Phase}b; the contribution of this work is to study the consequences of these constraints. Models of templated self assembly (fig. \ref{Phase}a) of equivalent complexity can be constructed, but they are not bound by these constraints and hence the underlying results and biophysical interpretation are distinct.

Having specified model thermodynamics, we now parameterize kinetics. We assume that there are no ``futile cycles'', such as appear in kinetic proofreading \cite{Hopfield}, each forward step must add either a correct or incorrect match, and each reverse step must remove one. 
Reactions are thus tightly coupled: each step requires a well defined input of free energy determined by \(\Delta G_{\rm{pol}}\) and \(\pm\Delta G_{\rm{TT}}\)  \cite{esposito2018}, and no free-energy release occurs without a step. 

A full kinetic treatment would be a continuous time Markov process  incorporating the intermediate states shown schematically in fig. \ref{Phase}b. To identify sequence output, however, we need only consider the state space in fig. \ref{Phase}b and the relative probabilities for transitions between these explicitly modelled states, ignoring the complexity of non-exponential transition waiting times \cite{Cady}. We define propensities \(\psi^{+}_{xy}\) as the rate per unit time in which a system in state \(\&x\) starts the process of becoming \(\&xy\) and \(\psi^{-}_{xy}\) as the equivalent quantity in the reverse direction (\(\&\) is an unspecified polymer sequence). Our system has eight of these propensities (\(\psi^{\pm}_{rr}\), \(\psi^{\pm}_{rw}\), \(\psi^{\pm}_{wr}\) and \(\psi^{\pm}_{ww}\)); the simplest templated self assembly models require four \cite{Bennett,Sartori1,Sartori2,esposito2010,Cady}.

Prior literature on templated self assembly \cite{Sartori1} has differentiated between purely ``kinetic'' discrimination, in which \(r\) and \(w\) have an equal template-binding free energy but different binding rates; and 
purely thermodynamic discrimination in which \(r\) and \(w\) bind at the same rate, but $r$ is stabilized in equilibrium by stronger binding interactions. Eventually, all discrimination is ``kinetic'' for persistent copying, since there is no lasting equilibrium bias (Fig.~\ref{Phase}\,(b)). However, by analogy with templated self assembly, we do consider two distinct mechanisms for discrimination -  a kinetic one, in which \(r\) is added faster than \(w\) to the growing tip, and one based on the temporary thermodynamic bias towards correct matches at the tip of the growing polymer due to short-lived favourable interactions with the template, quantified by \(\Delta  G_{\rm{TT}}>0\) (Fig.~\ref{Phase}b). The kinetic mechanism should not be conflated with fuel consuming ``kinetic proofreading'' cycles that can also enhance accuracy, which are not considered.

We parameterize the propensities as follows. Assuming, for simplicity, that the propensity for adding \(r\) or \(w\) is independent of the previous monomer, we have: \(\psi^{+}_{rr}=\psi^{+}_{wr}\) and \(\psi^{+}_{rw}=\psi^{+}_{ww}=1\), also defining the overall timescale. ``Kinetic'' discrimination is then quantified by
\(\psi^{+}_{xr} /\psi^{+}_{xw}=\exp(\Delta G_{\rm{K}}/k_BT)\). Forwards propensities are thus differentiated solely by \(\Delta G_{\rm{K}}\); backwards propensities are set by fixing the ratios  \(\psi^{+}_{xy}/ \psi^{-}_{xy}\) according to the free energy change of the reaction, which follow from \(\Delta G_{\rm{pol}}\) and \(\Delta G_{\rm TT}\) (Fig.~\ref{Phase}(b)). Thus, setting \(k_{B}T=1\),
\begin{align}
\begin{split}
  &\psi^{+}_{rr}=e^{\Delta G_{\rm{K}}},\,     \psi^{-}_{rr}=e^{-\Delta G_{\rm pol}}e^{\Delta G_{K}},\label{eq:rr}
\end{split}\\
\begin{split}
  &\psi^{+}_{rw}=1,\,     \psi^{-}_{rw}=e^{-\Delta G_{\rm pol}}e^{-\Delta G_{\rm{TT}}},\label{eq:rw}
\end{split}\\
\begin{split}
&\psi^{+}_{wr}=e^{\Delta G_{\rm{K}}},\,       \psi^{-}_{wr}=e^{-\Delta G_{\rm pol}}e^{\Delta G_{\rm{K}}}e^{-\Delta G_{\rm{TT}}},\label{eq:wr}
\end{split}\\
\begin{split}
   &\psi^{+}_{ww}=1,\,        \psi^{-}_{ww}=e^{-\Delta G_{\rm pol}}.\label{eq:ww}
\end{split}
\end{align}
For a given \(\Delta G_{\rm{K}}\), \(-\Delta G_{\rm pol}\) and \(\Delta G_{\rm{\rm{TT}}}\), eqs. \ref{eq:rr}-\ref{eq:ww} describe a set of models with distinct intermediate states that yield the same copy sequence distribution. We can thus analyze the simplest model in each set, which is Markovian at the level of the explicitly modelled states with \(\psi^{\pm}_{xy}\) as rate constants. 

\subsection{Model analysis} 
\subsubsection{Properties of the growing monomer}

{\it This section is adapted from the paper ``Kinetics and thermodynamics of first-order
Markov chain copolymerization"\cite{Gaspard}, for the specific case of the system being studied in this work.}

In order to solve this system we use a method developed by Gaspard\cite{Gaspard}; we assume that the growth process of a polymer is a Markov process, the addition of the monomer at site \(l+1\) is dependent only on the monomer at site \(l\). We work in the limit where the system is sufficiently dilute, so that while there are in fact multiple templates in solution, they are independent and the system can be considered at the level of a single template. This is the equivalent of considering a single stochastic trajectory, and so time evolution is described in terms of the probability \(P_{t}(M_{1}...M_{l-1},M_{l})\) of the specific polymer \(M_{1}M_{2}...M_{l-1}M_{l}\) attached to the template, existing in solution.

In this approach, we set up the kinetic equations
\begin{align}
    \frac{d}{dt}P_{t}(M_{1}...M_{l-1},M_{l}) = &\psi^{+}_{M_{l-1}M_{1}}P_{t}(M_{1}...M_{l-1})\\ \nonumber
    + &\psi^{-}_{M_{l}r}P_{t}(M_{1}...M_{l-1},M_{l},r) 
    + \psi^{-}_{M_{l}w}P_{t}(M_{1}...M_{l-1},M_{l},w)\\ \nonumber
    - &(\psi^{-}_{M_{l-1}M_{l}} - \psi^{+}_{M_{l}r} - \psi^{+}_{M_{l}w})P_{t}(M_{1}...M_{l-1},M_{1}),
\end{align}
which form an infinite hierarchy of coupled ordinary differential equations linking the probabilities all possible copies at different stages of growth. In a dilute solution, the concentrations follow the same kinetic equations as the probabilities.

It is possible to define a mean velocity of growth of the polymer, so that the polymer length is distributed around an average length \(\langle l \rangle \approx vt\) with variance \(\sigma \approx 2Dt\) where \(D\) is the the diffusivity. The sequence of the polymer measured back from the tip can be assumed to become stationary over long time. This allows the solution of the kinetic equations to be written as
\begin{align}
    P_{t}(M_{1}...M_{l-1},M_{l})\approx\mu(M_{1}...M_{l-1},M_{l})p_{t}(l),
\end{align}
where \(p_{t}(l)\) is the probability that a polymer has length \(l\) at time \(t\) found by marginalising over the probability of all possible polymers of length \(l\). \(\mu(M_{1}...M_{l-1},M_{l})\) is the stationary probability of the copy polymer having the specific sequence \(M_{1}...M_{l-1},M_{l}\), given the polymer is length \(l\). This assumption that \(\mu\) becomes stationary in the long time limit is the key assumption of this derivation.

The paper now define tip-monomer identity probabilities and joint tip and penultimate-monomer identity probabilities by marginalising over all earlier monomers giving
\begin{align}
    \mu(M_{1})\equiv \sum_{M_{1}...M_{l-1}}\mu(M_{1}...M_{l-1}M_{l}),\\
    \mu(M_{l-1}M_{l})\equiv \sum_{M_{1}...M_{l-2}}\mu(M_{1}...M_{l-1}M_{l}).
\end{align}

These can be inserted into the kinetic equations to give
\begin{align}
    \frac{d}{dt}p_{t}(l)=ap_{t}(l-1)+bp_{t}(l+1)-(a+b)p_{t}(l),\label{KE}
\end{align}
where
\begin{align}
   a&=(\psi_{rr}^{+}+\psi_{rw}^{+})\mu(r)+(\psi_{wr}^{+}+\psi_{ww}^{+})\mu(w)\\
   b&=\psi_{rr}^{-}\mu(rr)+\psi_{rw}^{-}\mu(rw)+\psi_{wr}^{-}\mu(wr)+\psi_{ww}^{-}\mu(ww).
\end{align}

We find the velocity by first postulating an ansatz, \(p_{t}(l)=e^{s_{q}t+iql}\), which we put into equation \ref{KE}. Here \(s_{q}=iqv-Dq^{2}+O(q^{3})\) is an exponential rate and \(q\) is a phase which can take values from \(-\pi\) to \(\pi\). On doing this we find;
\begin{align}
    s_{q}=&ae^{-iq}+be^{iq}-a+b\\
    =&(a+b)(\cos{q}-1)-i(a-b)\sin{q}\\
    =&(a+b)(\cancel{1}-\frac{q^2}{2}+\frac{q^4}{24}...-\cancel{1})-i(a-b)(q-\frac{q^3}{6}...).
\end{align}
From here it is straightforward to observe that \(v=a-b\) and \(D=\frac{a+b}{2}\) giving
\begin{align}
    v= (\psi_{rr}^{+}+\psi_{rw}^{+})\mu(r)+(\psi_{wr}^{+}+\psi_{ww}^{+})\mu(w)\\ \nonumber
    -\psi_{rr}^{-}\mu(rr)-\psi_{rw}^{-}\mu(rw)-\psi_{wr}^{-}\mu(wr)-\psi_{ww}^{-}\mu(ww).
\end{align}

In order to find the stationary probabilities we return to the equation \ref{KE} in the limit where \(P_{t}(m_{1},m_{2},...,m_{l})=\mu (m_{1},m_{2},...,m_{l})p_{t}(l)\). W first marginalise over all monomers except the tip monomer to find a first equation, and marginalise over all except the tip and penultimate monomer to find a second equation, marginalise over all except the tip, penultimate and single previous monomer for a third, carrying on until all monomers have been marginalised out. We then subsitute \(p_{t}(l)=e^{s_{q}t+iql}\) and take \(q\rightarrow 0\). This gives \(l\) equations of which the first three are;
\begin{align}
\label{ss1}
    0 = \sum_{M_{l-1}}&\psi^{+}_{M_{l-1}M_{l}}\mu(M_{l-1}) + \psi^{-}_{M_{l}r}\mu(M_{l},r) + \psi^{-}_{M_{l}w}\mu(M_{l},w)\\ \nonumber
    - \sum_{M_{l-1}}&\psi^{-}_{M_{l-1}M_{l}}\mu(M_{l-1}M_{l}) - (\psi^{+}_{M_{l}r} - \psi^{+}_{M_{l}w}))\mu(M_{l}),
\end{align}
\begin{align}
\label{ss2}
    0 = &\psi^{+}_{M_{l-1}M_{1}}\mu(M_{l-1}) + \psi^{-}_{M_{l}r}\mu(M_{l-1},M_{l},r) + \psi^{-}_{M_{l}w}\mu(M_{l-1},M_{l},w)\\ \nonumber
    - &\psi^{-}_{M_{l-1}M_{l}} - \psi^{+}_{M_{l}r} - \psi^{+}_{M_{l}w})\mu(M_{l-1}M_{l}),
\end{align}
\begin{align}
\label{ss3}
    0 = &\psi^{+}_{M_{l-1}M_{1}}\mu(M_{l-2},M_{l-1}) + \psi^{-}_{M_{l}r}\mu(M_{l-2},M_{l-1},M_{l},r) + \psi^{-}_{M_{l}w}\mu(M_{l-2},M_{l-1},M_{l},w)\\ \nonumber
    - &\psi^{-}_{M_{l-1}M_{l}} - \psi^{+}_{M_{l}r} - \psi^{+}_{M_{l}w})\mu(M_{l-2},M_{l-1}M_{l}).
\end{align}

Examining the form of these equations allows us to observe that solutions for the joint probabilities should be of the form;
\begin{align}
    \mu(M_{l-1}M_{l})=\mu(M_{l-1}|M_{l})\mu(M_{l})\label{markov1},\\
    \mu(M_{l-2}M_{l-1}M_{l})=\mu(M_{l-2}|M_{l-1})\mu(M_{l-1}|M_{l})\mu(M_{l}),\label{markov2}
\end{align}
and if these are replaced into equations \ref{ss1}-\ref{ss3} then the conditional probabilities can be found as a pair of coupled nonlinear equations;
\begin{align}
    \mu(M_{l-1}|M_{l})=\frac{\psi_{M_{l-1}M_{l}}^{+}\mu(M_{l-1})}
    {(\psi^{-}_{M_{l-1}M_{l}} + \psi^{+}_{M_{l}r} - \psi^{+}_{M_{l}w})\mu(M_{l})-\psi^{-}_{M_{l}r}\mu(M_{l}|r)\mu(r)-\psi^{-}_{M_{l}w}\mu(M_{l}|w)\mu(w)},\label{conditional1}
\end{align}
\begin{align}
    \mu(M_{l})=\frac{\sum_{M_{l-1}}\psi_{M_{l-1}M_{l}}^{+}\mu(m_{l-1})+\psi_{M_{l}r}^{-}\mu(M_{l}|r)\mu(r)+\psi_{M_{l}w}^{-}\mu(M_{l}|w)\mu(w)}{\psi_{M_{l}r}^{+}+\psi_{M_{l}w}^{+}+\sum_{M_{l-1}}\psi_{M_{l-1}M_{l}}^{-}\mu(M_{l-1}|M_{l})}.
\end{align}

We can also express the mean velocity in terms of our tip and conditional probabilities as
\begin{align}
     v= &(\psi_{rr}^{+}+\psi_{rw}^{+})\mu(r)+(\psi_{wr}^{+}+\psi_{ww}^{+})\mu(w)\\ \nonumber
     &-\psi_{rr}^{-}\mu(r|r)\mu(r)-\psi_{rw}^{-}\mu(r|w)\mu(w)-\psi_{wr}^{-}\mu(w|r)\mu(r)-\psi_{ww}^{-}\mu(w|w)\mu(w).
\end{align}

Equation \ref{markov1} and \ref{markov2} show that the stationary distribution of sequences is described by a first order Markov chain running backwards from the tip of the polymer
\begin{align}
    \mu(M_{1}...M_{l-1}M_{l})=\mu(M_{1}|M_{2})\mu(M_{2}|M_{3}) ... \mu(M_{l-2}|M_{l-1})\mu(M_{l-1}|M_{l})\mu(M_{l}).
\end{align}

We now define partial velocities, \(v_{r}\) and \(v_{w}\). The quantity \(v_{x}\mu(x)\) is the net rate at which monomers are added after an \(x\),
\begin{align}
 v_{x}\equiv\psi^{+}_{xr}-\frac{\mu(x|r)\mu(r)}{\mu(x)}\psi^{-}_{xr}+\psi^{+}_{xw}-\frac{\mu(x|w)\mu(w)}{\mu(x)}\psi^{-}_{xw}. \label{eq:vx1}
\end{align}
By inserting these velocities into equation \ref{conditional1} the conditional probabilities can be rewritten as
\begin{align}
    \mu(m|n)=\frac{\psi_{mn}^{+}\mu(m)}{(\psi_{mn}^{-}+v_{n})\mu(n)},\label{cond}
\end{align}
which in turn can be reinserted into the partial velocities to find
\begin{align}
 v_{x}=\frac{\psi^{+}_{xr} v_{r}}{\psi^{-}_{xr}+v_{r}} +\frac{\psi^{+}_{xw} v_{w}}{\psi^{-}_{xw}+v_{w}}. \label{eq:vx}
\end{align}

Allowing \(x=r,w\) gives a pair of simultaneous equations that 
can be solved via the tip and conditional probabilities to find \(v_{x}\) in terms of the propensities. In turn, the velocities and propensities determine tip probability \(\mu(M_{l})\) via 
\begin{align}
  \mu(n)= \sum_{m}\frac{\psi^{+}_{mn}\mu(m)}{\psi^{-}_{mn}+v_{n}}
\end{align}
which follows from multiplying equation \ref{cond} by \(\mu(n)\), summing over \(m\) and normalising. Conditional probability follows straightforwardly from equation \ref{cond}. 
It is therefore possible to solve for the steady state of the tip velocity by identifying our propensities \(\psi^{\pm}_{xy}\), and solving a pair of non-linear simultanious equations.

\subsubsection{Final state of the polymer}

Gaspard's method describes the chain while it is still growing through the stochastic variables \(M_{l}\) and  \(M_{l-1}\), with the index \(l\) being the current length of the polymer. We, however, are interested in the identity of the monomer at position \(n\) when \(l \gg n\). We label this ``final" state of the monomer at position \(n\) as \(M_{n}^{\infty}\). \(M_n^\infty\) is described by the error probability \(\epsilon\) and the conditional error probabilities \(\epsilon_{r}\) and \(\epsilon_{w}\), defined as the probability that \(M^\infty_{n+1} = w\) given that \(M^\infty_{n} = r\) or \(M^\infty_{n} = w\), respectively. 

It might not be immediately obvious why the properties of the growing chain described by \(\mu(M_{l})\) and \(\mu(M_{l}, M_{l-1})\) should be different from those of the final chain described by \(\epsilon\), \(\epsilon_r\) and \(\epsilon_w\) but the difference can be illustrated with a simple example. Consider a system in which incorrect monomers could be added to the end of the chain, but where nothing can be added after an incorrect match. In this case while the tip probability for an incorrect match µ(w) would be finite, the error of the final chain \(\epsilon\) would be vanishingly small, as all incorrect matches would have to be removed in order for the polymer to grow further. 

In order to demonstrate that \(\epsilon\), \(\epsilon_{r}\) and \(\epsilon_{w}\) are sufficient it is necessary to prove that the \(M^\infty\) is a Markov chain of \(r\) and \(w\) monomers. While the model has been designed with Markovian dynamics, it is not necessarily the case that the produced sequence itself is Markovian. The system will generate correlations as it creates a polymer and these correlations could hypothetically be long range. We note that \(\epsilon\neq\epsilon_{r}\neq\epsilon_{w}\) as a direct result of the dependence of the transition propensities on the current and previous tip monomers, which in turn arises from detachment.

We consider the process of stepping along the final chain, and ask what form the correlations between monomers take. Let \(M^\infty_{n}\) be the monomer at the \(n\)th site in the final chain. Let a polymer be represented by \(M^\infty_{1},...,M^\infty_{n}\). The probability of a given chain existing is then \(P^\infty(m_{1},...,m_{n})\). In order for the sequence of monomers moving along the chain (increasing \(n\)) to be able to be represented by a Markov chain, the condition 
\begin{equation}
 P^\infty(m_{n}|m_{n-1},...,m_{1})=P^\infty(m_{n}|m_{n-1})
\label{eq:M1}
\end{equation}
must hold.

In order to demonstrate that eq.~\ref{eq:M1} holds, we rewrite the final chain probability in terms of properties of the growing chain. Specifically we state that the probability of the sequence \(m_{1},...,m_{n}\) existing in the final chain is the product of the probability \(Q(m_{1},...,m_{n-1},t)\) that the chain is in the state \(m_{1},...,m_{n-1}\) at a time \(t\) during the growth process, and the propensity \(\nu(m_n;m_{1},...,m_{n-1})\) with which \(m_{n}\) is added to a chain \(m_{1},...,m_{n-1}\) and never removed, integrated over all time. This gives

\begin{equation}
P^\infty(m_{1},...,m_{n})=\int Q(m_{1},...,m_{n-1},t)\nu(m_{n};m_{1},...,m_{n-1})dt.
\label{eq:M2}
\end{equation}
It should be noted that \(\nu(m_n;m_{1},...,m_{n-1})\) is time-independent, giving
\begin{equation}
P^\infty(m_{1},...,m_{n})=\nu(m_{n};m_{1},...,m_{n-1}) \int Q(m_{1},...,m_{n-1},t)dt.
\label{eq:M3}
\end{equation}
Setting the integral equal to \(I(m_{1},...,m_{n-1})\) gives
\begin{equation}
P^\infty(m_{1},...,m_{n})=\nu(m_{n};m_{1},...,m_{n-1}) I(m_{1},...,m_{n-1}).
\label{eq:M4}
\end{equation}

Let's consider the probabilities of two sub-sequences, identical except for the final monomer. We call the two final monomers \(m_{n}\) and \(m^\prime_{n}\) and we can denote the ratios of the probabilities of these two chains as 
\begin{equation}
\frac{P^\infty(m_{1},...,m_{n})}{P^\infty(m_{1},...,m^\prime_{n})}=\frac{\nu(m_{n};m_{1},...,m_{n-1})I(m_{1},...,m_{n-1})}{\nu(m^\prime_{n};m_{1},...,m_{n-1})I(m_{1},...,m_{n-1})}.
\label{eq:M5}
\end{equation}
The \(I\) terms are independent of this final monomer and so cancel. Thus
\begin{equation}
\frac{P^\infty(m_{1},...,m_{n})}{P^\infty(m_{1},...,m^\prime_{n})}=\frac{\nu(m_{n};m_{1},...,m_{n-1})}{\nu(m^\prime_{n};m_{1},...,m_{n-1})}.
\label{eq:M6}
\end{equation}
The same relationship holds for the conditional probabilities
\begin{equation}
\frac{P^\infty(m_n|m_{1},...,m_{n-1})}{P^\infty(m^\prime_{n}|m_{1},...,m_{n-1})}=\frac{\nu(m_{n};m_{1},...,m_{n-1})}{\nu(m^\prime_{n};m_{1},...,m_{n-1})}.
\label{eq:M7}
\end{equation}

For our system, the propensity with which a monomer \(m_{n}\) is added and never removed, \(\nu(m_{n};m_{1},...,m_{n-1})\), is dependent on only on the final two monomers in the sequence fragment, \(m_n\) and \(m_{n-1}\). To see why, note that this propensity is determined by addition and removal of monomers at sites \(i \geq n\). The identities of monomers at positions \(j<n-1\), however, only influence addition and removal propensities at sites \(k<n\) (eq.~\ref{eq:rr}-\ref{eq:ww}).  
Thus we convert \(\nu(m_{n};m_{1},...,m_{n-1})\) to \(f(m_{n},m_{n-1})\) giving
\begin{equation}
\frac{P^\infty(m_n|m_{1},...,m_{n-1})}{P^\infty(m^\prime_{n}|m_{1},...,m_{n-1})}
=\frac{f(m_{n},m_{n-1})}{f(m^\prime_{n},m_{n-1})}.
\label{eq:M8}
\end{equation}
Multiplying both sides by \(P^\infty(m_{1},...,m_{n-2})\) and summing over all values of \(m_{1},...m_{n-2}\) (recall \(\sum_{c,d}P(a|b,c,d)P(c,d)=P(a|b)\) and \(\sum_{c,d}P(c,d)=1\)) gives
\begin{equation}
\frac{P(m_{n}|m_{n-1})}{P(m^\prime_{n}|m_{n-1})}=\frac{f(m_{n},m_{n-1})}{f(m^\prime_{n},m_{n-1})}.
\label{eq:M9}
\end{equation}
Comparing equations \ref{eq:M8} and \ref{eq:M9} yields:
\begin{equation}
\frac{P^\infty(m_{n}|m_{n-1},...,m_{1})}{P(m^\prime_{n}|m_{n-1},...,m_{1})}=\frac{P^\infty(m_{n}|m_{n-1})}{P^\infty(m^\prime_{n}|m_{n-1}).}
\label{eq:M10}
\end{equation}
Summing over the possible values of \(m^\prime_{n}\) and recalling that \(P^\infty(r|m_{n-1})+P^\infty(w|m_{n-1})=1\) and \(P^\infty(r|m_{n-1},...,m_{1})+P^\infty(w|m_{n-1},...,m_{1})=1\) yields:
\begin{equation}
\begin{split}
P^\infty(m_{n}|m_{n-1},...,m_{1})= P^\infty(m_{n}|m_{n-1}),
\label{eq:M11}
\end{split}
\end{equation}
thereby proving that the sequence of the final chain is Markovian. It would be straightforward to envisage a system with longer range correlations. If, instead of enforcing a single copy template bond at the end of each step, we had two or more bonds allowed, the correlations would be longer range and the final polymer would not be Markovian in sequence. This however is distinct from the Markovian dynamics of the model itself, which are designed in.

\subsubsection{Linking the growing polymer to the final sequence}

Now we have demonstrated that the sequence of the final chain is Markovian, we can use this fact to calculate \(\epsilon\), \(\epsilon_r\) and \(\epsilon_w\). We define currents \(J_{xy}\) that are related to both \(\psi^{\pm}_{xy}\) and \(\epsilon\), \(\epsilon_r\) and \(\epsilon_w\) separately. The current \(J_{xy}\) is the net rate per unit time at which transitions \(\&x\rightarrow \&xy\) occur: \(J_{xy}=\psi_{xy}^{+}\mu(x)-\psi_{xy}^{-}\mu(x,y)\). By considering the transitions in our system as a tree, as in Fig. \ref{Flux}, we can relate the current through a branch to the overall rate at which errors are permanently incorporated into a polymer growing at total velocity \(v=v_{r}\mu(r)+v_{w}\mu(w)\)
\begin{align}
\begin{split}
 J_{rr}=(1-\epsilon)(1-\epsilon_{r})v=\mu(r)\psi^{+}_{rr}-\mu(r,r)\psi^{-}_{rr},\label{eq:jrr}
\end{split}\\
\begin{split}
 J_{rw}=(1-\epsilon)\epsilon_{r}v=\mu(r)\psi^{+}_{rw}-\mu(r,w)\psi^{-}_{rw},\label{eq:jrw}
\end{split}\\
\begin{split}
J_{wr}=\epsilon(1-\epsilon_{w})v=\mu(w)\psi^{+}_{wr}-\mu(w,r)\psi^{-}_{wr},\label{eq:jwr}
\end{split}\\
\begin{split}
J_{ww}=\epsilon\epsilon_{w}v=\mu(w)\psi^{+}_{ww}-\mu(w,w)\psi^{-}_{ww}.\label{eq:jww}
 \end{split}
\end{align}
\begin{figure}
\begin{center}
\includegraphics[scale=0.3]{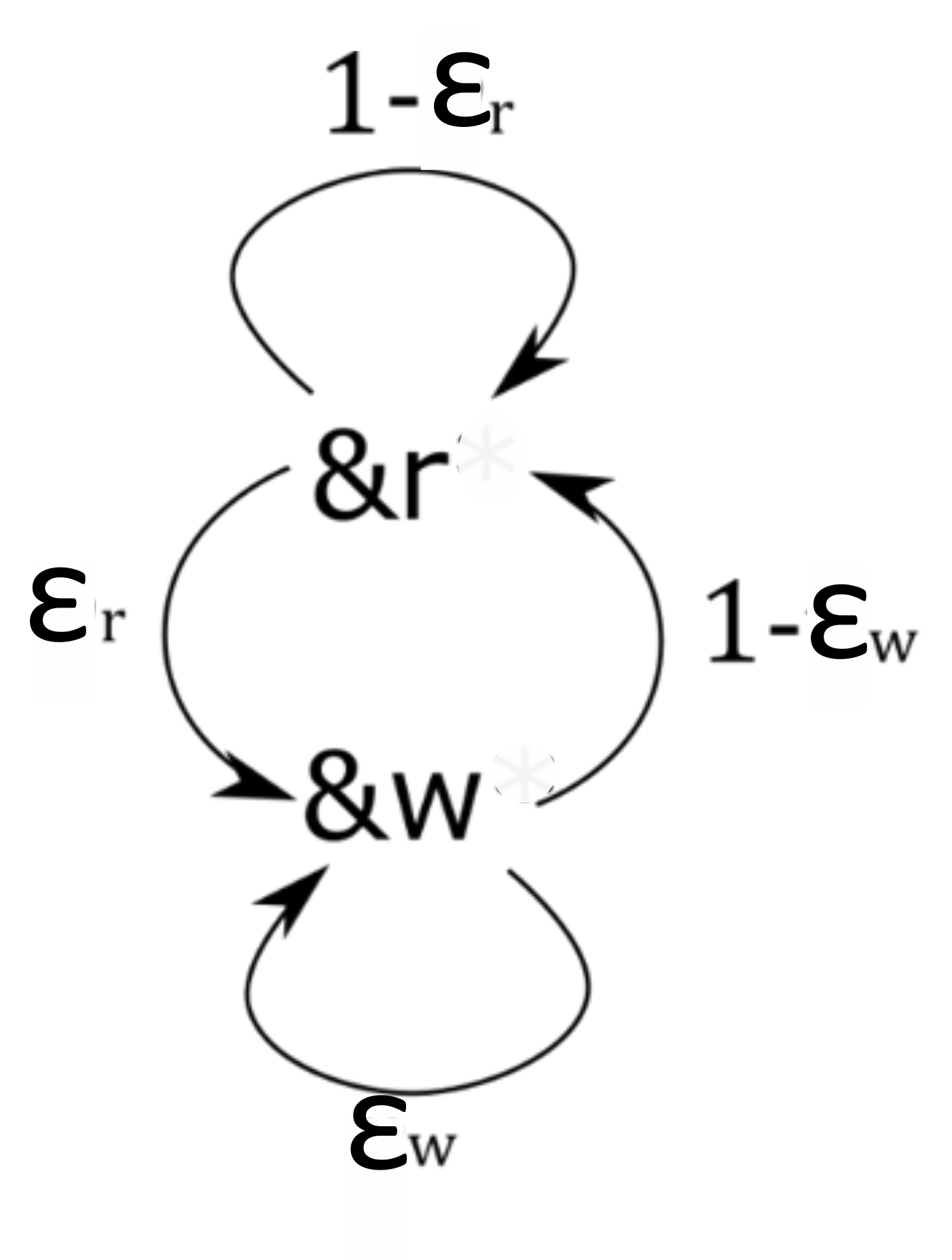}
\end{center}
\caption{The transition diagram for the Markov process describing the sequence of monomers found by stepping forward along a completed chain.}
\label{Trans}
\end{figure}
Eliminating \(\epsilon\) from the simultaneous equations \ref{eq:jrr}-\ref{eq:jww} yields \(\epsilon_r\) and \(\epsilon_w\) in terms of known quantities \(\psi\) and \(\mu\). To find \(\epsilon\), note that the final sequence is a Markov process which can be solved using transition matrix parameterized by \(\epsilon_r\) and \(\epsilon_w\), with the overall error \(\epsilon\) given by its dominant eigenvector. The transition matrix for this process is
\begin{align}
T=\begin{bmatrix}
   1-\epsilon_{r}       &1-\epsilon_{w} \\
    \epsilon_{r}       & \epsilon_{w} \\ 
\end{bmatrix}.\label{eq:matrix}
\end{align}
The first eigenvector of this transition matrix gives the steady state of the Markov process shown in figure \ref{Trans}. The component of the eigenvector corresponding to the probability of incorrect matches is \(\epsilon = {\epsilon_{r}}/{(1+\epsilon_{r}-\epsilon_{w})}\). From \(\epsilon\), \(\epsilon_r\) and \(\epsilon_w\) we calculate properties of the copy in terms of \(\psi^{\pm}_{xy}\) and thus the free energies. 

\subsubsection{Corroboration with simulations}

\begin{figure*}
\includegraphics[scale=0.35]{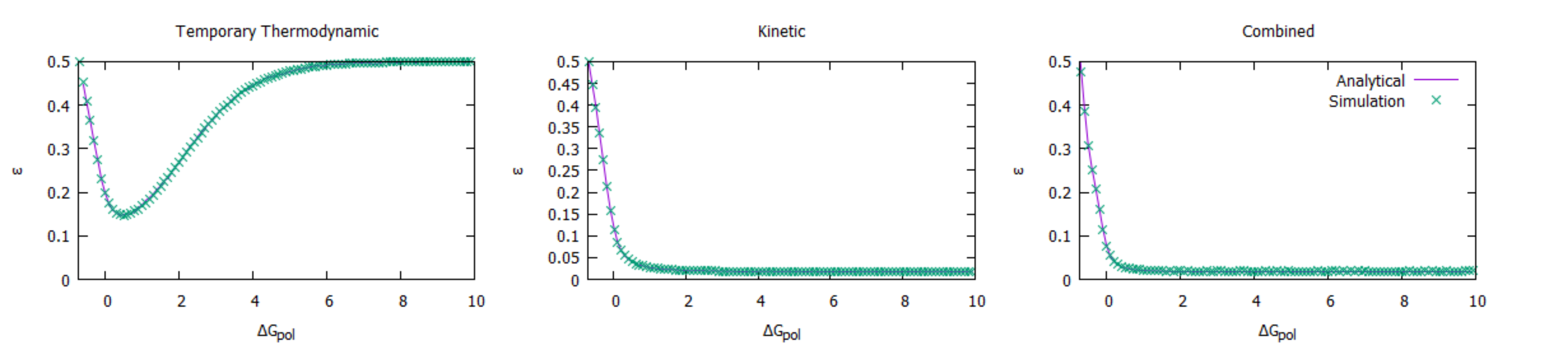}
\caption{Errors obtained from Gillespie simulations are indistinguishable from the analytical results obtained using Gaspard's method for all three mechanisms.}
\label{compareerr}
 \end{figure*}

To check the analytical methods used to solve the system we also simulated the growth of a polymer. We used a Gillespie simulation \cite{Gillespie}, with transition rates given by \(\psi^\pm_{xy}\). Simulations were initialised with a randomly determined two monomer sequence, and truncated as soon as the polymer reached 1000 monomers, repeated 100 times. We found that such a length rendered edge effects negligible in all but the most extreme cases for the calculation of \(\epsilon\). Polymer error probabilities were inferred directly from the simulations, and are compared to analytical results in fig. \ref{compareerr}.

We note in passing that the calculation of \(H\), \(H_{ss}\), and particularly the efficiency's \(\eta\) and \(\eta_{ss}\), are more vulnerable to random fluctuations in a simulation of finite length, and peculiar edge effects, than \(\epsilon\). Gaspard's solution is therefore invaluable in reaching robust conclusions for these quantities.

\begin{figure*}
\begin{center}
\includegraphics[scale=0.15]{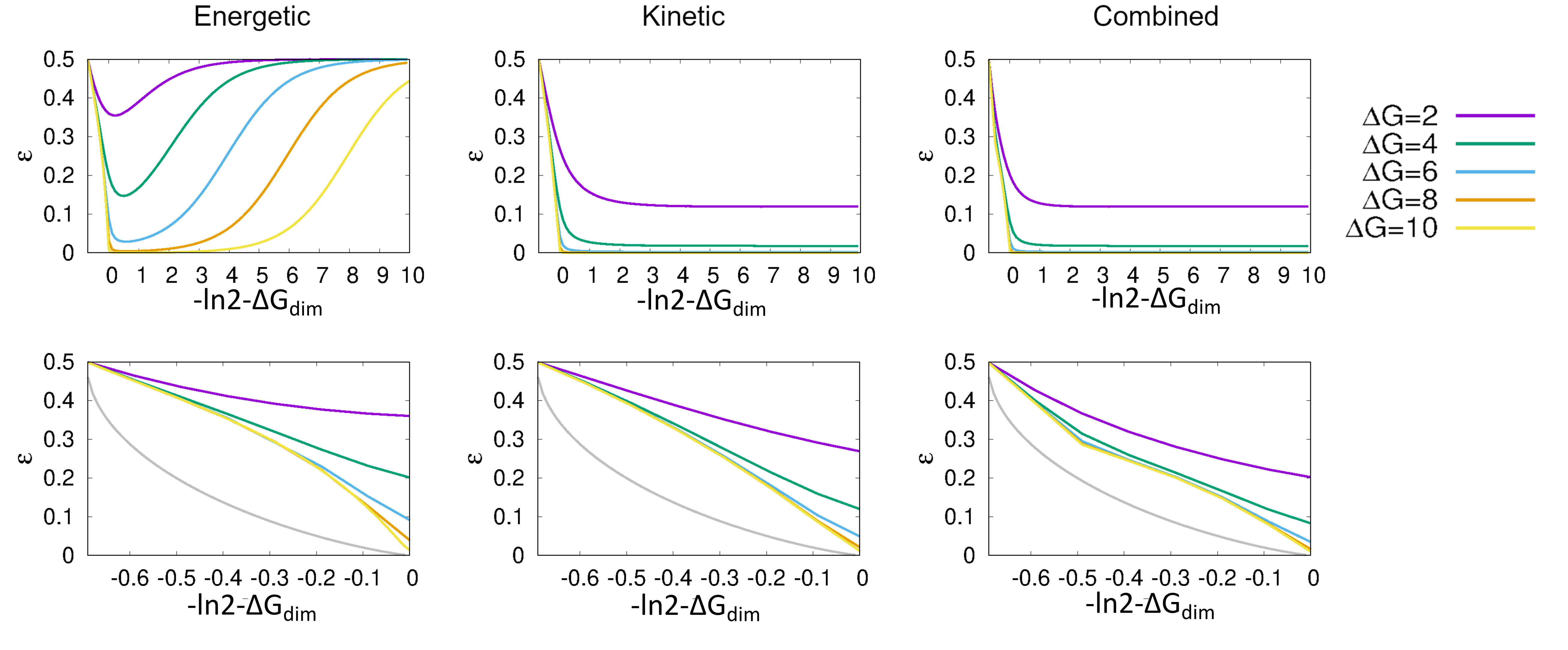}
\caption{Error probability \(\epsilon\) as a function of \(\Delta G_{\rm{\rm{pol}}}\) for all three mechanisms: (a) over a wide range of  \(\Delta G_{\rm pol}\), and (b) within the entropy-driven region \(\Delta G_{\rm pol} \leq 0\). The temporary thermodynamic mechanism is always least accurate, and the combined mechanism most accurate. All mechanisms have no accuracy in the limit of \(\Delta G_{\rm pol} \rightarrow -\ln{2}\), and are far from the fundamental bound on single-site accuracy imposed by \(H_{\rm ss} = -\epsilon \ln \epsilon - (1-\epsilon) \ln (1-\epsilon) \geq -\Delta G_{\rm pol}\) except at \(-\Delta G_{\rm pol}\approx 0\).}
\label{Error}
\end{center}
\end{figure*}

\section{Results}
\subsection{General thermodynamic bounds}
The free energy of the combined bath and polymer system decreases over time while the polymer is growing on the template. There are two contributions to the free-energy change per added monomer: the chemical free energy  \(\Delta G_{\rm{pol}}\), and a contribution from the uncertainty of the final polymer sequence \cite{Ouldridge}. 
The latter is quantified by the entropy rate \(H\) \cite{Crutchfield, Cover}:
\begin{align}
    H({M^\infty})=\lim_{n\rightarrow\infty}\frac{1}{n}H(M^\infty_{1},M^\infty_{2},...,M^\infty_{n}),
\end{align}
which in our case is given by \cite{Cover}
\begin{align}
    H({M^\infty})=&-\epsilon\left(\epsilon_{w}\ln{\epsilon_{w}}+(1-\epsilon_{w})\ln{(1-\epsilon_{w})}\right)\nonumber\\&-(1-\epsilon)\left(\epsilon_{r}\ln{\epsilon_{r}}+(1-\epsilon_{r})\ln{(1-\epsilon_{r})}\right).
\end{align}
The overall free energy change per added monomer is then \(\Delta G_{\rm{tot}}=-\Delta G_{\rm pol}-H\), which must be negative for growth giving a limit of \(H \geq-\Delta G_{\rm pol}\). Since copy-template interactions are not extensive in the copy length, they do not contribute. Given that \(H\geq0\), growth is possible in the region where \(\Delta G_{\rm pol}<0\), corresponding to the ``entropically driven'' regime \cite{esposito2010,Bennett}.

The entropy rate is bounded by the single site entropy \(H \leq H_{ss}=-\epsilon\ln{\epsilon}-(1-\epsilon)\ln{(1-\epsilon)}\). \(H_{ss}\) quantifies the desired correlations {\it between} copy and template. For previous models of templated self assembly with uncorrelated monomer incorporation, \(H=H_{ss}\) \cite{Bennett,Cady,Andrieux,Sartori1,Sartori2,Gaspard,esposito2010}.  In our model,  the necessary complexity of \(\psi^{\pm}_{xy}\) generates correlations {\it within} the copy. A stronger constraint on the single site entropy, and hence accuracy, then follows: \(H_{ss}\geq H\geq -\Delta G_{\rm pol}\), \(H_{ss}\geq -\Delta G_{\rm pol}\) is the bound in fig. \ref{Error}, we use \(H_{ss}\) rather than \(H\) because that corresponds correlations only due to the accuracy of the polymer, whereas \(H\) includes wasted within change correlations.

Fundamentally, a persistent copy is a high free energy state, as the entropic cost of copy-template correlations cannot be counteracted by stabilizing copy-template interactions. Thus the process moves a system between two high free energy states, converting chemical work into correlations. 
In general, only a fraction of the chemical work done by the monomer bath is retained in the final state, implying dissipation, and so it is natural to introduce an efficiency. The overall free energy stored in the polymer has contributions both from the creation of an equilibrium (uncorrelated) polymer and from correlations within the copy and with the template. We are interested only in the contributions above equilibrium. The efficiency \(\eta\) is then the proportion of the additional free energy expended in making a copy above the minimum required to grow a random equilibrium polymer that is successfully converted into the non-equilibrium free energy of the  copy sequence rather than being dissipated. In our simple case, 
\begin{equation}
\eta\equiv\frac{H_{\rm eq}-H}{H_{\rm eq}+\Delta G_{\rm pol}} \leq 1. \label{eq:eff}
\end{equation}
Here, \(\Delta G_{\rm pol}+H_{\rm eq} = \Delta G_{\rm pol}+\ln2\) is the extra chemical work done by the buffer above that required to grow an equilibrium polymer, \(-\ln{2}-\Delta G^{\rm eq}_{\rm{dim}} = -H_{\rm eq} = -\ln 2\). The free energy stored in the copy sequence, above that stored in an equilibrium system, is \(H_{\rm eq}-H\). \(\eta \leq1\) follows from \(\Delta G_{\rm pol}+ H \geq 0\). Similarly, since \(H_{\rm ss}\geq H\) we can define a single-site efficiency 
\begin{equation}
\eta_{\rm ss}\equiv\frac{H_{\rm eq}-H_{\rm ss}}{H_{\rm eq}+\Delta G_{\rm pol}} \leq \eta \leq 1. \label{eq:effss}
\end{equation}
Unlike \(\eta\), the single site efficiency \(\eta_{\rm ss}\) discounts the free energy stored in  ``useless" correlations {\it within} the copy.

\subsection{Behaviour of specific systems}
We consider three representative models consistent with eqs.~\ref{eq:rr}-\ref{eq:ww}. First, the purely kinetic mechanism obtained by setting \(\Delta G_{\rm{TT}}=0\) and \(\Delta G_{\rm{K}}=\Delta G\) in  equations \ref{eq:rr}-\ref{eq:ww}. Originally proposed
 by Bennett for templated self assembly~\cite{Bennett}, it is coincidentally a limiting case of persistent copying since there is no equilibrium bias. We also consider two new mechanisms:
pure ``temporary thermodynamic discrimination'' with \(\Delta G_{\rm{K}}=0\) and \(\Delta G_{\rm{TT}}=\Delta G\), and a ``combined discrimination mechanism'', in which both template binding strengths and rates of addition favour \(r\) monomers: \(\Delta G_{\rm{K}}=\Delta G_{\rm{TT}}=\Delta G\).


All three mechanisms have two free parameters, the overall driving strength \(\Delta G_{\rm pol}\) and the discrimination parameter \(\Delta G\). We plot error probability against \(\Delta G_{\rm pol}\) for various \(\Delta G\) in  fig. \ref{Error}. Also shown is the thermodynamic lower bound on \(\epsilon\) implied by \(H_{\rm ss}\geq H\geq -\Delta G_{\rm pol}\). All three cases have no accuracy (\(\epsilon=0.5\)) in equilibrium (\(\Delta G_{\rm pol}\rightarrow-\ln{2}\))
since an accurate persistent copy is necessarily out of equilibrium \cite{Ouldridge}. By contrast, templated self assembly allows for  accuracy in equilibrium \cite{Sartori1,Sartori2,Johansson}.

The temporary thermodynamic mechanism is always the least effective. It has no accuracy for high \(\Delta G_{\rm pol}\) as the difference between \(r\) and \(w\) is only manifest when stepping backwards, and for large positive \(\Delta G_{\rm pol}\) back steps are rare \cite{Sartori1,Sartori2}. More interestingly, temporary thermodynamic discrimination is also inaccurate as \(\Delta G_{\rm pol}\rightarrow-\ln{2}\), when the system takes so many back steps that it fully equilibrates. Low \(\epsilon\)  only occurs when \(\Delta G_{\rm pol}\) is sufficient to inhibit the detachment of \(r\) monomers, but not the detachment of \(w\) monomers. This trade-off region grows with \(\Delta G\). By contrast, both the combined case and the kinetic case have accuracy in the limit of \(\Delta G_{\rm pol}\rightarrow \infty\), since they allow \(r\) to bind faster than \(w\).
 Considering \(\Delta G_{\rm pol}\leq0\) closely (fig. \ref{Error}b) shows the combined case to be superior.

\begin{figure*}
\begin{center}
\includegraphics[scale=0.15]{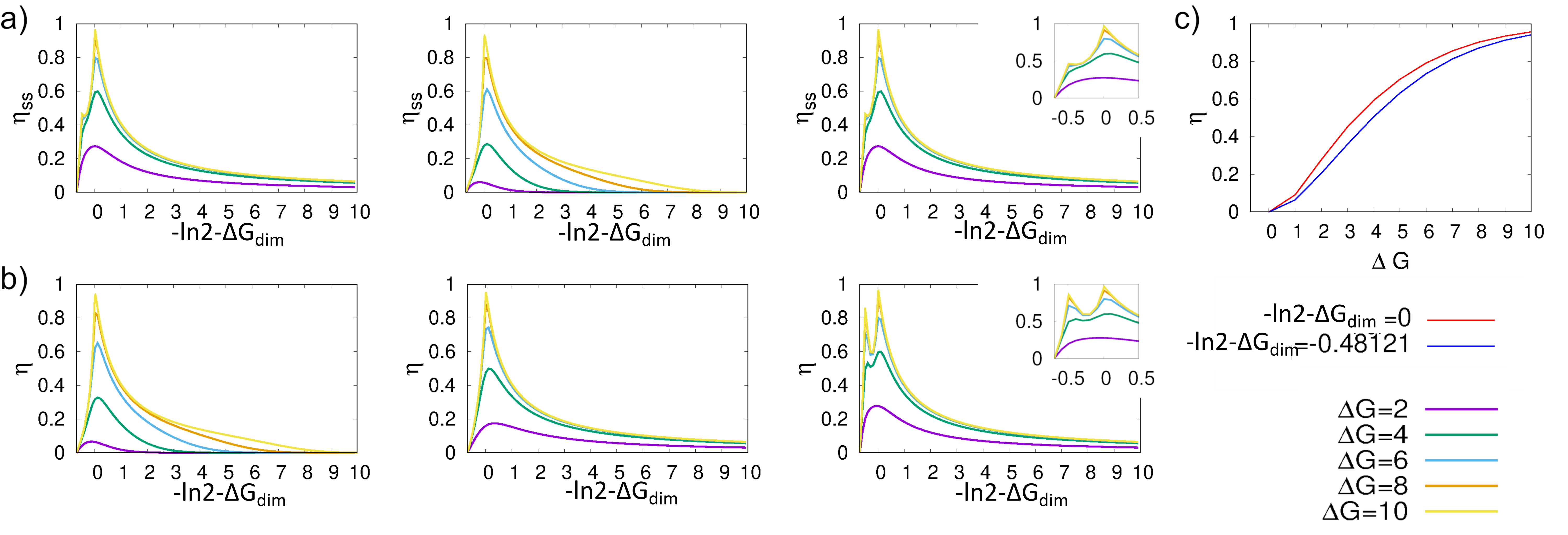}
\caption{Efficiencies \(\eta_{\rm ss}\) (a) and \(\eta\) (b) plotted against \(\Delta G_{\rm pol}\) show sharp peaks at \(\Delta G_{\rm pol}=0\) as \(\Delta G\rightarrow \infty\) in all three cases. In the combined case we see a second peak in \(\eta\), and a shoulder in \(\eta_{\rm ss}\) at \(\Delta G_{\rm pol}=-0.48121\). (c) Both of these peaks in \(\eta\) tend to unity at \(\Delta G \rightarrow \infty\).}
\label{Efficiency}
\end{center}
 \end{figure*}

Intriguingly, all three mechanisms are far from the fundamental bound on \(\epsilon\) implied by \(H_{\rm ss}\geq H\geq -\Delta G_{\rm pol}\) as \(\Delta G_{\rm pol}\rightarrow-\ln{2}\), and there is an apparent cusp in \(\epsilon\) at \(\Delta G_{\rm pol} \approx 0.48\) as \(\Delta G \rightarrow \infty\) in the combined case.
The performance relative to the bound is quantified by \(\eta_{\rm ss}\) (fig. \ref{Efficiency}). Surprisingly, we observe in fig.~\ref{Efficiency}a that not only does \(\epsilon\) go to zero as \(\Delta G_{\rm pol}\rightarrow-\ln{2}\), so does \(\eta_{ss}\) in all cases. For small non-equilibrium driving, none of the extra chemical work input is stored in correlations with the template. Mathematically, this inefficiency arises because \(\epsilon-0.5 \propto \Delta G_{\rm pol}-\ln{2}\) as \(\Delta G_{\rm pol}\rightarrow-\ln{2}\) (as observed in fig.~\ref{Error}), and \(H_{\rm ss}-\ln2 \propto (\epsilon-0.5)^{2}\) for \(\epsilon\approx0.5\) by definition. Thus, from equation \ref{eq:effss}, \(\eta_{\rm ss}\propto \Delta G_{\rm pol}-\ln{2}\) as \(\Delta G_{\rm pol}\rightarrow-\ln{2}\). That this result only depends on error probability decreasing proportionally with \(\Delta G_{\rm pol}\) for small driving suggests that a vanishing \(\eta_{ss}\) in equilibrium may be quite general.

In all cases, the single-site efficiency \(\eta_{\rm ss}\) increases from 0 at \(\Delta G_{\rm pol}=-\ln{2}\) to a peak near \(\Delta G_{\rm pol}=0\), with \(\eta_{\rm ss} \rightarrow 1\) as \(\Delta G\rightarrow\infty\). Beyond this peak, \(\eta_{\rm ss}\) drops as the stored free energy  is bounded by \(\ln{2}\) per monomer.
To understand the peak,  note that for every  \(\Delta G_{\rm pol}\leq0\) there is a hypothetical highest accuracy copy with \(\epsilon\) fixed by \(H_{\rm ss}=-\Delta G_{\rm pol}\) that is marginally thermodynamically permitted. However, this copy is not usually kinetically accessible. At \(\Delta G_{\rm pol}=0\) the marginal copy has 100\% accuracy and, unusually,  all three mechanisms can approach it kinetically, causing a peak. The efficiency  approaches its limit of unity even for moderate values of \(\Delta G\). We note that as \(\Delta G\rightarrow\infty\), growth is slow for \(\Delta G_{\rm pol}\leq 0\): the total number of steps taken diverges.

A related argument explains the apparent cusp in the error \(\epsilon\) for the combined mechanism at \(\Delta G_{\rm{pol}}\approx-0.48\) and high \(\Delta G\). On the plot of \(\eta_{\rm ss}\) (fig.~\ref{Efficiency}\,a) this cusp manifests as a shoulder. The full efficiency \(\eta\) (fig.~\ref{Efficiency}\,b) has a prominent second peak. Uniquely, the combined mechanism's kinetics strongly disfavour chains of consecutive \(w\)s for high \(\Delta G\). A final copy with no consecutive \(w\)s  has \(\epsilon_{w}=0\) but \(\epsilon_{r}\neq0\). Maximizing the entropy rate of such a Markov chain over \(\epsilon_r\) gives \(H_{\rm max}= 0.48121\); \(\Delta G_{\rm pol}= -H_{\rm max}\) matches the location of our peak/cusp. Thus the combined mechanism initially eliminates consecutive \(w\)s, and at \(\Delta G_{\rm pol} =-0.48121 \) a state with \(\eta_w=0\) is thermodynamically permitted for the first time. For large \(\Delta G\), this polymer is kinetically accessible and grows with thermodynamic efficiency approaching unity (fig.~\ref{Efficiency}\,c). In this limit, the overall entropy generation is zero.


The above behaviour is a striking example of correlations being generated within the copy sequence, as well as with the template. Notably, whilst \(\eta\) approaches unity at this point, \(\eta_{\rm ss}\) does not (fig.~\ref{Efficiency}). Correlations {\it within} the copy sequence limit the chemical work that can be devoted to improving the single site accuracy of the copy polymer, since they prevent the bound \(H_{\rm ss}\geq H\geq -\Delta G_{\rm pol}\) from being saturated. 

\section{Conclusion}
The thermodynamic constraint on copying that underlies this work is deceptively simple: unlike templated self assembly, the overall chemical contribution to the free-energy of a copy must be independent of the match between template and copy sequences. By studying the simplest mechanisms satisfying this constraint, however, we can draw important conclusions for copying mechanisms and thermodynamics more generally.

For copying, the most immediate contrast with  previous work on templated self assembly \cite{Bennett,Cady,Andrieux,Sartori1,Sartori2,Gaspard,esposito2010} is that accuracy is necessarily zero when the copy assembles in equilibrium, since equilibrium correlations between physically separated polymers are impossible~\cite{Ouldridge}. Consequently, unlike self-assembly, no autonomous copying system can rely on relaxation to  near-equilibrium; fundamentally different paradigms are required.   

A direct result of the temporary nature of thermodynamic discrimination in persistent copying is that relying solely on the strong binding of correct copies is ineffective in ensuring accuracy. At \(\Delta G= 6k_BT\),  comparable to the cost of a mismatched base pair \cite{SantaLucia2004}, the temporary thermodynamic discrimination model never improves upon \(\epsilon = 0.0285\), which is more than ten times the equilibrium error probability based on energetic discrimination obtainable in templated self assembly, \(1/(1+\exp(\Delta G))\). This performance would degrade further if many competing monomers were present. Mechanisms for copying must therefore be more carefully optimized than those for templated self assembly. Either some degree of direct ``kinetic'' discrimination (with correct monomers incorporated more quickly), or as an alternative, fuel-consuming proofreading cycles, appear necessary. It is well-established that proofreading cycles can enhance discrimination above equilibrium in templated self assembly  \cite{Sartori2,Bennett,EHRENBERG1980333}, and the challenges in achieving direct kinetic discrimination in diffusion-influenced reactions via the details of the microscopic sub-steps may explain the ubiquity of such cycles in true copying systems.



Correlations within the copy, as well as between copy and template, arise naturally in persistent mechanisms. Indeed, in one case, pairs of mistakes are eliminated well before individual mistakes. These correlations contribute to the non-equilibrium free energy of the final state, reducing the single-site copy accuracy achievable for a given chemical work input. Biologically, however, it is arguably the accuracy of whole sequences, rather than individual monomers, that matters. In this case, positive correlations could advantageously  increase the number of 100\% correct copies for a given average error rate. It remains to be seen whether this tactic, which indeed may arise in real systems \cite{Rao}, is feasible. Regardless, we predict that within-copy correlations may be significant, particularly in simple systems with low accuracy. These correlations may change significantly if the requirement to remain bound with one exactly bond is relaxed, and exploiting correlations to extend functionality  beyond simple copying is an intriguing prospect. 

Relaxing this requirement also raises the possibility of early copy detachment. This risk is likely to be worse for genome replication than for transcription and translation, which may explain why the latter proceed by mechanisms analogous to our model, while DNA replication does not: here, the copy of a single DNA strand from the double helix is first completely assembled on the template and separated only much later.

Thermodynamically, a persistent copying mechanism  converts the high free energy of the input molecules into a high free-energy copy state; we have defined a general efficiency of this free-energy transduction for copying. In typical physical systems with tight coupling of reactions, high efficiency occurs when the load is closely matched to the driving, either in autonomous systems operating near the stall force, or in quasistatically manipulated systems. For the polymer copying mechanisms studied here, however, we find that the thermodynamic efficiency of information transfer, and not just the accuracy, approaches zero as polymer growth stalls. We predict that this result is general, since the alternative would require a sub-linear convergence of the error rate on 50\% as thermodynamic driving tends towards the stall point. 

Fundamentally, the copy process transduces free energy into a complex system with many  degrees of freedom (the sequence), and not just the polymer length. To be accurate, the sequence must be prevented from equilibrating. Thus, whilst weak driving leads to polymer growth with little overall entropy generation, it does a poor job of pushing the polymer sequence out of equilibrium. We predict that similar behaviour will arise in other systems intended to create an output in which a subset of the degrees of freedom are out of equilibrium.

Away from the equilibrium limit, the efficiency shows one or more peaks as the polymerization free energy is varied. At these peaks, the system transitions between two non-equilibrium states with remarkably little dissipation. These particular values of \(\Delta G_{\rm pol}\) are sufficient to stabilise non-equilibrium distributions that happen to be especially kinetically accessible, rendering the true equilibrium particularly inaccessible. This alignment of kinetic and thermodynamic factors is most evident in the combined mechanism that efficiently produces a state with few adjacent mismatches  at \(\Delta G_{\rm pol} = -0.48121\). These results slightly qualify the prediction of Ref. \cite{Ouldridge} that accurate copying is necessarily entropy generating, since entropy generation can be made arbitrarily small, whilst retaining finite copy accuracy, by taking \(\Delta G\rightarrow \infty\) at these specific values of \(\Delta G_{\rm pol}\).

The behaviour of the efficiency in these models emphasizes the importance, in both natural and synthetic copying systems, of kinetically preventing equilibration. Our work emphasizes that this paradigm should be applied not only to highly-evolved systems with kinetic proofreading mechanisms \cite{Hopfield}, but also the most basic mechanisms imaginable. 

Extending our analysis to consider fuel-consuming proofreading cycles would be natural. However these cycles will not change the fundamental result that the entropy of the copy sequence is thermodynamically constrained, in this case by \(H_{\rm ss} \geq H \geq -\Delta G_{\rm pol} - \Delta G_{\rm fuel}\), where the final term is the additional free energy expended per step to drive the system around proofreading cycles. We predict that non-equilibrium proofreading cycles, by their very nature, are unlikely to approach efficiencies of unity.

The next step in our research however is to consider the full polymerisation process, including the first step in which a single polymer comes out of solution, and the final step in which the polymer fully detaches from the template and mixes with the surrounding baths. We predict that these ``edge effects" will have the largest impact on shorter oligomers, and so the research will start with a dimerisation process.

\newpage
\setcounter{equation}{0}
\setcounter{figure}{0}
\setcounter{table}{0}
\setcounter{section}{0}

\section*{Edge-effects dominate copying thermodynamics for finite-length molecular oligomers}

\setcounter{equation}{0}
\setcounter{figure}{0}
\setcounter{table}{0}
\setcounter{section}{0}

\begin{figure*}
    \centering
    \includegraphics[scale=0.3]{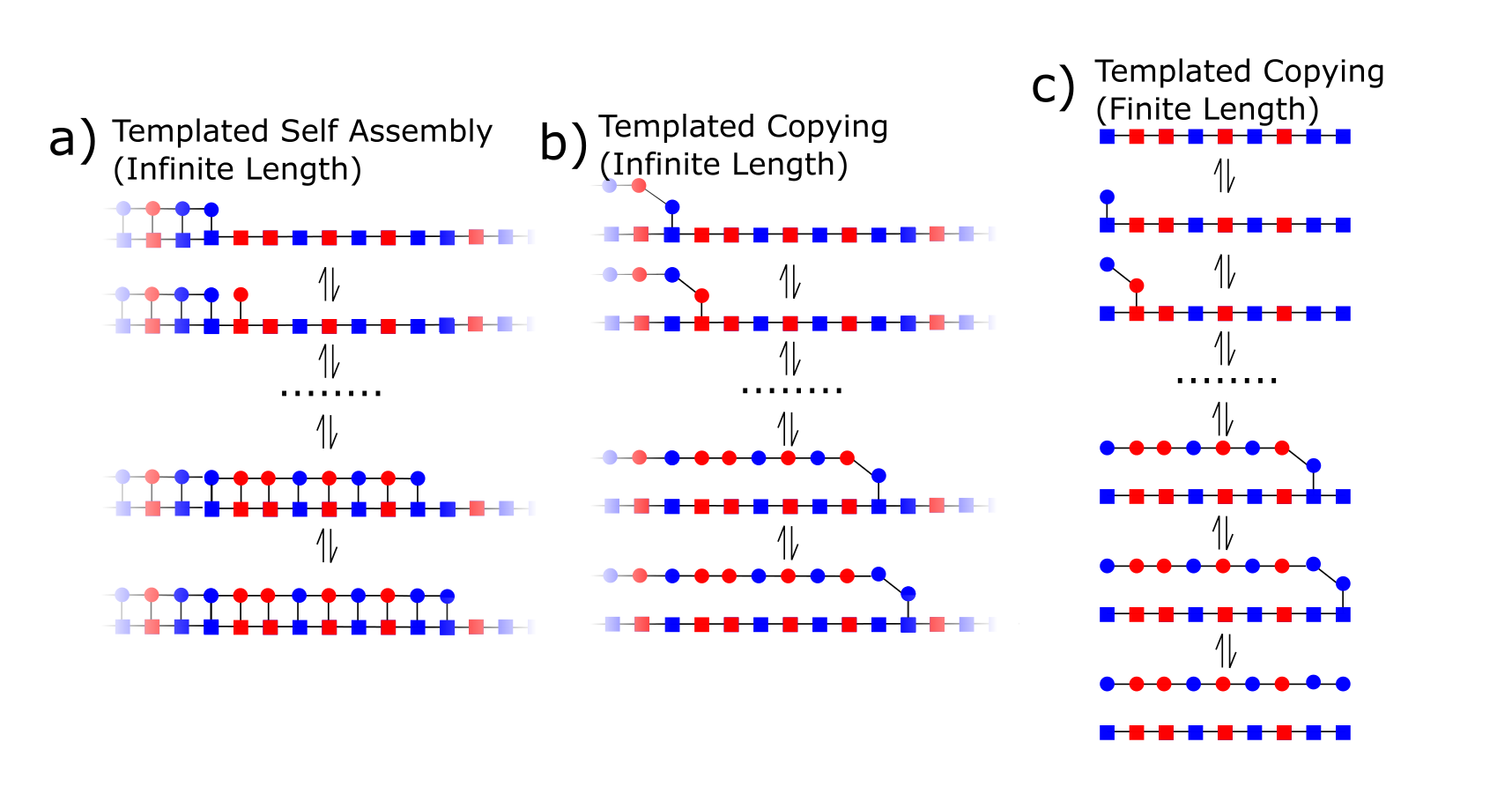}
    \caption{(a) A model for templated self-assembly on an infinitely long template. Here bonds retained between copy and template stabilise the low entropy state. (b) A model of persistent copying on an infinitely long template. Association of the first monomer, and complete dissociation of the complete polymer are omitted. (c)  A model of persistent copying on a finite oligomer template, including complete dissociation of full length oligomers and single monomers into surrounding baths. In all cases the polymer/oligomer grows sequentially and sequence-specifically on the template, with the copy process oligomer separating from the template behind its leading edge \cite{Poulton}.}
    \label{Diagrams}
\end{figure*}

\begin{figure}
    \centering
    \includegraphics[scale=0.35]{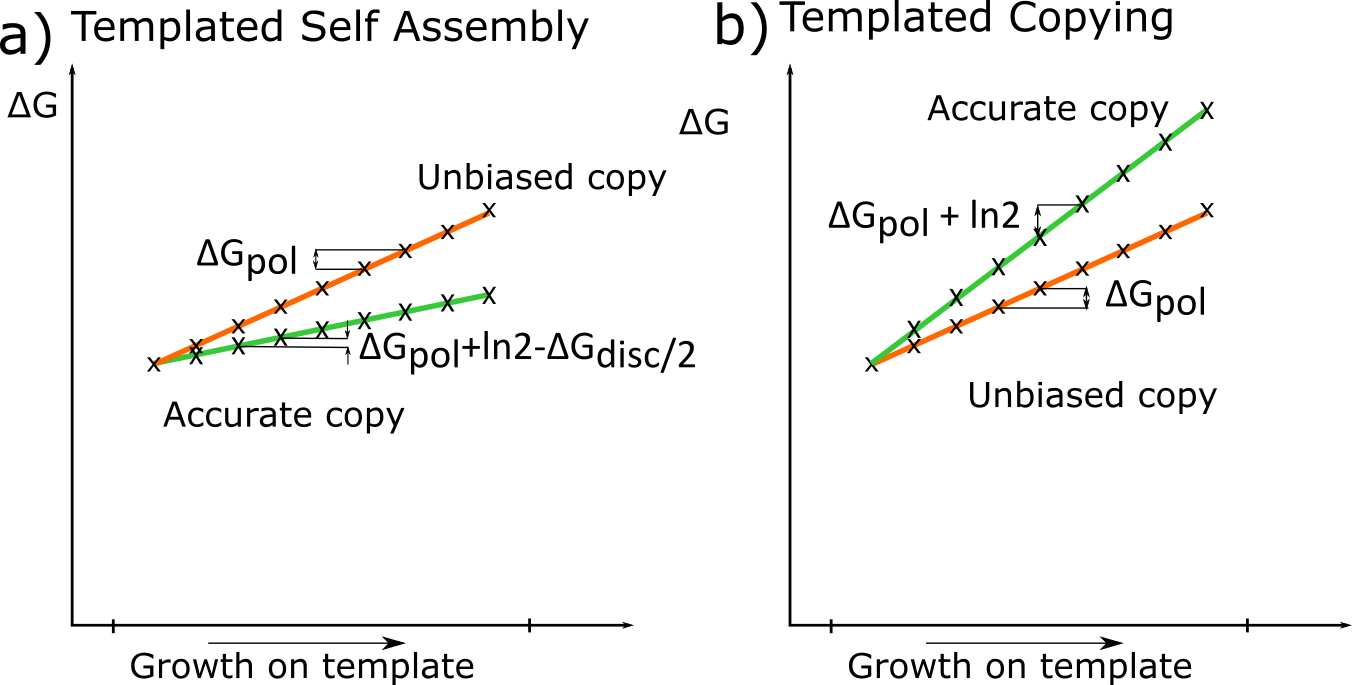}
    \caption{(a) Templated self-assembly on an infinite length polymer and (b) templated copying with sequential separation. Growth on the template in all cases has a a constant slope with a gradient that depends on the average free-energy of adding an arbitrary monomer to the growing chain (\(\Delta G_{\rm pol}\)) and the accuracy. The entropy cost between a perfectly accurate copy and an unbiased copy is \(\ln{2}\), increasing the gradient of the slope. In templated self-assembly, favourable copy/template bonds can overcome the cost of correlating copy and template, meaning that accuracy is favoured with a shallower gradient. For templated copying, accuracy increases the gradient due to the cost of correlating the two sequences without compensating interactions.}
    \label{NewFig1}
\end{figure}

So far, this thesis\cite{Ouldridge,Poulton}, has made some crucial observations about the consequences of explicitly considering separation of polymer copies from a template, considering a system shown in fig. \ref{Diagrams}(b) in which a growing polymer separates sequentially from the template as it grows. We observed that for a theoretical system with two monomer types, the entropic cost of creating a perfectly accurate polymer is \(\ln{2}\) above that of creating an unbiased polymer (\(\Delta G_{\rm pol}\) per monomer). In a model which omits separation, known as templated self-assembly (illustrated in fig. \ref{NewFig1}(b)), the stronger copy/template interaction of binding a correct match can compensate for this. 

Throughout this work we define the free-energy change of a matching monomer (type \(1\)) to be \(\Delta G^\plimsoll_{r} -\ln[1]\) and a mismatching monomer (here type \(2\)) to be \(\Delta G^\plimsoll_{w} -\ln[2]\), in units of \(k_{\rm B}T\). The difference between the two at the reference concentration, \(\Delta  G_{\rm disc}=\Delta G_{w}^\plimsoll-\Delta G^\plimsoll_{r}\), is the discrimination free-energy. In a model that omits separation, for an arbitrarily large \(\Delta G_{\rm disc}\), the polymer can be arbitrarily accurate at equilibrium. Explicitly considering separation changes this. With the stabilising bonds broken, the cost of creating the accurate polymer must be paid, and accuracy requires the system to be out of equilibrium (fig. \ref{NewFig1}(b)).

Additionally, the need to separate the copy and template fundamentally changes the detailed mechanics of the process, as copy/template bonds that favour accuracy are temporary. Therefore, the system has no permanent free-energetic biases towards accuracy and sequence discrimination becomes a purely kinetic phenomenon~\cite{Ouldridge,Poulton}. In ch. 1\cite{Poulton}, we further observe that for finite binding free energies, excess polymerisation free-energy could not be fully transformed into free-energy stored in sequence information by the model systems considered. Producing an accurate copy was therefore seen to be necessarily thermodynamically irreversible.

These previously-obtained results, both for templated self-assembly \cite{Bennett,Cady,Andrieux,Sartori1,Sartori2,esposito2010,EHRENBERG1980333,Johansson} and for a polymer that continuously separates from its template \cite{Poulton}, were derived for infinite-length polymers. The tip of the growing copy is assumed to reach a steady velocity along the template, and the identity of monomers measured relative to this tip reach a stationary distribution. Initiation and termination of the polymerization process were ignored, and even in ch. 1~\cite{Poulton} the copy remains attached by a single bond - complete detachment was neglected. This omits the significant difference between the entropy of a copy bound to a template by a single monomer and the much larger entropy of a free polymer in solution.

This approximation might be reasonable for some long biopolymers {\it in vivo}, but it is a poor approximation for the copying of shorter oligomers and dimers.  Given that synthetic systems are currently limited to short oligomers, and that early life is likely to have created short oligomers before the origin of complex enzyme-based copying machinery, it is worth studying initiation and termination in more detail. Equally, ``charged" tRNA -  hybrid molecules of one tRNA codon and its matching amino acid that are necessary for the process of RNA translation into proteins - are dimerised via a specific template enzyme called a synthetase~\cite{gomez2020aminoacyl}.  This process is essentially the copying of a dimer template, and so the creation of short copies is highly relevant even to extant biology.


In this chapter we probe the consequences of the ``edge-effects" of initial attachment and final detachment on the copying of oligomer sequences.
We first consider the free-energy change for the production of a single finite-length copy under constant external conditions, separating it into chemical and informational terms.  Using  dimerisation as an example,  we show that the overall thermodynamic constraints on information transfer are fundamentally altered relative to infinite-length polymers: in general there is no entropic cost to accurately reproducing the template sequence, in direct contrast to previous work on infinite length templates\cite{Poulton,Bennett,Cady,Andrieux,Sartori1,Sartori2,esposito2010,EHRENBERG1980333,Johansson}. Instead, the template acts as an information engine, selectively coupling to a large number of out-of-equilibrium molecular reservoirs; the relationship between the copy sequence and these reservoirs sets the overall thermodynamics. Although these results hold for oligomers of arbitrary finite length, we nonetheless observe a gradual cross-over to the previously predicted constraints on accuracy  \cite{Poulton} for longer oligomers produced by a particular dynamical model of copying. The thermodynamic constraints on accuracy in the infinite length limit instead become kinetic barriers for finite length templates.

%
\begin{figure}
    \includegraphics[scale=0.47]{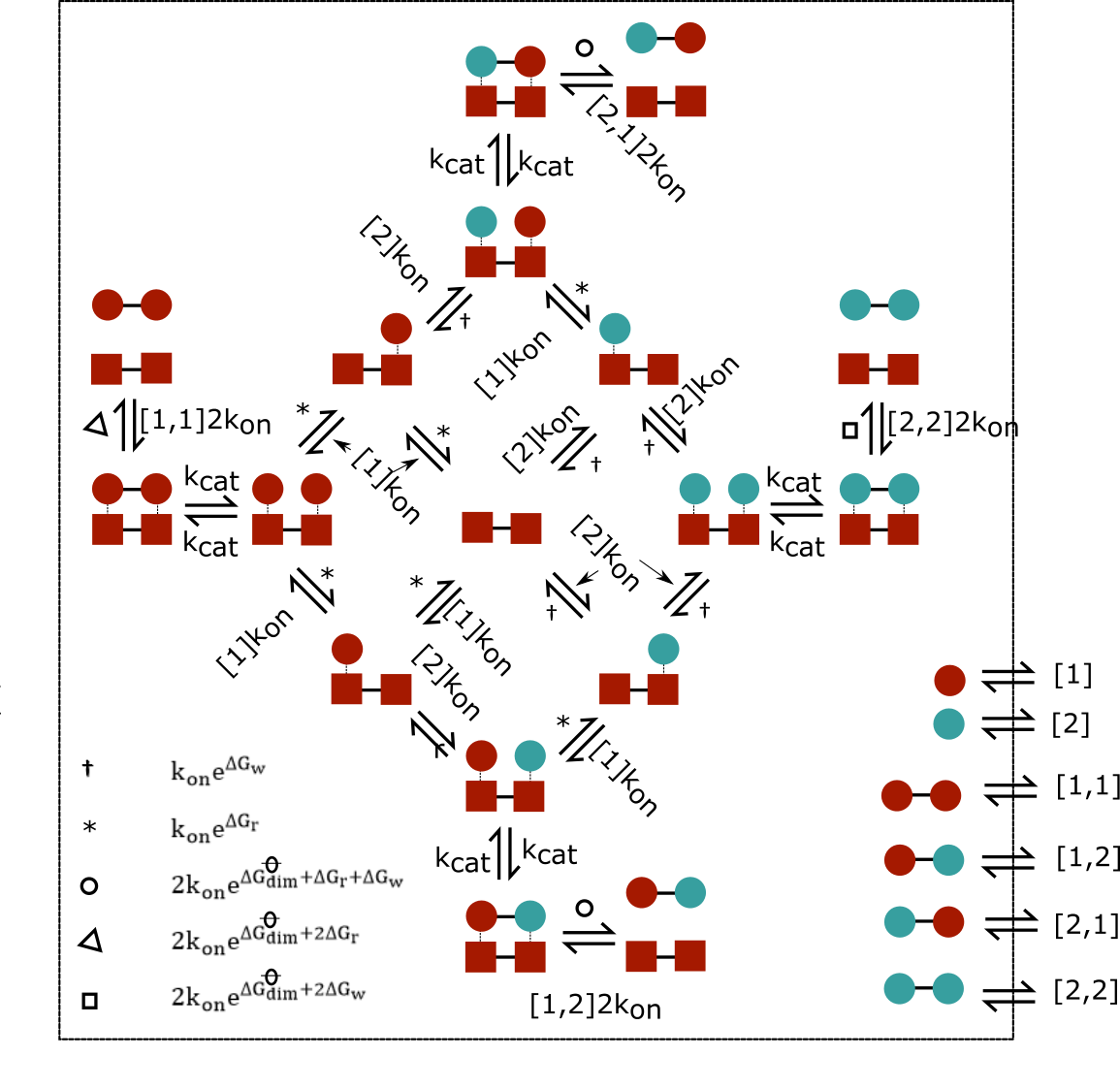}
    \caption{{\it Minimal model of oligomerisation illustrated with dimers.} A template (squares) interacts with baths of monomers and dimers of a second species (circles). Monomers can bind to the template and dimerize, while dimers binding to the template can be destroyed or interconverted. 
    The standard dimerisation free-energy is \(\Delta G^\plimsoll_{\rm dim}\) for all sequences, but the free-energy change of matching and non-matching varieties of monomer coming out of solution and binding to the template at the reference concentration are \(\Delta G^\plimsoll_r-[1]\) and \(\Delta G^\plimsoll_w-[2]\), respectively,  allowing selectivity. Rate constants $k_{\rm cat}$ and $k_{\rm on}$ define the dynamics.}
    \label{Fig1}
\end{figure}

\section{Model of oligomerisation} 
Fig.~\ref{Fig1} shows a dimerisation system, a prototype for a broader class of oligomerisation systems. A solvated template dimer carries information in its sequence of monomer units; in this case, the sequence is 1,1. The template is coupled to large baths of monomers and oligomers (in this case dimers) of a distinct second type of molecule, like a DNA template in a bath of RNA nucleotides and oligomers. This second type of molecule also  comes in multiple varieties -- in this case two -- and can interact with the template in a sequence-specific way. In  Fig.~\ref{Fig1}, we propose a particular thermodynamically-consistent model for dimer production. Dimers from the baths can also be broken down into their component monomers by the template.

We start by defining \(\Delta G^{\plimsoll}_{\rm dim}\) as the free-energy change of dimerisation at the reference concentration. The concentrations of the two monomer types  are \([1]\) and \([2]\) (defined with respect to a standard reference concentration), and the four oligomer types are \([1,1]\), \([1,2]\),\([2,1]\) and \([2,2]\).

 The total free-energy change of the baths upon creating a single dimer of sequence \({\bf s}=s_1,s_2\) is therefore
\(
 \Delta G({\bf s})=\Delta G_{\rm{dim}}^\plimsoll+ \left(\ln{[{\bf s}]}-\ln{[s_{1}][s_{2}]}\right)
\).
Since the template itself acts catalytically \cite{Ouldridge}, its free-energy is unchanged.
In addition, for simplicity, we do not allow for kinetic proofreading cycles \cite{Hopfield} in our analysis. This means that the process is tightly coupled; it doesn't have an unknown and variable number of fuel-consuming futile cycles. The free-energy change of the oligomerisation process is therefore unambiguously defined as \(\Delta G({\bf s})\). This free-energy change is not dependent on any specific templating mechanism and our first result is derived for a generalised oligomerisation system.

\section{Properties of general oligomerisation systems}

For a generalized system with oligomers of length \(|{\bf{s}}|\), %
\begin{equation}
 \Delta G({\bf s})=(|{\bf s}|-1)\Delta G^\plimsoll_{\rm{dim}}+ \left(\ln{[{\bf s}]}-\ln{\prod_{i=1}^{|{\bf s}|}[s_{i}]}\right).
\end{equation}\\
Let \(J({\bf s})\) be the expected net rate at which sequence \({\bf s}\) is produced by the the system when the template reaches a steady state.
The average rate of change of free-energy is then \(\Delta \dot{G}=\sum_{\bf s} J({\bf s})\Delta G({\bf s})\). We define the normalised flux; \(q({\bf s})=J({\bf s})/J_{\rm{tot}}\). We further define the following probability distributions: \(p({\bf s})=[{\bf s}]/[S_{\rm{tot}}]\), the probability of picking an oligomer of sequence \(\bf{s}\) from the oligomers with total concentration \([S_{\rm{tot}}]\); \(m({s})=[{ s}]/[M_{\rm tot}]\), the probability of picking a monomer of type \({s}\) from the monomers with total concentration of \([M_{\rm tot}]\); and \(t({\bf s})= \prod_{i}m(s_{i})\), which corresponds to the probability of the sequence \({\bf s}\) occurring by selecting monomers randomly from the monomer pools. In these terms, 
\begin{align*}
&\Delta \dot{G}=J_{\rm{tot}}\left(\sum_{{\bf s}}q({\bf s})\left(|{\bf s}|-1\right)\Delta G^\plimsoll_{\rm{dim}}\right)\\
+&J_{\rm{tot}}\left( \sum_{{\bf s}}q({\bf s})\ln{\frac{p({\bf s})}{t(\bf s)}}+\sum_{{\bf s}}q({\bf s})\ln{\frac{[S_{\rm{tot}}]}{[M_{tot}]^{|\bf s|}}} \right).
\end{align*}
This expression can be re-written as
\begin{equation}
 \Delta \dot{G}=J_{\rm{tot}}\Delta G_{\rm{chem}} + J_{\rm{tot}}\Delta G_{\rm{inf}},
 \label{INFCHEM}
\end{equation}\\
with
\begin{align}
\Delta G_{\rm chem} &=  (|{\bf s}|-1) \Delta G^\plimsoll_{\rm dim} + \ln{[S_{\rm tot}]/[M_{\rm tot}]^{|{\bf s}|}}, \nonumber \\
\Delta G_{\rm{inf}} &= \sum_{{\bf s}}q({\bf s})\ln{\frac{p({\bf s})}{t(\bf s)}},
\label{dG_inf}
\end{align}
assuming for simplicity that all oligomers are of the same length. The first term in Eq.~\ref{INFCHEM} is the average chemical free-energy change of oligomerisation ignoring sequence, multiplied by the net rate of oligomer production. The second term is
information-theoretic: for non-negative net production of all oligomers \(J({\bf s}) = q({\bf s})J_{\rm tot} \geq 0\), \( q({\bf s})\) is the probability of picking a sequence \({\bf s}\) from the pool of net products and \(\Delta G_{\rm inf} = D(q||t) -D(q||p)\), where \(D(q||p) = \sum_{\bf s}q({\bf s})\log{\frac{q({\bf s})}{p({\bf s})}}\) is the  Kullback-Leibler divergence between \(q({\bf s})\) and \(p({\bf s})\). \(\Delta G_{\rm inf}\) reflects the sequence statistics of monomer and oligomer baths, and the sequence-dependence of net oligomer production. This splitting into chemical and informational terms holds for arbitrary oligomer lengths and sequence alphabets, and is the first result of this chapter.

\section{An explicit model of dimerisation to explore the consequences of the general result in Eq.~\ref{INFCHEM}}

We return to our model, shown in Fig.~\ref{Fig1}, with two varieties of monomer in the template and copy. Here \(k_{\rm on}\) is the underlying rate constant for the process of adding or removing a monomer or dimer from the template. As above, \([1]\) and \([2]\) are the monomer concentration of monomer types 1 and 2 respectively (defined with respect to a standard reference concentration) and \([1,1]\), \([1,2]\), \([2,1]\) and \([2,2]\) are the concentrations of the four types of dimers. The free-energy change when a matching monomer binds to the template (here both matching monomers are of type \(1\) and the template is type \(1,1\)) is \(\Delta G^\plimsoll_{r} -\ln[1]\) and a mismatching monomer (here type \(2\)) is \(\Delta G^\plimsoll_{w} -\ln[2]\), in units of \(k_{\rm B}T\). The difference between the two at the reference concentration, \(\Delta  G_{\rm disc}=\Delta G_{w}^\plimsoll-\Delta G^\plimsoll_{r}\), is the discrimination free-energy. As above, \(\Delta G^{\plimsoll}_{\rm dim}\) is the free-energy change of dimerisation at the reference concentration. In the model of Fig.~\ref{Fig1}, we assume that any free-energy released by dimerisation is used to destabilise the bonds between dimer and template. The state with two monomers and the states with a dimer bound to the template therefore have equal free-energy, with \(k_{\rm cat}\) the underlying rate of the formation or breaking of the backbone bond. Such a free-energy landscape has been proposed to be optimal for minimising product inhibition\cite{deshpande2019optimizing}. Here, matching and non-matching monomers bind to the template with the same rate constant \(k_{\rm on}\), but since \(\Delta G^\plimsoll_{r}<\Delta G^\plimsoll_{w}\), mismatches detach faster. Since we use an unbiased \(m( s)\) in all cases, the symmetry of the problem gives identical physics for all template sequences; we shall use 1,1 for clarity.
We assume constant bath concentrations, and calculate the flux \(J({\bf s})\) in steady state by analysing a single template as a Markov process.

\section{Finding the fluxes through the network using a transition matrix}

Fig. \ref{Fig1} defines a Markov process for the states of a single template of type \(1,1\). Note that we assume the dimers have a directionality (like biopolymers such as nucleic acids and polypeptides), so that 1,2 is distinct from 2,1. Below, we use ``left" to refer to the first site and ``right" to the second, for consistency with Fig. \ref{Fig1}. 

The available states are as follows; 
\small
\begin{itemize}
    \item State 0: the empty template (shown in five different locations in Fig. \ref{Fig1}, the four outer edges and the centre).
    \item State 1: the template with an incorrect monomer (2) on its left side.
    \item State 2: the template with an incorrect monomer (2) on its right side.
    \item State 3: the template with an correct monomer (1) on its left side.
    \item State 4: the template with an correct monomer (1) on its right side.
    \item State 5: the template with an incorrect monomer (2) on its left side and a correct monomer (1) on its right side.
    \item State 6: the template with an incorrect monomer (2) on its left side and an incorrect monomer (2) on its right side.
    \item State 7: the template with a correct monomer (1) on its left side and an incorrect monomer (2) on its right side.
    \item State 8: the template with a correct monomer (1) on its left side and a correct monomer (1) on its right side.
    \item State 9: the template with a \(2,2\) dimer attached to it.
    \item State 10: the template with a \(1,2\) dimer attached to it.
    \item State 11: the template with a \(1,1\) dimer attached to it.
    \item State 12: the template with a \(2,1\) dimer attached to it.
\end{itemize}
\normalsize

Using these states we can set up a rate matrix $K$ where $K_{xy}$ gives the transitions out of state \(x\) and into state \(y\):
\small
\begin{frame}{}
\footnotesize
\setlength{\arraycolsep}{2.5pt}
\medmuskip = 1mu 
\[\left(
\begin{array}{ccccccccccccccccc}
-X_{0} & [2]k_{\rm on} & [2]k_{\rm on} & [1]k_{\rm on} & [1]k_{\rm on} & 0 & 0 & 0 & 0 & 2[2,2]k_{\rm on} & 2[1,2]k_{\rm on} & 2[1,1]k_{\rm on} & 2[2,1]k_{\rm on}\\
k_{\rm on}e^{\Delta G_{w}} & -X_{1} & 0 & 0 & 0 & [1]k_{\rm on} & [2]k_{\rm on} & 0 & 0 & 0 & 0 & 0 & 0\\
k_{\rm on}e^{\Delta G_{w}} & 0 & -X_{2} & 0 & 0 & 0 & [2]k_{\rm on} & [1]k_{\rm on} & 0 & 0 & 0 & 0 & 0\\
k_{\rm on}e^{\Delta G_{r}^{\plimsoll}} & 0 & 0 & -X_{3} & 0 & [2]k_{\rm on} & 0 & 0 & [1]k_{\rm on} & 0 & 0 & 0 & 0\\
k_{\rm on}e^{\Delta G_{r}^{\plimsoll}} & 0 & 0 & 0 & -X_{4} & 0 & 0 & [1]k_{\rm on} & [2]k_{\rm on} & 0 & 0 & 0 & 0\\
0 &k_{\rm on}e^{\Delta G_{r}^{\plimsoll}} & 0 & k_{\rm on}e^{\Delta G_{w}^{\plimsoll}} & 0 & -X_{5} & 0 & 0 & 0 & 0 & 0 & 0 & k_{\rm cat}\\
0& k_{\rm on}e^{\Delta G_{w}^{\plimsoll}} & k_{\rm on}e^{\Delta G_{w}^{\plimsoll}} & 0 & 0 & 0 & -X_{6} & 0 & 0 & k_{\rm cat} & 0 & 0 & 0\\
0 & 0 & k_{\rm on}e^{\Delta G_{r}^{\plimsoll}} & 0 & k_{\rm on}e^{\Delta G_{w}^{\plimsoll}} & 0 & 0 & -X_{7} & 0 & 0 & k_{\rm cat} & 0 & 0\\
0 & 0 & 0 & k_{\rm on}e^{\Delta G_{r}^{\plimsoll}} & k_{\rm on}e^{\Delta G_{r}^{\plimsoll}} & 0 & 0 & 0 & -X_{8} & 0 & 0 & k_{\rm cat} &  0\\
2k_{\rm on}e^{\ddagger} & 0 & 0 & 0 & 0 & 0 & k_{\rm cat} & 0 & 0 & -X_{9} & 0 & 0 & 0\\
2k_{\rm on}e^{*} & 0 & 0 & 0 & 0 & 0 & 0 & k_{ \rm cat} & 0 & 0 & -X_{10} & 0 & 0\\
2k_{\rm on}e^{\dagger} & 0 & 0 & 0 & 0 & 0 & 0 & 0 & k_{\rm cat} & 0 & 0 & -X_{11} & 0\\
2k_{\rm on}e^{*} & 0 & 0 & 0 & 0 & k_{\rm cat} & 0 & 0 & 0 & 0 & 0 & 0 & -X_{12}\\
\end{array}\right),\]
\end{frame}
\normalsize
where \(* = G^\plimsoll_{\rm dim} + \Delta G^\plimsoll_{r} + \Delta G^\plimsoll_{w}\), \(\dagger = G^\plimsoll_{\rm dim} + 2\Delta G^\plimsoll_{r}\) and \(\ddagger=\Delta G^\plimsoll_{\rm dim} + 2\Delta G_{w}^\plimsoll\). \(X_{x}\) is the sum over all the other terms in row \(x\). Now we can solve for the steady state \(\pi\) by finding the appropriate left-eigenvector \(\pi K = 0\). 

Next we consider the probability of the system creating either a \(1,1\), \(1,2\), \(2,1\) or \(2,2\) dimer, or destroying a dimer into component monomer parts, given an initial state with either a monomer or dimer bound to the template. We thus split the empty state (state 0) into five destination states, as follows: 13) the empty state having just released a monomer (corresponding to destruction), 14) the empty state having just released a \(1,1\) dimer, 15) the empty state having just released a \(1,2\) dimer, 16) the empty state having just released a \(2,1\) dimer and 17) the empty state having just released a \(2,2\) dimer. We put these states on the end of the list of states above, and delete the original empty state 0. Treating those states as distinct absorbing states, we can calculate the probability of reaching any one of them first, given a specific staring point\cite{peters2017reaction}. To do this we use the following  transition matrix that describes the discrete-time process embedded in the continuous-time model of dimerisation. This embedded discrete time process describes the sequence of states visited, without reference to the time taken. The matrix takes the form

\begin{frame}{}
\footnotesize
\setlength{\arraycolsep}{2.5pt}
\medmuskip = 1mu 
\[\left(
\begin{array}{c|c}
M & A\\
\cline{1-1}\cline{2-2}
0 & I
\end{array}\right),\]
\end{frame}
\normalsize

where \(M\) is
\begin{frame}{}
\footnotesize
\setlength{\arraycolsep}{2.5pt}
\medmuskip = 1mu 
\[\left(
\begin{array}{cccccccccccc}
0 & 0 & 0 & 0 & \frac{[1]k_{\rm on}}{N_{1}} & \frac{[2]k_{\rm on}}{N_{1}} & 0 & 0 & 0 & 0 & 0 & 0\\
0 & 0 & 0 & 0 & 0 & \frac{[2]k_{\rm on}}{N_{2}} & \frac{[1]k_{\rm on}}{N_{2}} & 0 & 0 & 0 & 0 & 0\\
0 & 0 & 0 & 0 & \frac{[2]k_{\rm on}}{N_{3}} & 0 & 0 & \frac{[1]k_{\rm on}}{N_{3}} & 0 & 0 & 0 & 0\\
0 & 0 & 0 & 0 & 0 & 0 & \frac{[1]k_{\rm on}}{N_{4}} & \frac{[2]k_{\rm on}}{N_{4}} & 0 & 0 & 0 & 0\\
\frac{k_{\rm on}e^{\Delta G_{r}^{\plimsoll}}}{N_{5}} & 0 & \frac{k_{\rm on}e^{\Delta G_{w}^{\plimsoll}}}{N_{5}} & 0 & 0 & 0 & 0 & 0 & 0 & 0 & 0 & \frac{k_{\rm cat}}{N_{5}}\\
\frac{k_{\rm on}e^{\Delta G_{w}^{\plimsoll}}}{N_{6}} & \frac{k_{\rm on}e^{\Delta G_{w}^{\plimsoll}}}{N_{6}} & 0 & 0 & 0 & 0 & 0 & 0 & \frac{k_{\rm cat}}{N_{6}} & 0 & 0 & 0\\
0 & \frac{k_{\rm on}e^{\Delta G_{r}^{\plimsoll}}}{N_{7}} & 0 & \frac{k_{\rm on}e^{\Delta G_{w}^{\plimsoll}}}{N_{7}} & 0 & 0 & 0 & 0 & 0 & \frac{k_{\rm cat}}{N_{7}} & 0 & 0\\
0 & 0 & \frac{k_{\rm on}e^{\Delta G_{r}^{\plimsoll}}}{N_{8}} & \frac{k_{\rm on}e^{\Delta G_{r}^{\plimsoll}}}{N_{8}} & 0 & 0 & 0 & 0 & 0 & 0 & \frac{k_{\rm cat}}{N_{8}} &  0\\
0 & 0 & 0 & 0 & 0 & \frac{k_{\rm cat}}{N_{9}} & 0 & 0 & 0 & 0 & 0 & 0\\
0 & 0 & 0 & 0 & 0 & 0 & \frac{k_{ \rm cat}}{N_{10}} & 0 & 0 & 0 & 0 & 0\\
0 & 0 & 0 & 0 & 0 & 0 & 0 & \frac{k_{\rm cat}}{N_{11}} & 0 & 0 & 0 & 0 \\
0 & 0 & 0 & 0 & \frac{k_{\rm cat}}{N_{12}} & 0 & 0 & 0 & 0 & 0 & 0 & 0
\end{array}\right),\]
\end{frame}
\normalsize
and \(T\) is
\begin{frame}{}
\footnotesize
\setlength{\arraycolsep}{2.5pt}
\medmuskip = 1mu 
\[\left(
\begin{array}{ccccc}
\frac{k_{\rm on}e^{\Delta G_{w}^{\plimsoll}}}{N_{1}} & 0 & 0 & 0 & 0 \\
\frac{k_{\rm on}e^{\Delta G_{w}^{\plimsoll}}}{N_{2}} & 0 & 0 & 0 & 0 \\
\frac{k_{\rm on}e^{\Delta G_{r}^{\plimsoll}}}{N_{3}} & 0 & 0 & 0 & 0 \\
\frac{k_{\rm on}e^{\Delta G_{r}^{\plimsoll}}}{N_{4}} & 0 & 0 & 0 & 0 \\
0 & 0 & 0 & 0 & 0 \\
0 & 0 & 0 & 0 & 0 \\
0 & 0 & 0 & 0 & 0 \\
0 & 0 & 0 & 0 & 0 \\
0 & \frac{2k_{\rm on}e^{\ddagger}}{N_{9}} & 0 & 0 & 0 \\
0 & 0 & \frac{2k_{\rm on}e^{*}}{N_{10}} & 0 & 0 \\
0 & 0 & 0 & \frac{2k_{\rm on}e^{\dagger}}{N_{11}} & 0 \\
0 & 0 & 0 & 0 & \frac{2k_{\rm on}e^{*}}{N_{12}}
\end{array}\right),\]
\end{frame}
\normalsize
where \(N_{x}\) normalises each row across both \(M\) and \(T\) \(x\) to unity. 

The sub matrix \(M\) quantifies the transitions between non-absorbing states. The sub matrix \(T\) quantifies transitions into the absorbing states. The bottom left quadrant is a zero matrix and the bottom right quadrant is the identity matrix representing the probability of leaving the absorbing states. From here we can find the fundamental matrix \(W=(I-M)^{-1}\), which quantifies the number of times the system visits a non-absorbing state on the way to being absorbed, given that it started in a particular state. From here we can calculate \(Z=W.T\). \(Z_{xy}\) gives the probability of eventually reaching absorbing state $12+y$ given a starting state $x$.

The overall rate in steady state at which the system produces dimers with the sequence \(1,1\) (absorbing state \(y=2\)) from monomers is simply \(\Phi_{1,1}^{\rm create} = \pi_0 \sum_{x=1}^{4} k_{0x} Z_{x2}\). Other rates for production, degradation and interconversion of dimers can be calculated similarly, giving
\small
\begin{equation}
    \Phi^{\rm create}_{2,2}/\pi_0 = [2]k_{\rm on}Z_{1,2} +[2]k_{\rm on}Z_{2,2} + 
    [1]k_{\rm on}Z_{3,2} + [1]k_{\rm on}Z_{4,2},\notag
\end{equation}
\begin{equation}
    \Phi^{\rm create}_{1,2}/\pi_0 = [2]k_{\rm on}Z_{1,3} +[2]k_{\rm on}Z_{2,3} + 
    [1]k_{\rm on}Z_{3,3} + [1]k_{\rm on}Z_{4,3},\notag
\end{equation}
\begin{equation}
    \Phi^{\rm create}_{2,1}/\pi_0  = [2]k_{\rm on}Z_{1,4} +[2]k_{\rm on}Z_{2,4} + 
    [1]k_{\rm on}Z_{3,4} + [1]k_{\rm on}Z_{4,4},\notag
\end{equation}
\begin{equation}
    \Phi^{\rm create}_{1,1}/\pi_0  = [2]k_{\rm on}Z_{1,5} +[2]k_{\rm on}Z_{2,5} + 
    [1]k_{\rm on}Z_{3,5} + [1]k_{\rm on}Z_{4,5},
\end{equation}
\footnotesize
\begin{equation}
    \Psi^{\rm destroy}_{2,2}/\pi_0  = 2k_{/\pi_0 \rm on}Z_{9,1},\hspace{5mm}
    \Psi^{\rm destroy}_{2,1}/\pi_0  = 2k_{\rm on}Z_{10,1},\hspace{5mm}
    \Psi^{\rm destroy}_{1,2}/\pi_0  = 2k_{\rm on}Z_{12,1},\hspace{5mm}
    \Psi^{\rm destroy}_{1,1}/\pi_0  = 2k_{\rm on}Z_{11,1},
\end{equation}
\begin{equation}
\Psi^{\rm switch}_{1,1\rightarrow1,2}/\pi_0 =2k_{\rm on}Z_{11,3},\hspace{5mm}
\Psi^{\rm switch}_{1,1\rightarrow2,1}/\pi_0 =2k_{\rm on}Z_{11,5},\hspace{5mm}
\Psi^{\rm switch}_{1,1\rightarrow2,2}/\pi_0 =2k_{\rm on}Z_{11,2},\notag
\end{equation}
\begin{equation}
\Psi^{\rm switch}_{1,2\rightarrow1,1}/\pi_0 =2k_{\rm on}Z_{10,4},\hspace{5mm}
\Psi^{\rm switch}_{1,2\rightarrow2,1}/\pi_0= k_{\rm on}Z_{10,5},\hspace{5mm}
\Psi^{\rm switch}_{1,2\rightarrow2,2}/\pi_0 =2k_{\rm on}Z_{10,2}, \notag
\end{equation}
\begin{equation}
\Psi^{\rm switch}_{2,1\rightarrow1,2}/\pi_0 =2k_{\rm on}Z_{12,3},\hspace{5mm}
\Psi^{\rm switch}_{2,1\rightarrow1,2}/\pi_0 =2k_{\rm on}Z_{12,4},\hspace{5mm}
\Psi^{\rm switch}_{2,1\rightarrow1,2}/\pi_0=2k_{\rm on}Z_{12,2}, \notag
\end{equation}
\begin{equation}
\Psi^{\rm switch}_{2,2\rightarrow1,2}/\pi_0=2k_{\rm on}Z_{9,3},\hspace{5mm}
\Psi^{\rm switch}_{2,2\rightarrow1,2}/\pi_0=2k_{\rm on}Z_{9,5},\hspace{5mm}
\Psi^{\rm switch}_{2,2\rightarrow1,2}/\pi_0=2k_{\rm on}Z_{9,4},
\end{equation}
\normalsize
where \(\Psi_{x,y}\) is a rate per unit concentration of dimer \(x,y\), and \(\Phi\) is an absolute rate.
From here we can identify the total fluxes \(J({\bf s})\):
\footnotesize
\begin{equation}
   J(1,1)= \Phi^{\rm create}_{1,1} + [1,2]\Psi^{\rm switch}_{1,2\rightarrow1,1} + [2,1]\Psi^{\rm switch}_{2,1\rightarrow1,1} + [2,2]\Psi^{\rm switch}_{2,2\rightarrow1,1}- [1,1]\left( \Psi^{\rm destroy}_{1,1} + \Psi^{\rm switch}_{1,1\rightarrow1,2} + \Psi^{\rm switch}_{1,1\rightarrow2,1} + \Psi^{\rm switch}_{1,1\rightarrow2,2}\right),\notag
\end{equation}
\begin{equation}
    J(1,2)=\Phi^{\rm create}_{1,2} + [1,1]\Psi^{\rm switch}_{1,1\rightarrow1,2} + [2,1]\Psi^{\rm switch}_{2,1\rightarrow1,2} + [2,2]\Psi^{\rm switch}_{2,2\rightarrow1,2} - [1,2]\left( \Psi^{\rm destroy}_{1,2} + \Psi^{\rm switch}_{1,2\rightarrow1,1} + \Psi^{\rm switch}_{1,2\rightarrow2,1} + \Psi^{\rm switch}_{1,2\rightarrow2,2}\right),\notag
\end{equation}
\begin{equation}
    J(2,1)=\Phi^{\rm create}_{2,1} + [1,2]\Psi^{\rm switch}_{1,2\rightarrow1,1} + [1,1]\Psi^{\rm switch}_{1,1\rightarrow2,1} + [2,2]\Psi^{\rm switch}_{2,2\rightarrow2,1} -  [2,1]\left( \Psi^{\rm destroy}_{2,1} + \Psi^{\rm switch}_{2,1\rightarrow1,2} + \Psi^{\rm switch}_{2,1\rightarrow1,1} + \Psi^{\rm switch}_{2,1\rightarrow2,2}\right),\notag
\end{equation}
\begin{equation}
    J(2,2)=\Phi^{\rm create}_{2,2} + [2,2]\Psi^{\rm switch}_{2,2\rightarrow1,1} + [2,2]\Psi^{\rm switch}_{2,2\rightarrow1,2} + [2,2]\Psi^{\rm switch}_{2,2\rightarrow2,1} - [2,2]\left( \Psi^{\rm destroy}_{2,2} + \Psi^{\rm switch}_{2,2\rightarrow1,2} + [2,1]\Psi^{\rm switch}_{2,2\rightarrow2,1} + \Psi^{\rm switch}_{2,2\rightarrow1,1}\right).
\end{equation}
\normalsize


%

\begin{figure}
    \centering
    \includegraphics[scale=0.35]{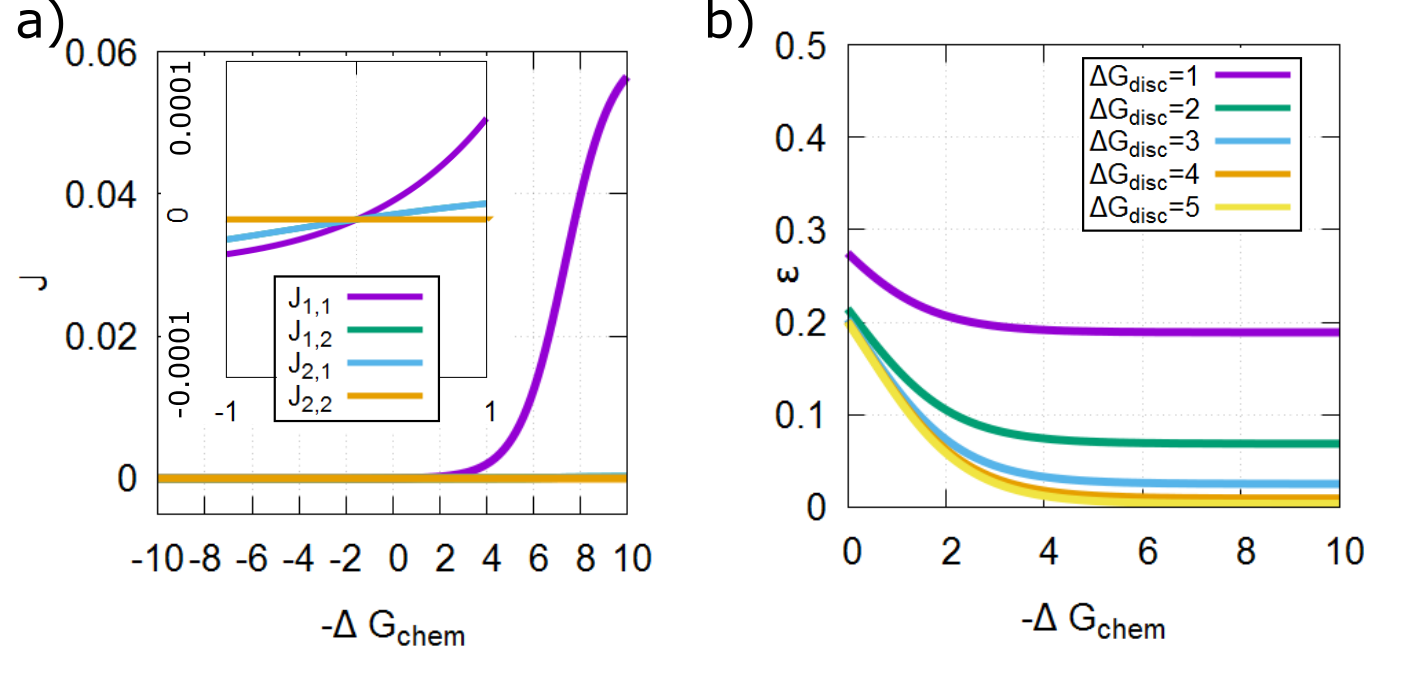}
        \caption{{\it A dimer copying model shows finite accuracy in the equilibrium limit}. (a)  The net production rate \(J({\rm s})\) against chemical driving \(-\Delta G_{\rm chem}\) for each of the four dimers 1,1, 1,2, 2,1 and 2,2 relative to a template of sequence 1,1, with \(\Delta G_{\rm disc}=\Delta G^\plimsoll_{w}- \Delta G^\plimsoll_{r}=10, \Delta G^\plimsoll_r+ \Delta G^\plimsoll_w = 0\) and unbiased monomer and oligomer baths. All \(J({\rm s})\) pass through zero at the equilibrium point \(\Delta G_{\rm chem}=0\) but quickly separate when \(\Delta G_{\rm chem}\neq0\) (inset). \(J({1,2})\) and \(J({2,1})\) overlap. At larger driving, accurate copies are preferentially produced. (b) Error fraction at which incorrect type 2 monomers are incorporated into dimers, \(\epsilon\), against \(\Delta G_{\rm chem}\) for various \(\Delta G_{\rm disc}\) and \(\Delta G_r+ \Delta G_w = 0\).  The system has finite accuracy, $\epsilon \neq 0.5$, as \(\Delta G_{\rm chem} \rightarrow 0^-\), with the accuracy dependent on \(\Delta G_{\rm disc}\).}
    \label{Fig2}
\end{figure}

\section{Is there necessarily a thermodynamic cost to accuracy?}

In all previous work on infinite length templates there is an entropic cost to creating an accurate, low entropy copy. In templated self-assembly systems\cite{Bennett,Cady,Andrieux,Sartori1,Sartori2,esposito2010,EHRENBERG1980333,Johansson}, low entropy sequences can be compensated by favourable bonds between copy and template. In our previous work on templated copying which incorporates separation\cite{Ouldridge,Poulton} this entropic cost has to be paid for directly by the system. The general result above, however, suggests the accuracy of the created oligomer distribution doesn't directly affect the free energy required to create the distribution. 

To illustrate, let us first consider a template of sequence 1,1 coupled to  baths 
where all oligomers have the same  concentration \([S_{\rm tot}]/4\), and all monomers have the same concentration \([M_{\rm tot}]/2\). Without loss of generality we may choose our standard concentration so that \([S_{\rm tot}]/[M_{\rm tot}]^2=1\), and thus \(\Delta G_{\rm{chem}}=\Delta G^\plimsoll_{\rm{dim}}\).

In Fig.~\ref{Fig2}(a) we plot the net production rate of each dimer as a function of \(\Delta G_{\rm chem}\) at set \(\Delta G_{\rm disc}=\Delta G^\plimsoll_{r}-\Delta G^\plimsoll_{w}\). Equilibrium is at \(\Delta G_{\rm chem}=0\), with net creation for all dimers for \(\Delta G_{\rm chem}<0\)  and net destruction for \(\Delta G_{\rm chem}>0\). In Fig. 2(b) we plot \(\epsilon = \left(J(2,2)+\frac{1}{2}(J(1,2)+J(2,1))\right)/J_{\rm tot}\), the proportional rate at which incorrect monomers are incorporated into dimers. It is noticeable that while at exactly \(\Delta G_{\rm{chem}}=0\), \(\epsilon\) is undefined, as \(\Delta G_{\rm{chem}}\rightarrow 0^-\) we obtain \(\epsilon < 0.5\), implying non-zero accuracy.

\begin{figure}
    \centering
    \includegraphics[scale=0.35]{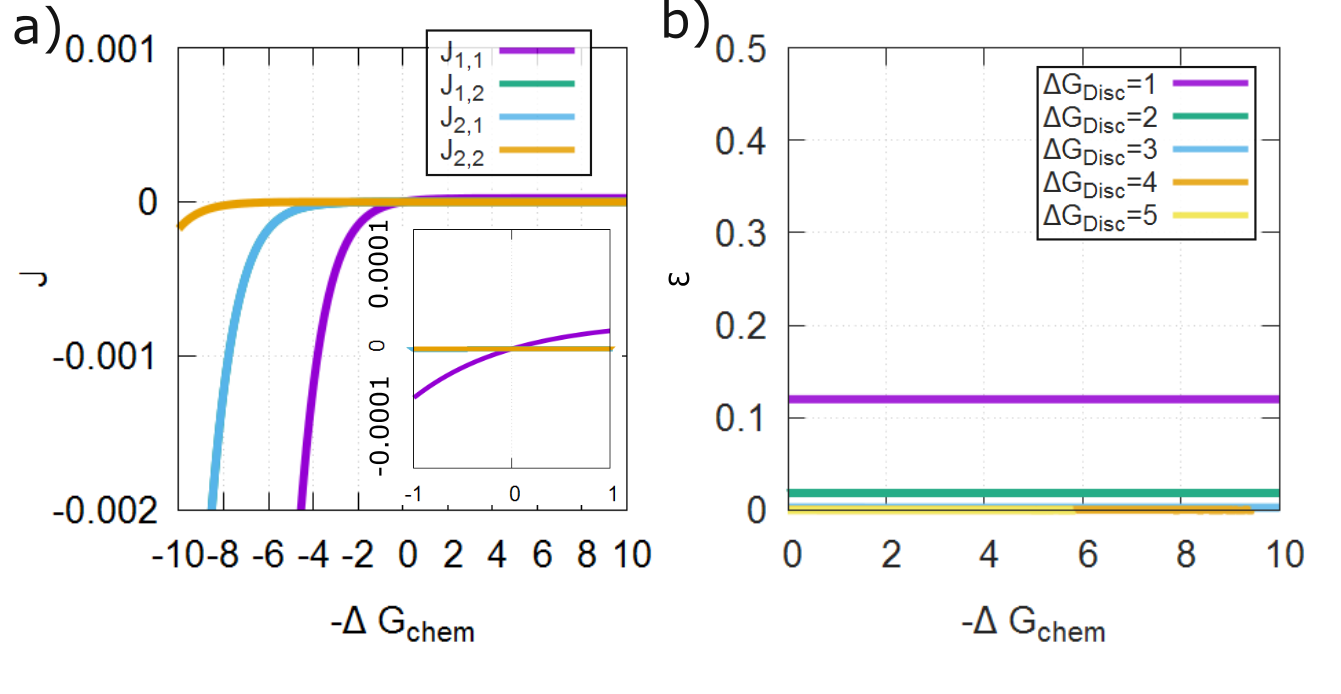}
        \caption{{\it The dimer copying model can show perfect accuracy in the equilibrium limit}. (a)  We plot the net production rate \(J_{\rm s}\) against chemical driving \(-\Delta G_{\rm chem}\) for each of the four dimers 1,1, 1,2, 2,1 and 2,2 relative to a template of sequence 1,1 and with \(\Delta G_{\rm disc}=\Delta G_{w}^{\plimsoll}- \Delta G_{r}^{\plimsoll}=2\), but \(\frac{\Delta G_{r}^{\plimsoll} + \Delta G_{w}^{\plimsoll}}{2}=5\). All oligomers have the same (fixed) concentration \([S_{\rm tot}]/4\), and all monomers have the same (fixed) concentration \([M_{\rm tot}]/2\). All \(J_{\rm s}\) pass through zero at the equilibrium point \(\Delta G_{\rm chem}=0\) but separate quickly (inset) at non-zero driving. Here accurate copies are preferentially produced but all fluxes are very low. (b) Error fraction at which incorrect type 2 monomers are incorporated into dimers, \(\epsilon\), against \(-\Delta G_{\rm chem}\) for various \(\Delta G_{\rm disc}\) with \(\frac{\Delta G_{r}^{\plimsoll} + \Delta G_{w}^{\plimsoll}}{2}=5\).  The system tends towards perfect accuracy, $\epsilon \rightarrow 0$, as \(\Delta G_{\rm disc} \rightarrow \infty\), even as the system tends to the equilibrium point \(\Delta G_{\rm chem}\rightarrow0^{-}\).}
    \label{SI1}
\end{figure}

Fig.~\ref{SI1} shows that when the average template binding free-energy of right and wrong monomers is increased, at fixed \(\Delta G_{\rm disc}=\Delta G_w - \Delta G_r\), the error remains low as the system tends towards equilibrium (\(\Delta G_{\rm chem}\rightarrow0^-\) for unbiased \(p({\bf s})\), \(t({\bf s})\) and \([S_{\rm tot}]/[M_{\rm tot}] =1\)). Indeed, \(\epsilon \rightarrow 0\) (perfect accuracy) as \(\Delta G_{\rm disc} \rightarrow \infty\) and \(\Delta G_{\rm chem} \rightarrow 0^{-}\). The error at \(\Delta G_{\rm chem}=0\) is still undefined, but the fluxes separate quickly after this point (inset Fig. \ref{SI1}a) to keep the error low. Due to the unstable bonds between copy and template, the system has a low flux for $\Delta G_{\rm chem}<0$.

The  minimal value of \(-\Delta G_{\rm{chem}}\) required for growth gives non-zero accuracy, and reversible processes can create a low entropy dimer sequence distribution at finite discrimination free-energy \(\Delta G_{\rm disc}\).
However, interactions between template and product do not persist; \(\Delta G_{\rm disc}\) does not feature in the overall thermodynamics, and cannot compensate for a low entropy state in equilibrium, as in templated self-assembly.
So is there no thermodynamic cost to accuracy in this setting? Consider \(\Delta G_{\rm inf}\) in Eq.~\ref{dG_inf}. In this case, \(p({\bf s})\) and \(t({\bf s})\) are unbiased and equal, and thus \(\Delta G_{\rm inf}=0\) for any \(q({\bf s})\) - even if only accurate copies are produced. Whenever the surrounding oligomers and monomers have no (or the same) sequence bias there is no extra thermodynamic cost to producing sequences of arbitrary accuracy. We emphasise the surprising result that you can get arbitrary accuracy at no extra cost {\it even if} the monomer distribution is unrelated to the template sequence.

\(\Delta G_{\rm inf}\) is generally non-zero, however, for systems with \(p({\bf s})\neq t({\bf s})\), {\it i.e.,} where the monomer and oligomer baths have different distributions. \(\Delta G_{\rm inf}\) is positive if the system produces sequences that are common in the oligomer bath \(p({\bf s})\) and rare in the monomer bath \(t({\bf s})\). The alternative representation \(\Delta G_{\rm inf} = D(q||t) - D(q||p)\) makes this fact particularly clear: the probability distribution of the creation fluxes \(q({\bf s})\) being similar to the monomer bath makes the first term less positive and \(q({\bf s})\) being unlike the oligomer bath makes the second term more negative.  Accuracy is therefore not directly constrained by thermodynamics in a general description of the full process of oligomer copying. Instead, there is a thermodynamic cost  to producing sequence distributions \(q({\bf s})\) that are closer to the oligomer sequences in the environment than a distribution of sequences obtained by randomly sampling monomers from the environment. 
This argument is the second main result of this chapter.


\section{Template copying as an inherently non-equilibrium information engine}
\begin{figure}
    \centering
    \includegraphics[scale=0.35]{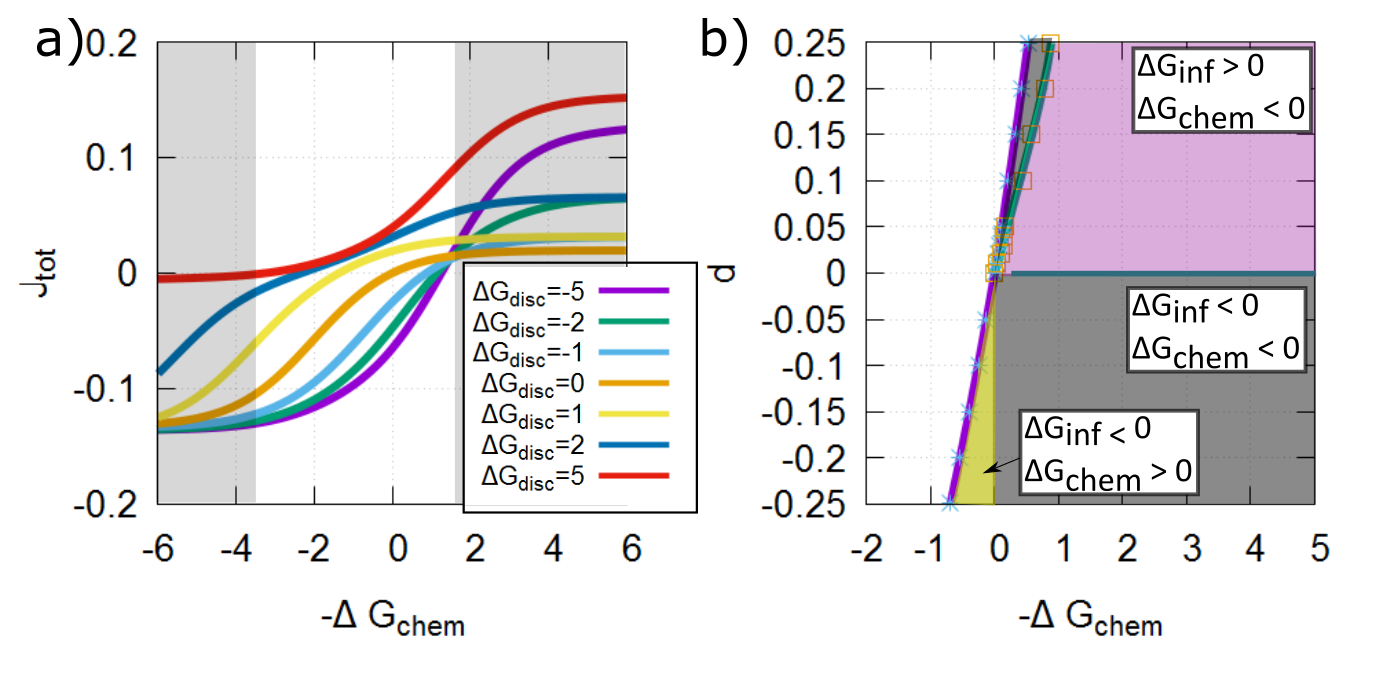}
    \caption{{\it Oligomer copying as an information engine with no equilibrium point}. (a) Total flux \(J_{\rm{tot}}\) against driving \(-\Delta G_{\rm chem}\) for a range of discrimination free energies \(\Delta G_{\rm disc}=\Delta G^\plimsoll_{w}-\Delta G^\plimsoll_{r}\), with
   \([1]=[2]=0.1\), \([1,1]=[1,2]=[2,1]=0.001\) and \([2,2]=0.1\).
    \(\Delta G_{\rm disc}\) is varied with \(\Delta G^\plimsoll_{r}+\Delta G^\plimsoll_{w}=0\) fixed. The point \(J_{\rm tot}=0\) at which there is no net dimerisation varies within the allowed white range despite the fact that  the overall dimerisation free-energy is independent of \(-\Delta G_{\rm disc}\). Specificity for  \({\bf s}_{\rm min} = 1,1\) 
    makes growth easier and pushes $J_{\rm tot}=0$ to the lower limit, and specificity for \({\bf s}_{\rm max}=2,2\) has the opposite effect. (b) Phase plot of the information engine. Here the leftmost purple boundary is the transition from \(J_{\rm tot}\) negative to positive. Here we fix \(\Delta G_{\rm disc}=5\), \([S_{\rm tot}]=1\), \([M_{\rm tot}]=1\), \(t(s)=0.25\) for all \(s\) and vary \(\Delta G_{\rm chem}\). We further vary \(p(s)=0.25+d,0.25,0.25,0.25-d\) by varying \(d\). There is a regime in which chemical work is used to specifically produce sequences of high free-energy and a regime in which specific production of low free-energy sequences is used to drive oligomerisation against a chemical load.}
    \label{Fig3.2}
\end{figure}

Unlike templated self-assembly systems\cite{Bennett,Cady,Andrieux,Sartori1,Sartori2,esposito2010,EHRENBERG1980333,Johansson} or infinite length templated copying systems\cite{Poulton}, in this system the template is coupled to multiple baths that are well out-of-equilibrium with each other. The fact that this system couples to so many baths, means that it is generally impossible to find an equilibrium point. \(J_{\rm tot}=0\) occurs over a range, with the value determined by the details of the templates interaction with the baths. To illustrate, consider a set of systems with \(p({\bf s})\neq t({\bf s})\), ie where the monomer and oligomer baths are different. Here, there is no equilibrium point at which all fluxes are zero because the baths are out of equilibrium with each other. There is instead a range of \(\Delta G_{\rm chem}\)  over which \(J_{\rm tot} =0\) could occur, depending on which sequences best couple to the template. The most positive possible \(\Delta G_{\rm chem}\) at which \(J_{\rm tot} =0\) occurs when a system  specifically produces the sequence \({\bf s}_{\rm min}\), where \({\bf s_{\rm min}}\) minimises \(t({\bf s})/p({\bf s})\). The most negative is when the system specifically produces the sequence \({\bf s}_{\rm max}\), where \({\bf s_{\rm max}}\) maximises \(t({\bf s})/p({\bf s})\). \({\bf s}_{\rm min}\) is intuitively the sequence most like the monomer baths and least like the oligomer baths and \({\bf s}_{\rm max}\) is the opposite. In Fig.~\ref{Fig3.2}(a), we vary \(\Delta G_{\rm disc}\) for a system heavily thermodynamically biased towards creating accurate copies by the baths. When \(\Delta G_{\rm disc}>0\), and the system is also kinetically biased towards creating accurate copies and \(J_{\rm tot}=0\) for a more positive \(\Delta G_{\rm chem}\) than if \(\Delta G_{\rm disc}<0\).

Copying accurately can thus either make production of oligomers thermodynamically easier or harder, depending on the environment. This is true even for an unbiased pool of monomers. The parameter which dictates accuracy is the discrimination free-energy \(\Delta G_{\rm disc}\), and changing this quantity  changes the point when the net production of oligomers becomes positive. For a system without full detachment, this point would be independent of \(\Delta G_{\rm disc}\). Detaching full oligomers into oligomer baths clearly fundamentally changes the thermodynamics of the system. The selective coupling to the template does more than fix the point of \(J_{\rm tot}=0\), due to full detachment, the system can now be thought of as an engine in the conventional thermodynamic sense, like a carnot cycle. Here, the engine operates between the monomer and oligomer reservoirs, and its thermodynamics is set by the relationship between its input/output and the chemical potential of those reservoirs. This is the physics encapsulated by our main result.

The system acts as a chemical/information engine without an equilibrium point, transducing free-energy between a number of out of equilibrium baths. In order to explore this we define \(q_{\rm min}({\bf s})\) which is the \(q({\bf s})\) which results in the most negative value of \(\Delta G_{\rm inf}\), and \(q_{\rm max}({\bf s})\) which maximises \(\Delta G_{\rm inf}\). When the probability distribution of fluxes \(q({\bf s})\) is close to \(q_{\rm min}({\bf s})\) it is possible for a negative \(\Delta G_{\rm inf}\) to overcome a positive \(\Delta G_{\rm chem}\). Equally, a more negative \(\Delta G_{\rm chem}\) allows for a \(q({\bf s}) \approx q_{\rm max}({\bf s})\) with positive \(\Delta G_{\rm inf}\). Therefore this system is a chemical/information engine, in which chemical and information-based free-energy can be traded against each other \cite{mcgrath2017biochemical,PhysRevLett.111.010602}. 

The second law implies that the rate of change of free-energy \(\Delta \dot{G}\) is negative. Thus, from Eq.~\ref{INFCHEM}, there are three possible regimes for this informational engine, illustrated in Fig. \ref{Fig3.2}(b). If \(\Delta G_{\rm chem}<0\) and \(\Delta G_{\rm inf}>0\) then the system channels chemical work through a specific copying mechanism to store free-energy in a distribution of outputs closer to the oligomer bath \(p({\bf s})\) than the monomer bath \(t({\bf s})\), with an efficiency \(\eta=\frac{\Delta G_{\rm inf}}{-\Delta G_{\rm chem}} \leq 1\). In our case, \(\eta\) reaches a maximum of \(\sim 0.3\) when \(p({\bf s})\) is heavily biased towards accurate copies of the template and \(\Delta G_{\rm chem}\) is small and negative. In the case where \(\Delta G_{\rm chem}>0\) and \(\Delta G_{\rm inf} <0\), the system generates outputs closer to the monomer baths \(t({\bf s})\) than the oligomer baths \(p({\bf s})\), expending information to compensate for an unfavourable chemical work term. Here the efficiency \(\eta=\frac{\Delta G_{\rm chem}}{-\Delta G_{ \rm inf}} \leq 1\) reaches a maximum of \(~0.15\) when \(p({\bf s})\) is heavily biased against accurate copies of the template and \(\Delta G_{\rm chem}\) is small and positive. The final case, in which both \(\Delta G_{\rm chem}\leq 0\) and \(\Delta G_{\rm inf} \leq 0\), is less interesting as the system both spends chemical free-energy and generates outputs close to \(q({\bf s})=q_{min}({\bf s})\).

%


%

\section{Kinetic convergence on thermodynamic constraints for infinite-length polymers}

\begin{figure}
    \centering
    \includegraphics[scale=0.35]{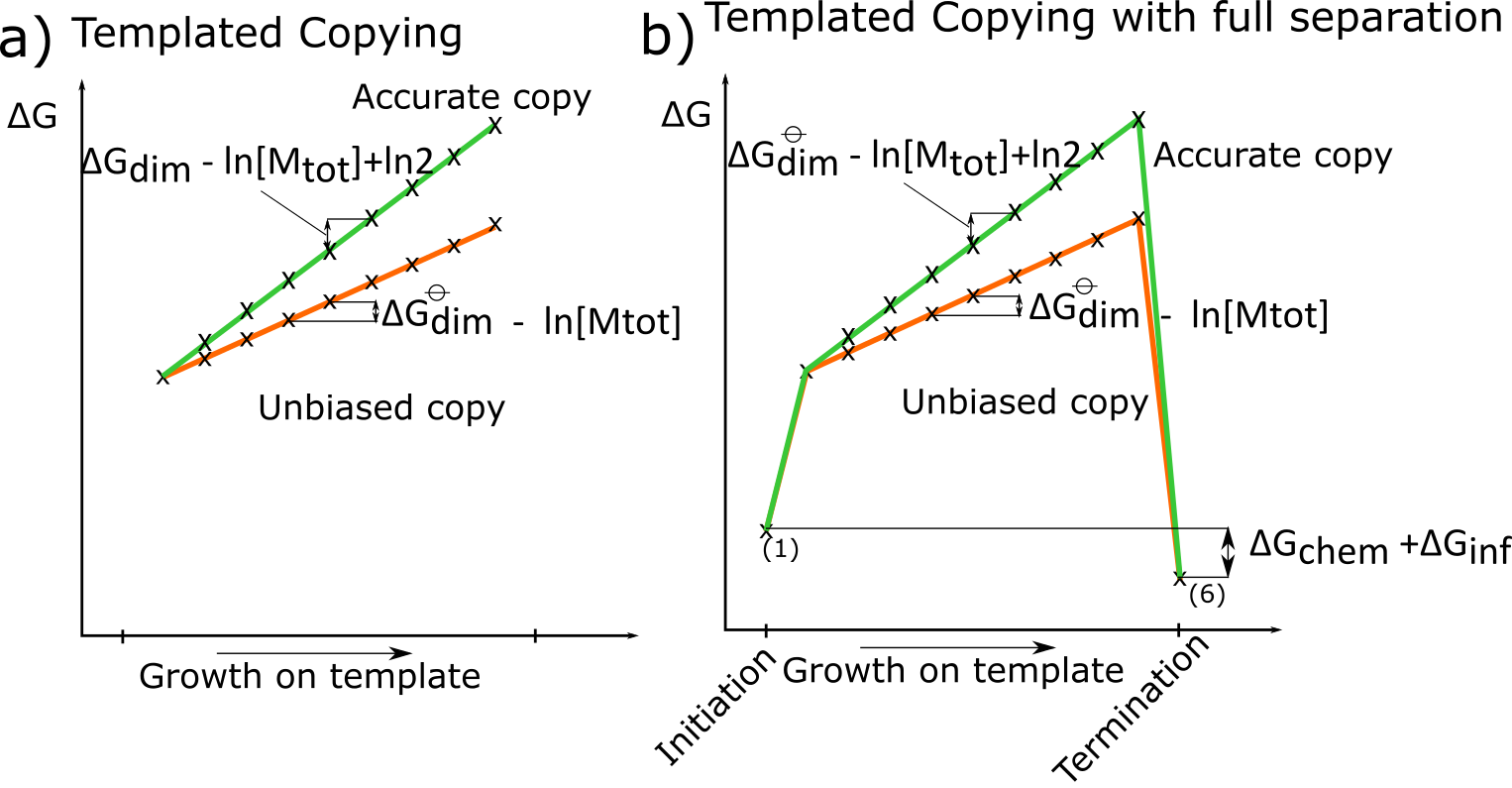}
    \caption{free-energy profiles for (a) a templated copying process on an infinite lengths template and (b) a finite length template. The gradient of the free-energy profile during the bulk of the copying process is set by the free-energy of dimerisation, the chemical driving due to the relative concentrations of the monomer and oligomer baths and the accuracy of the oligomer; an accurate oligomer required an extra \(\ln{2}\) to add a  monomer in the bulk more than a unbiased oligomer. However in the case of a finite length template, the initiation and termination steps can provide a theoretically unlimited adjustment to the overall free-energy of the process. It is intuitive that the initial rise and final drop off represent the entropy reduction due to isolating a free monomer on the template and the entropy increase due to releasing a polymer into solution. What is less intuitive is how these two steps depend on the composition of the monomer and polymer baths.  In some cases creating the most accurate oligomer can be thermodynamically favourable compared to other oligomers.}
    \label{Lastfig}
\end{figure}

Our results apply to oligomers of arbitrary finite length. So are the thermodynamic constraints derived neglecting initiation and termination, {\it eg.} extra chemical work is required for accuracy \cite{Poulton} in the infinite length, invalid? When initiation and termination are neglected, the physics is set by whether the slope of the free-energy profile of template-attached copy growth is favourable. As shown in  Fig.~\ref{Lastfig}(a), higher accuracy gives a more positive slope. However, fig.\ref{Lastfig}(b) illustrates that initiation and termination provide a theoretically unlimited adjustment to the overall \(\Delta G\). teo{It is intuitive that the initial rise and final drop off represent the entropy reduction due to isolating a free monomer on the template and the entropy increase due to releasing a polymer into solution. What is less intuitive is how these two steps depend on the composition of the monomer and polymer baths.} An arbitrarily unfavourable oligomerisation process on the template can be made favourable with the right concentration of products, despite an uphill free-energy profile for both accurate and unbiased copies- {\it i.e.}, when \(\Delta G_{\rm chem}>0\). 

Importantly, however, if template-attached growth is unfavourable ({\it i.e.}, the free-energy profile is uphill), it will be kinetically suppressed by a large free-energy barrier even if oligomer production is favourable overall. Here, barrier height grows proportionally to oligomer length, suggesting that the {\it thermodynamic} constraints derived for infinite-length polymers should become {\it kinetic} constraints for sufficiently long oligomers.
To probe this hypothesis we consider a kinetic model for the growth and destruction of oligomers of arbitrary fixed length, using the same parameters as the dimerisation model of Fig.~\ref{Fig1}, by extending the model of ch. 1\cite{Poulton}.

\section{Model of copy production for oligomers of length \(|s|>2\)}

The model, schematically illustrated in Fig.~\ref{Diagrams}\,(c), is adapted from temporary thermodynamic discrimination model in ch. 1 \cite{Poulton}.

We consider a copy oligomer \({\bf s}=s_{1},...,s_{l}\), made up of a series of sub-units or monomers \(s_{x}\), growing with respect to a template \({\bf n}=n_{1},...,n_{L}\) (\(l\leq L\)). Inspired by transcription and translation, we consider a copy that detaches from the template as it grows. We consider whole steps in which a single monomer is added or removed, encompassing many individual chemical sub-steps. As illustrated in Figure \ref{Diagrams}(c), after each step there is only a single inter-oligomer bond at position \(l\), between $s_l$ and $n_l$. As a new monomer joins the copy at position \(l+1\), the bond position \(l\) is broken.

As in the dimerisation model considered earlier in the text, we shall consider a template oligomer \({\bf n}\) made entirely of monomers of type 1. Given the assumed symmetry between the interactions of the two monomer types, and equal concentrations of the monomer baths as used throughout this work, the results apply equally well to any template sequence. Monomers of type 1 can simply be interpreted as correct matches and monomers of type 2 as incorrect matches for any template sequence \({\bf n}\). 

Having defined the model's underlying states, we now consider state free energies. By analogy with the dimerisation model, we define \(\Delta G^\plimsoll_{\rm{dim}}\) as the free-energy change of adding a specific monomer to the end of the copy oligomer at standard concentration. The environment contains baths of monomers; a monomer of type \(s\) has a constant concentration \([s]\) relative to the standard concentration. The chemical free-energy change for the transition between any specific sequence \(s_{1},...,s_{l}\) and any specific sequence \(s_{1},...,s_{l+1}\), ignoring any contribution from interactions with the template, is then \(\Delta G^\plimsoll_{\rm{dim}} - \ln [s_{l+1}]\).

We then consider the effect of specific interactions with template. We again define $\Delta G_{r/w}^{\plimsoll}$ as the standard binding free energies for matched/mismatched monomers and the template. Overall, each forward  step makes and breaks one copy-template bond. There are four possibilities for forward steps:  either adding 1 or 2 at position \(l+1\) to a copy with \(s_l = 1\); or adding 1 or 2 in position \(l+1\) to a copy with \(s_l = 2\). The first and last of these options preserve the same interaction with the template, so the total free-energy change for monomer addition is \(\Delta G^\plimsoll_{\rm{dim}} - \ln[s_{l+1}]\). For the second case a correct bond is broken and an incorrect bond added, implying a  free-energy change of  \(\Delta G^\plimsoll_{\rm{dim}}-\Delta G_{r}^{\plimsoll}+\Delta G_{w}^{\plimsoll}-\ln[2]\).  Conversely, for the third case, an incorrect bond is broken and a correct bond added, giving a free-energy change of \(\Delta G^\plimsoll_{\rm{dim}}+\Delta G_{r}^{\plimsoll}-\Delta G_{w}^{\plimsoll}-\ln[1]\). 

These free energies constrain the kinetics of transitions between the various states, but are compatible with a range of kinetic models. In the temporary thermodynamic discrimination model of ch. 1~\cite{Poulton}, all forwards steps are assumed to occur with the same rate, and sequence-based discrimination occurs in the backwards step. For simplicity, in this case we further assume that each step can be modelled as a single transition with an exponential waiting time, yielding:
\begin{align}
\nu_{1,1}^{+}&=[1]k,\\
\nu_{1,2}^{+}&=[2]k,\\
\nu_{2,1}^{+}&=[1]k,\\
\nu_{2,2}^{+}&=[2]k,\\
\nu_{1,1}^{-}&=ke^{\Delta G_{\rm dim}^\plimsoll},\\
\nu_{1,2}^{-}&=ke^{\Delta G_{dim}^\plimsoll-\Delta G_{r}^{\plimsoll}+\Delta G_{w}^{\plimsoll}},\\
\nu_{2,1}^{-}&=ke^{\Delta G_{\rm dim}^\plimsoll-\Delta G_{w}^{\plimsoll}+\Delta G_{r}^{\plimsoll}},\\
\nu_{2,2}^{-}&=ke^{\Delta G_{\rm dim}^\plimsoll}.
\end{align}\\
Here, \(k\) is a rate constant that sets the overall timescale (we take \(k=1\) in reduced units without loss of generality). \(\nu_{i,j}^{+}\) is the rate for adding a monomer of type \(j\) to a copy with a monomer of type \(i\) at the leading edge, and  \(\nu_{i,j}^{-}\) is the reverse process. 

To allow for initiation and termination of copying, we include two additional transitions. Unbinding transitions, whether of the initial monomer or the final monomer of a complete copy, are parameterised by:
\begin{align}
\nu_{1}^{\rm off}=ke^{\Delta G_{r}^{\plimsoll}},\\
\nu_{2}^{\rm off}=ke^{\Delta G_{w}^{\plimsoll}}.
\end{align}\\

Binding of either a monomer to the initial site, or oligomer to the final site, is assumed to have a rate
\begin{align}
\nu_{s}^{\rm on}=k[s],\\
\nu_{\bf{s}}^{\rm on}=k[\bf{s}].
\end{align}\\

Note that, for simplicity, we ignore the (challenging) question of how partial fragments are prevented from binding to or detaching from the template, or the possibility of multiple copies being bound to the template at once.

We simulate the system repeatedly using a Gillespie simulation \cite{Gillespie}, with the system initiated with either a monomer \(s\) sampled from \(t(s)\) seeded at the first site, or an oligomer \({\bf s}\) sampled from \(p({\bf s})\) 
attached to the final position \(L\) of the template. The simulation is allowed to run, and terminates either when a complete oligomer detaches from the final site of the template, or a single monomer detaches from the first site of the template. 

By running many simulations it is possible to calculate the probability of creating a full length oligomer given that a single monomer binds to the template, \(P^{\rm create}\), and the probability of destroying an oligomer given that one binds to the final site on the template, \(P^{\rm destroy}\). Finally \(P^{\rm transform}=1-P^{\rm destroy}\) is the probability of a full oligomer detaching from the template given that an oligomer previously attached at the final site (including those where attached and detached template are identical). This oligomer will have had some subset of its initial monomers transformed through removal of old and addition of new monomers.
 
We set the oligomer concentration 
\begin{align}
    [S_{\rm tot}]=[M_{\rm tot}]^{n}e^{(\Delta G_{\rm dim}^\plimsoll-F)(n-1)}.
\end{align}
Here \(F\) sets the equilibrium position; for \(F=0\), the equilibrium is \(\Delta G_{\rm dim}^\plimsoll=0\), for \(F=3\), the equilibrium is at \(\Delta G_{\rm dim}^\plimsoll=3\) etc. In our model, \(F=5\).

\begin{figure}
\includegraphics[scale=0.25]{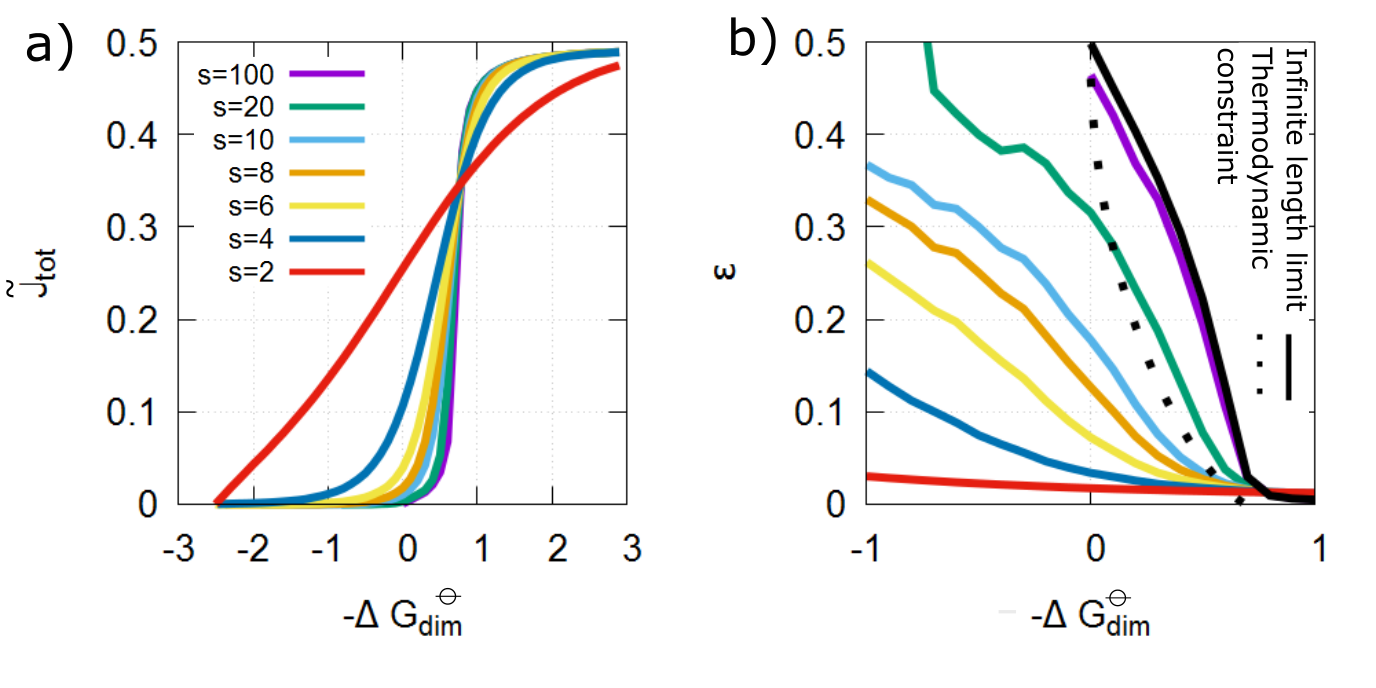}
\caption{{\it Thermodynamic restrictions in an infinite-length model become kinetic restrictions for longer oligomers.} (a) Net flux per unit empty template \(\tilde{J}_{tot}\) of oligomer production and (b) the net fraction of error creation  for a range of lengths \(|{\bf s}|\). We vary \(\Delta G_{\rm dim}^\plimsoll\), with \([M_{\rm tot}]=1\), \([S_{\rm tot}]\) chosen to give \(\Delta G_{\rm chem} =0\) at \(\Delta G_{\rm dim}^{\plimsoll} = -5\), \(p(s)=t(s)\) unbiased and \(\Delta G_{\rm disc}=8\) at \(\Delta G_r+\Delta G_w =0\). Also plotted in (d) is the  thermodynamic constraint on accuracy for an infinite length polymer, set by requiring a non-positive slope of the free-energy profile, \(-\epsilon\ln{\epsilon}-(1-\epsilon)\ln{(1-\epsilon)}\geq  \Delta G_{\rm dim}^\plimsoll+\ln2\), and the actual error rate obtained for an infinite-length copy for these parameters \cite{Poulton}. Short oligomers  overcome kinetic barriers to produce copies with \(\Delta G_{\rm dim}^\plimsoll>0\) and \(\epsilon\) below the infinite-length limit and thermodynamic constraint. At each length the simulation is averaged over \(10,000,000\) trajectories.}
\label{fig4}
\end{figure}

The rate of oligomer creation per unit time in which the template is in an empty state is given by \(\tilde{k}_{\rm create}=[M_{\rm Tot}]kP^{\rm create}\). The rate of oligomer destruction per unit empty template is \(\tilde{k}_{\rm destroy}=k[S_{\rm tot}]P^{\rm destroy}\). The net flux per empty template, \(\tilde{J}_{\rm tot}=\tilde{k}_{\rm create}-\tilde{k}_{\rm destroy}\), is plotted in Fig. \ref{fig4}(a).

We can also calculate the average fraction of  the monomers incorporated during a creation event that are mismatches with the template, \(\epsilon_{\rm create}\). Similarly, the average fraction of incorrect monomers destroyed during a destruction event, \(\epsilon_{\rm destroy}\), can be extracted from simulations. Finally, \(\epsilon_{\rm transform}\) is defined as the difference in average error density between the sequences at the end and start of a transformation event: \(\epsilon_{\rm transform}=\epsilon_{\rm final}-\epsilon_{\rm initial}\). From these quantities we calculate the overall error rate as the proportion of the net number of monomers added to oligomers that are incorrect matches to the template:
\begin{equation}
    \epsilon= \frac{k_{\rm create}\epsilon_{\rm create} - k_{\rm destroy}\epsilon_{\rm destroy} + k[S_{\rm tot}]P^{\rm transform}\epsilon_{\rm transform}}{\tilde{J}_{\rm tot}},
\end{equation}
which is plotted in Fig. \ref{fig4}(b).

%

To analyse the model we perform stochastic simulations for a range of lengths  \(|{\bf s}|\), varying the free-energy released by backbone formation \(\Delta G_{\rm dim}^\plimsoll\) while keeping all other parameters fixed. We calculate the total flux per empty template \(\tilde{J}_{\rm tot}=\frac{J_{\rm tot}}{P_{\rm empty}}\) in the steady state. We use unbiased \(m({s})\) and \(p({\bf s})\), set $[M_{\rm tot}] =1$ and choose $[S_{\rm tot}]$ so that \(\Delta G_{\rm chem}=0\) at \(\Delta G_{\rm dim}^{\plimsoll}=5\); growth is thermodynamically favourable for all sequences when \(\Delta G_{\rm dim}^{\plimsoll}<5\). However, the slope of the on-template free-energy profile of an unbiased sequence, \(\Delta G_{\rm dim}^\plimsoll -\ln [M_{\rm tot}]\),  is positive for \(\Delta G_{\rm dim}^{\plimsoll}>0\). For \(5> \Delta G_{\rm dim}^{\plimsoll} > 0\), on-template oligomerisation is thus a kinetic barrier to formation of a thermodynamically favourable product. For short oligomers, non-negligible \(\tilde{J}_{\rm tot}\) are observed in the region \(5> \Delta G_{\rm dim}^{\plimsoll} > 0\) nonetheless, but as oligomers get longer the kinetics is slowed and both forward and backwards contributions to \(\tilde{J}_{\rm tot}\) are vanishingly small unless \(\Delta G_{\rm dim}^{\plimsoll} < 0\)


Kinetic barriers not only control overall the production flux per empty template \(\tilde{J}_{\rm tot}\), but also error incorporation. The on-template production of an accurate copy  has a more positive slope in its free-energy profile than an unbiased sequence (Fig.~\ref{Lastfig}(b)). For an infinite-length polymer (Fig.~\ref{Lastfig}(a)), this fact provides a thermodynamic constraint on accuracy for \(0>\Delta G_{\rm dim}^{\plimsoll}>-\ln 2\) \cite{Poulton}. We plot the fraction of net incorporated monomers that do not match the template, \(\epsilon\), in Fig.~\ref{fig4}(b), alongside this thermodynamic constraint for infinite-length polymers and the actual error rate obtained for our specific model in the infinite-length limit \cite{Poulton}. Short oligomers can overcome kinetic barriers and beat both the thermodynamic bound and the accuracy obtained in the infinite-length limit; longer oligomers approach the limiting behaviour slowly, with significant differences even at length 20.

\section{Conclusion} 

In this chapter we have investigated the copying of finite-length oligomers, with explicit focus on initiation and termination. Copying creates correlations between copy and template sequences, but the mixing of the products with oligomers in the environment means that the information between copy and template sequences is not thermodynamically exploitable \cite{Ouldridge}. 

Accurate copying creates a low entropy sequence. In templated self-assembly, this low entropy is costly, but can be compensated for and even favoured in equilibrium due to stabilising interactions of specific copy/template bonds. If these bonds are transient, however, as in the production of a true copy, this compensation is impossible in the overall thermodynamics of the process. In an infinite-length model of continuous detachment from behind the leading edge of the copy \cite{Poulton}, the low entropy of the product implies a non-equilibrium state. A reduction in the entropy of the copy of $\sim \ln 2$ per monomer relative to a random sequence must be formed without a compensating discrimination free energy that scales with the length of the copy.

The argument that transient bonds prevent the template from biasing the system towards a low entropy state in equilibrium \cite{Poulton} remains valid once full initiation and termination are considered. However, once the copy has fully detached and mixed with an environment, including oligomers of other sequences, the low entropy of the specific copy oligomer is thermodynamically irrelevant. Since that individual copy can no longer be identified without prior measurement of its sequence, what matters thermodynamically is the contribution of that copy to the entropy of the oligomer distribution as a whole, not the sequence entropy of the sequence actually produced. This change in the thermodynamic significance of the copy sequence itself is manifest as the arbirtarily-large offset in the final step of the free energy profile, shown in Fig. \ref{Lastfig}.

Thus, as we have shown, the overall thermodynamics of the full copy process does not explicitly depend on accuracy. Instead, the surrounding concentrations of oligomers and monomers set the thermodynamic constraints. Creating outputs that resemble the surrounding oligomers is costly, as is creating outputs unlike the input monomer baths. Arbitrary accuracy can be free-energetically neutral or even actively favourable if the oligomer baths are biased towards other sequences.

However, accuracy does play a role indirectly. Firstly, mixing with other oligomers is the final step, and therefore its thermodynamic consequences are irrelevant whilst a copy is growing on the template. Whilst attached to the template -- even if only by the leading site -- the copy is distinct from the surround oligomer pool and its low entropy is exploitable. For an infinite-length polymer, the associated costs \cite{Poulton} set absolute limits on what is possible. For finite length oligomers, they instead manifest as kinetic barriers; longer oligomers  have larger barriers and thus their kinetics converges on the behaviour dictated by the thermodynamic constraints.

Secondly, templates will typically influence their environment. If a template sets its own oligomer environment, \(p({\bf s}) = q({\bf s})\), \(\Delta G_{\rm inf}= D(q||t) \), which reduces to the entropy difference between \(t(\bf{s})\) and  \(q({\bf s})\) if \(t(\bf{s})\) is unbiased. In this case no information is lost upon mixing and accurate copying incurs a cost; the limits derived in \cite{Poulton} hold exactly. In general, there is no reason to suppose that \(p({\bf s}) = q({\bf s})\). As in a cell, other templates and differential degradation rates may be relevant in setting \(p({\bf s}) \). Nonetheless, particularly for longer oligomers, sequences common in q({\bf s}) are likely to be over-represented in \(p({\bf s}) \). 

 If many identical templates are present, then the environmental \(p({\bf s})\) will likely be more strongly peaked, and the cost of accuracy higher, than in a system with many distinct templates. Moreover, any template in an environment dominated by the copies of another will experience a relative thermodynamic advantage. This effect would act as a form of ``rubber banding" in evolutionary competition among minimal replicators, and favour virus-like templates invading new environments.
 
 In this work we have derived general results, assuming that there are no kinetic proofreading cycles that would consume a variable amount of fuel for each polymer produced and contribute an extra term to the overall free-energy change (Eq.~\ref{INFCHEM}). The central conclusions in this manuscript would be largely unaffected, however. The existence of kinetic proofreading cycles would effectively provide an unpredictable adjustment to $\Delta G_{\rm chem}$ for each oligomer produced; the role of $\Delta G_{\rm inf}$ and the copy accuracy in determining overall thermodynamics would be largely unchanged.

\newpage
\setcounter{equation}{0}
\setcounter{figure}{0}
\setcounter{table}{0}
\setcounter{section}{0}

\section*{A broader look at biological copying system; the addition of a degradation pathway}

\setcounter{equation}{0}
\setcounter{figure}{0}
\setcounter{table}{0}
\setcounter{section}{0}

\begin{figure}
    \centering
    \includegraphics[scale=0.3]{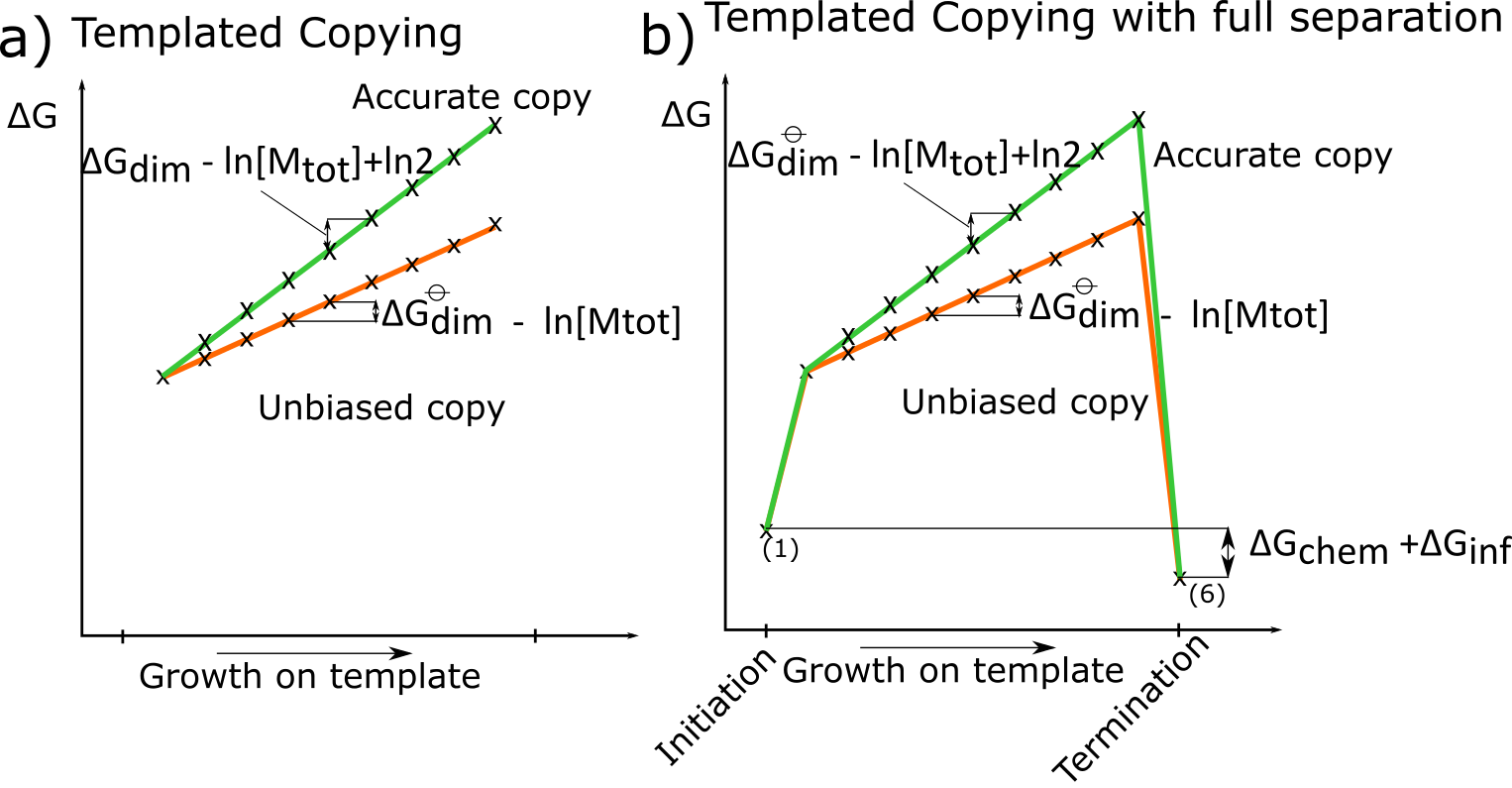}
    \caption{free-energy profiles for an infinite length polymer (a) and a finite length polymer (b). The free-energy required for a polymer to grow on a template (the gradient) is dependent on the free-energy change for the extension of the polymer by one, and the accuracy of the polymer (an entropic consideration). The gradient is steeper for a more accurate polymer. However, depending on the composition of the surrounding monomer and polymer baths, the process of final detachment can provide a theoretically unlimited adjustment to the overall free-energy of the process. Therefore, it is possible for a finite length polymer to be produced, through leveraging thermal fluctuation, at a free-energy per monomer below that of the minimum required for a infinite length copying process to progress.}
    \label{freeenergy}
\end{figure}

In the first two chapters, this thesis has analysed templated copying systems which explicitly consider separation in two different limits, the infinite length template\cite{Poulton} and the finite length template. We observe that strict thermodynamic limits on the accuracy of the created copy polymer in the infinite length limit become kinetic barriers in the finite length case. As shown in fig. \ref{freeenergy}, theoretically, the final detachment of the last monomer from the template can provide an unlimited adjustment to the overall free-energy change of the process. Consequently, under certain circumstances, the copying process can thermodynamically favour the creation of the completely accurate oligomer, relative to other sequences. However, as the length of the copy polymer increases, the intermediate steps that the system must undergo in order to create the copy start to dominate the dynamics of the growth process. During these intermediate steps, as in the infinite length case, the system must spend free-energy in order to systematically grow an accurate copy polymer on the template. Thus as the length increases, the dynamics of the finite length oligomerisation system undergo a kinetic transition to the dynamics of the infinite length case.

In ch. 2 we further observe that a finite length polymer which fully detaches from the template can act as an information engine with no fixed equilibrium point. By splitting the free-energy of oligomerisation into two terms, a chemical term and an informational term, we can observe that when the baths surrounding the oligomerisation process are heavily biased towards a specific oligomer distribution the system can convert between chemical and information free-energy by selectively producing a specific distribution of oligomers. Chemical free-energy can be converted to information free-energy if the system maintains an oligomer creation flux that is like the oligomer baths and unlike the monomer baths. Conversely, information free-energy can be converted to chemical free-energy if the flux of oligomer creation is like the monomer baths and unlike the polymer baths.

All of the work up until now has considered an isolated template coupled to pools of monomers and oligomers at fixed concentrations. In the biological context, the point of accurate copying is to {\em create} and {\em maintain} highly-specific pools of products. Therefore, we need to consider systems in which the oligomer concentrations are not fixed externally, but by the templates themselves. However, in biological systems, the concentrations of the output aren't just determined by the properties of the template, they are also determined by the degradation mechanisms that exist in the cell. In the context of information being translated from mRNA into proteins, these can include dilution due to cell growth or reproduction, or the constant degradation of proteins due to the cellular environment\cite{Alberts2002}. Equally in RNA transcription, all forms of RNA are degraded quickly by the cell\cite{Alberts2002}. In the absence of degradation, the concentrations of oligomers in the system would be unbounded. Therefore a system needs degradation to reach a steady-state. 

Previous work\cite{OuldridgePRX} has analysed a related system. In this work, a complex, which is in either an activated or inactivated state \(X^{*}\) or \(X\), is coupled to a receptor which detects the binding of ligands. When the receptor is bound to a ligand, the system catalyses the activation of the complex \(X\xrightarrow{RL} X^{*}\). When the receptor is vacant, the system catalyses the deactivation of the complex \(X^{*}\xrightarrow{R\emptyset} X\). A net flux around the system is achieved when the pathway catalysed by the empty receptor is driven backwards \(X^{*}\xrightarrow{R\emptyset} X\) relative to the pathway catalysed by the receptor bound to a ligand \(X\xrightarrow{RL} X^{*}\). This paper showed that in order to generate a far-from-equilibrium steady-state, transferring information about the state of the receptor to the complex, it is vital that there is a fuel-driven non-equilibrium flux around the cycle. Without a non-equilibrium flux, reaching a steady state would require the system to equilibrate. Therefore, the degradation pathway must be distinct from the creation pathway and fuel-consuming.

It is clear that in order to maintain a non-equilibrium flux around the cycle it is necessary to expend fuel which is lost into the environment as generated entropy. This entropy generation, known as ``housekeeping heat"\cite{speck2005integral,chetrite2019martingale} but which can more generally be known as ``housekeeping entropy" as the entropy generated does not have to be heat, scales with the size of the flux round the cycle. This is indisputable; in order to force a system to cycle, the system must generate a non-zero amount of entropy, but this dissipation can be arbitrarily small.  In his seminal paper ``The thermodynamics of computation"\cite{Bennett1982}, Charles Bennett suggested in order to create an accurate copy and then destroy it by a second pathway, it is essential to generate \(\ln{2}\) of entropy, likely dissipated as heat, per bit of information in the copy. This is because the destruction is not carried out by the original pathway, meaning the system is acting in a thermodynamically irreversible fashion. This suggests that the housekeeping entropy generated should scale with the properties of the baths rather than just being required to maintain the flux around the cycle.  In this chapter we argue that producing and degrading highly accurate copies via a cyclic process does not require a dissipation that scales with the information content of the copy. 

Bennett's conjectured constraint of \(\ln{2}\) per bit is proposed for a system that copies a template with a random sequence with perfect accuracy. We can generalise this constraint to a system in which a distribution of copies are created due to an imperfect templating system; ie one in which errors can be incorporated. We can calculate the amount of free-energy stored in the distribution of copies created, above that required to create an equilibrium distribution. This is given by the entropy change \(\Delta S=S_{\rm fin}-S_{\rm eqb}\) where \(S_{\rm fin}\) is the Shannon entropy of the created oligomer distribution and \(S_{\rm eqb}\) is the Shannon entropy of the equilibrium distribution for that system. Bennett's constraint suggests that \(\Delta S\) should be less than the free-energy dissipated to force the system to cycle. We will search for a system which breaks this constraint.

\begin{figure}
    \centering
    \includegraphics[scale=0.25]{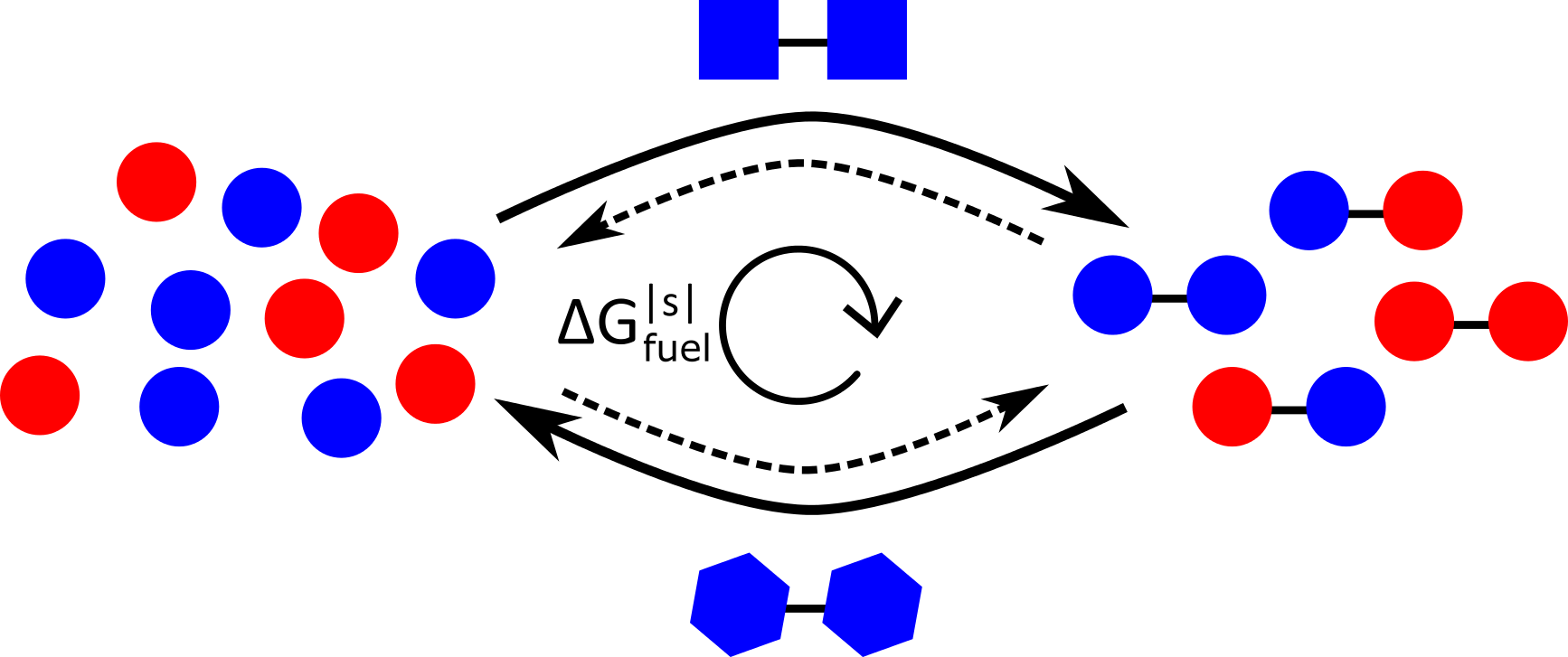}
    \caption{Our system now has two pathways for creating an oligomer, a template pathway (circles) and a destroyer pathway (hexagons). The free-energy required to create a polymer via the template path is \(\Delta G^{|{\bf s}|}_{\rm dim}\) and via the destroyer pathway \(\Delta G^{|{\bf s}|}_{\rm dim}-\Delta G^{|{\bf s}|}_{\rm fuel}\). Therefore, the system can exhibit cycles, with oligomers on average created by the template pathway, and destroyed by the destroyer pathway when \(\Delta G^{|{\bf s}|}_{\rm fuel}<0\). }
    \label{cycle}
\end{figure}

This chapter uses the same techniques used in ref.\cite{OuldridgePRX} for the receptor-ligand system, but applies them to a system which makes copies via oligomer assembly, like that which Bennett discusses in ref.\cite{Bennett1982}. We start by adapting the derivation from ch.2 to explicitly allow for multiple catalysts, some of which are coupled to fuel (templates and destroyers). We apply the second law of thermodynamics at the level of the whole system, or at the level of each template, or at the level of each copy sequence giving us four fundamental bounds on generic template/destroyer systems. We find no thermodynamic constraint consistent with Bennett's conjectures constraint. Then we analyse these results in the specific context of a single template/destroyer system, initially with fixed oligomer concentrations. Then we analyse the system in which the catalysts themselves are allowed to fix the population level of oligomers. We have not yet found a system which beats Bennett's suggested constraint, although we have found signatures of systems which could beat the constraint at longer lengths.

\section{Fundamental constraints on a generic cyclical oligomerisation system}

First, as illustrated in fig. \ref{cycle} we consider a system of template and destroyer species surrounded by baths of monomers and oligomers. There are two types of monomers, \(1\) and \(2\), and four types of oligomers \({\bf s}=1,1\), \({\bf s}=1,2\), \({\bf s}=2,1\) and \({\bf s}=2,2\). The concentrations of the two monomer types  are \([1]\) and \([2]\) (defined with respect to a standard reference concentration), and the four oligomer types are \([1,1]\), \([1,2]\),\([2,1]\) and \([2,2]\).

As in ch. 2 we define \(\Delta G^{\plimsoll}_{\rm dim}\) as the free-energy change of dimerisation at the reference concentration. Dimerisation catalysed by a destroyer is coupled to a fuel turnover of \(\Delta G_{\rm fuel}\), so that the free-energy change of a dimerisation reaction on the destroyer is \(\Delta G^{\plimsoll}_{\rm dim}-\Delta G_{\rm fuel}\). Within each process, the energy exchanges in the system remain tightly coupled to the reaction pathways, so destruction via a destroyer {\em always} consumes fuel. This encourages the system to cycle, with dimer creation more likely to be catalysed by a template, and destruction by a destroyer when \(\Delta G_{\rm fuel}<0\). For oligomers, $\bf s$, with length $|{\bf s}|>2$, the total free-energy change of oligomerisation is \(\Delta G^{|{\bf s}|}_{\rm dim}=(|{\bf s}|-1)\Delta G^{\plimsoll}_{\rm dim}\). The coupling to fuel has a similar relationship so that the free-energy change of a reaction on a destroyer for \(|{\bf s}|>2\) is \(\Delta G^{\bf |{\bf s}|}_{\rm dim}-\Delta G^{|{\bf s}|}_{\rm fuel}=(|{\bf s}|-1)(\Delta G^{\plimsoll}_{\rm dim}-\Delta G_{\rm fuel})\).

Initially, we consider a single, isolated template or destroyer. As in ch. 2, equation 1, the total free-energy change of the baths resulting from production of a single oligomer of sequence, ${\bf s}$, via an isolated template, $\Delta G^{T}({\bf s})$, or destroyer, $\Delta G^{D}({\bf s})$, is:
\begin{equation}
    \Delta G^{T}({\bf s})=\Delta G_{\rm{\rm dim}}^{|{\bf s}|}+ \left(\ln{[{\bf s}]}-\ln{\Pi^{|{\bf s}|}_{i}[s_{i}]}\right),
\end{equation}
\begin{equation}
    \Delta G^{D}({\bf s})=\Delta G_{\rm{\rm dim}}^{|{\bf s}|}-\Delta G^{|{\bf s}|}_{\rm fuel}+\left(\ln{[{\bf s}]}-\ln{\Pi^{|{\bf s}|}_{i}[s_{i}]}\right).
\end{equation}

Let \(J^{T}({\bf s})\) be the average net rate at which sequence \({\bf s}\) is produced by a template, either from its component monomers, or by transformation from a different oligomer. \(J^{D}({\bf s})\) is the equivalent net rate for the production of $\bf s$ by a destroyer. We define \(q^{T(D)}({\bf s})=J^{T(D)}({\bf s})/J^{T(D)}_{\rm{tot}}\), where \(J^{T}_{\rm{tot}}=\sum_{{\bf s}}J^{T}({\bf s})\) and \(J^{D}_{\rm{tot}}=\sum_{{\bf s}}J^{D}({\bf s})\). We define the following probability distributions: \(p({\bf s})=[{\bf s}]/[S_{\rm{tot}}]\), the probability of picking an oligomer of sequence \(\bf{s}\) from the oligomers with total concentration \([S_{\rm{tot}}]\); \(m({s})=[{ s}]/[M_{\rm tot}]\), the probability of picking a monomer of type \({s}\) from the monomers with total concentration of \([M_{\rm tot}]\); and \(t({\bf s})= \prod_{i}m(s_{i})\), which corresponds to the probability of the sequence \({\bf s}\) occurring by selecting monomers randomly from the monomer pools. As in ch. 2 section 2, we can define a rate of change of free-energy \(\Delta \dot{G^{T(D)}}=\sum_{\bf s} J^{T(D)}({\bf s})\Delta G^{T(D)}({\bf s})\) for the template and destroyer in terms of these quantities,
\begin{align}
&\Delta \dot{G^{T}}=J^{T}_{\rm{tot}}\sum_{{\bf s}}q^{T}({\bf s})\Delta G^{|{\bf s}|}_{\rm{\rm dim}}+&J^{T}_{\rm{tot}}\left( \sum_{{\bf s}}q^{T}({\bf s})\ln{\frac{p({\bf s})}{t(\bf s)}}+\sum_{{\bf s}}q^{T}({\bf s})\ln{\frac{[S_{\rm{tot}}]}{[M_{tot}]^{|\bf s|}}} \right)\leq0.\label{template}
\end{align}
\begin{align}
&\Delta \dot{G^{D}}=J^{D}_{\rm{tot}}\sum_{{\bf s}}q^{D}({\bf s})\left(\Delta G^{|{\bf s}|}_{\rm{\rm dim}}-\Delta G^{|{\bf s}|}_{\rm fuel}\right)+&J^{D}_{\rm{tot}}\left( \sum_{{\bf s}}q^{D}({\bf s})\ln{\frac{p({\bf s})}{t(\bf s)}}+\sum_{{\bf s}}q^{D}({\bf s})\ln{\frac{[S_{\rm{tot}}]}{[M_{tot}]^{|\bf s|}}} \right)\leq0.\label{destroyer}
\end{align}
Here, the inequalities follow directly from the second law which tell us that for a reaction to systematically progress, it must cause a negative change in free-energy\cite{crooks1999entropy,seifert2005entropy,Esposito,ouldridge2018importance,seifert2011stochastic, parrondo2015thermodynamics, jarzynski2011equalities}. It applies to each reaction separately because the reactions on a template and destroyer are not coupled together; a reaction on the template does not force a reaction on the destroyer.

Assuming that all oligomers are of the same length, these equations can be re-written as 
\begin{equation}
 \Delta \dot{G^{T}}=J^{T}_{\rm{tot}}\Delta G_{\rm{\rm chem}} + J^{T}_{\rm{tot}}\Delta G^{T}_{\rm{\rm inf}}\leq0,
 \label{INFCHEMT}
\end{equation}
\begin{equation}
 \Delta \dot{G^{D}}=J^{D}_{\rm{tot}}\Delta G_{\rm{\rm chem}} - J^{D}_{\rm{tot}}\Delta G^{|{\bf s}|}_{\rm{fuel}} + J^{D}_{\rm{tot}}\Delta G^{D}_{\rm{\rm inf}}\leq0,
 \label{INFCHEMD}
\end{equation}
with
\begin{align}
\Delta G_{\rm chem} &=  \Delta G^{|{\bf s}|}_{\rm dim} + \ln{[S_{\rm tot}]/[M_{\rm tot}]^{|{\bf s}|}}
\label{dG_chem}
\end{align}
\begin{align}
\Delta G^{T(D)}_{\rm{\rm inf}} &= \sum_{{\bf s}}q^{T(D)}({\bf s})\ln{\frac{p({\bf s})}{t(\bf s)}}.
\end{align}
Equations \ref{INFCHEMT} and \ref{INFCHEMD} are the first two fundamental constraints on the system, here considering the fluxes through template and destroyer in isolation. While these results are derived for the template and destroyer in isolation, the result also apply to a system with both templates and destroyers.

Now, we consider a system that consists of one template and one destroyer. We define the total oligomer production flux as \(J_{\rm tot}=\sum_{{\bf s}}J({\bf s})\) where \(J({\bf s})= J^{T}({\bf s})+ J^{D}({\bf s})\) is the total creation flux for sequence \({\bf s}\). By defining \(q({\bf s})=\frac{J({\bf s})}{J_{\rm tot}}\), we can write the information free-energy component for the combined template/destroyer system as \(\Delta G_{\rm inf}=\sum_{{\bf s}}q({\bf s})\log{\frac{p({\bf s})}{t({\bf s})}}\). Combining free-energy contributions from the template and destroyer for the entire system gives
\begin{align}
    \Delta \dot{G}=\Delta\dot{G^{T}}+\Delta\dot{G^{D}}=J_{\rm tot}\Delta G_{\rm chem}+J_{\rm tot}\Delta G_{\rm inf}-J^{D}_{\rm tot}\Delta G^{|{\bf s}|}_{\rm fuel}\leq0.
    \label{entire}
\end{align}
Finally, we consider a single sequence of the product oligomer, ${\bf s}$, giving
\begin{align}
    \Delta \dot{G}_{\bf s}=J({\bf s})\left(\Delta G_{\rm chem} + \log{\frac{p({\bf s})}{t({\bf s})}}\right)-J^{D}({\bf s})\Delta G^{|{\bf s}|}_{\rm fuel}\leq 0.
    \label{gsdot}
\end{align}

Equations \ref{INFCHEMT}, \ref{INFCHEMD}, \ref{entire} and \ref{gsdot} are four fundamental constraints on the thermodynamics of oligomer production in a copying system containing a single template and destroyer. In the next section, we explore the role of these constraints in specific contexts with the aid of an explicit model.

\section{A specific model of dimerisation}

\begin{figure}
    \centering
    \includegraphics[scale=0.47]{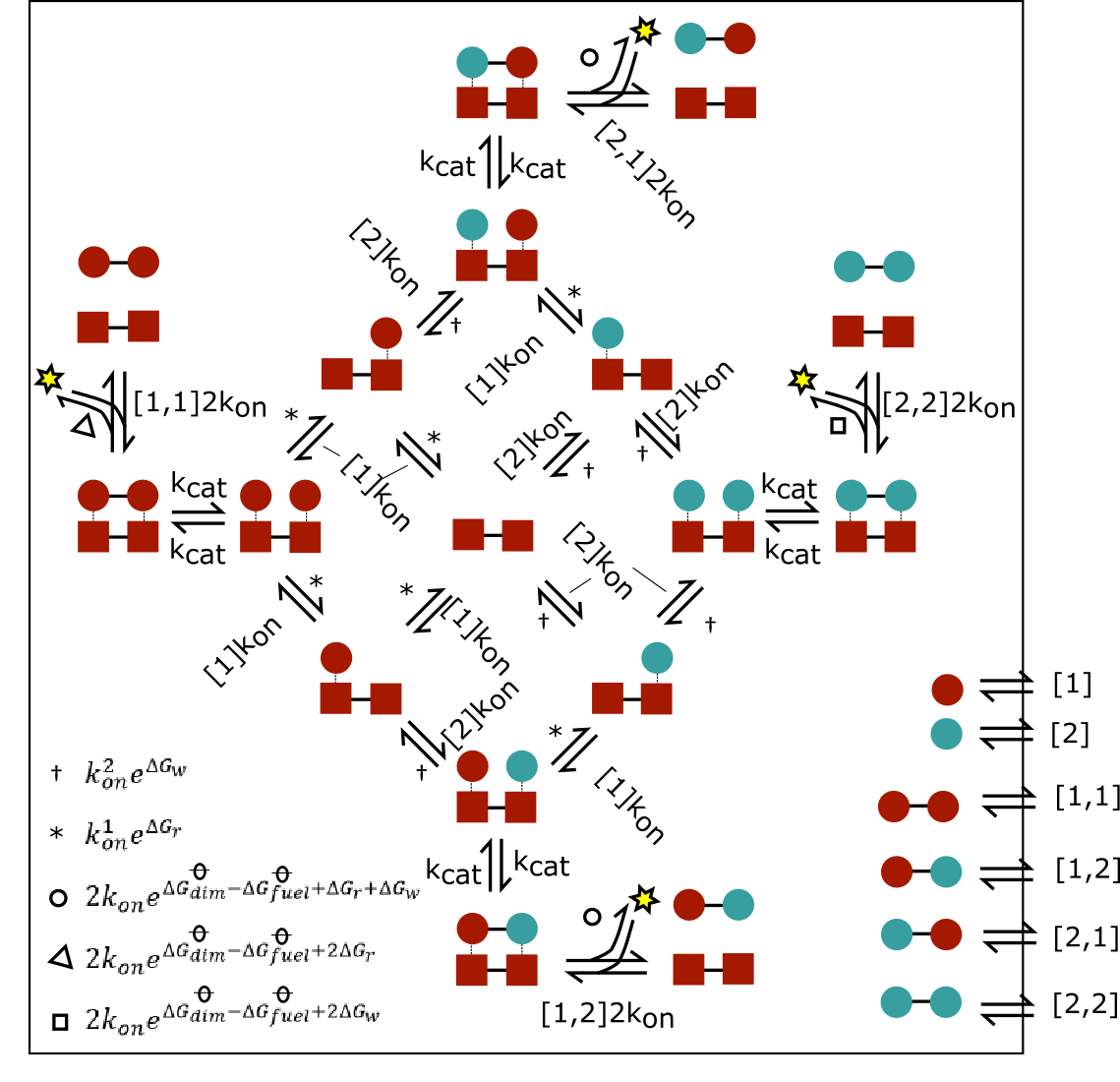}
    \caption{A specific example of a minimal model of the mechanism of the destroyer type catalysts which are coupled to fuel. The destroyer (squares) interacts with baths of monomers and dimers of a second species (circles). Monomers can bind to the template and dimerize, while dimers binding to the template can be destroyed or inter-converted. 
    The standard dimerisation free-energy is \(\Delta G^\plimsoll_{\rm dim}\) for all sequences, but the free-energy of matching and non-matching varieties of monomer coming out of solution and binding to the template are \(\Delta G^\plimsoll_r+[1]\) and \(\Delta G^\plimsoll_w+[2]\), respectively,  allowing selectivity. \(\Delta G^\plimsoll_r\) and \(\Delta G^\plimsoll_w\) are the free energies of forming a copy template bond that is a match or mismatch respectively, measured at the reference concentration. Rate constants $k_{\rm cat}$ and $k_{\rm on}$ set the speed of the dynamics. The destroyer has identical dynamics to the template, as described in ch. 2 except the final step in which a dimer is forced off the template is coupled to the consumption of a fuel molecule which causes a free-energy change of \(\Delta G^\plimsoll_{\rm fuel}\) at the reference concentration.}
    \label{dimerisation}
\end{figure}

In ch. 2 section 2, we outlined a thermodynamically-consistent model for dimer production on a template, a prototype for a broader class of oligomerisation systems. We assume that two such systems exist, a template which is identical to that outlined in ch. 2, and a destroyer (illustrated in fig. \ref{dimerisation}) which is identical except that the final step in which a dimer is forced off the template is coupled to the consumption of a fuel molecule causes a free-energy change of \(\Delta G^\plimsoll_{\rm fuel}\) at the reference concentration. Solvated template dimers and solvated ``destroyer" dimers carry information in their sequence of monomer units; throughout this chapter, the sequence is 1,1 for both the template and destroyer without loss of generality. The template and destroyer dimers are coupled to large baths of monomers and oligomers (in this case dimers) of a distinct second type of molecule, like a DNA template in a bath of RNA nucleotides and oligomers. This second type of molecule also comes in multiple varieties -- in this case two -- and can interact with the template and destroyer in sequence-specific ways. Reactions on both template and destroyer are microscopically reversible, dimers from the baths can be created and also be broken down into their component monomers by both the template and destroyer.

As in ch. 2 section 2, we define the free-energy change when a correct or incorrect match binds to the template as \(\ln{[1]}+\Delta G_{r}^\plimsoll\) and \(\ln{[2]}+\Delta G_{w}^\plimsoll\) respectively, where \(\Delta G_{r}^\plimsoll\) and \(\Delta G_{w}^\plimsoll\) are the standard free energies for the formation of the correct or incorrect copy/template bond at the reference concentration. We define \(\Delta G_{\rm diff}=\Delta G_{r}^\plimsoll-\Delta G_{w}^\plimsoll\). In this system, matching and non-matching monomers bind to the template/destroyer with the same rate constant \(k_{\rm on}\). The free-energy released by the dimerisation of monomers, and fuel consumption in the case of the destroyer, is used to weaken the bonds between the dimer and template, as was found to be optimal in ref. \cite{deshpande2019optimizing}. \(\Delta G_{\rm dim}^\plimsoll\) or \(\Delta G_{\rm dim}^\plimsoll-\Delta G_{\rm fuel}^\plimsoll\)  thus appear in the dimer off-rate, with the on-template dimerisation/undimerisation having a fixed rate \(k_{\rm cat}\). Unless stated otherwise, throughout this chapter \(k_{\rm cat}=k_{\rm on}=1\) for the template and destroyer systems. Since we use an unbiased \(m({\bf s})\) in all cases, the symmetry of the problem gives identical physics for all template sequences; we shall use 1,1 for clarity.

\begin{figure*}
\includegraphics[scale=0.3]{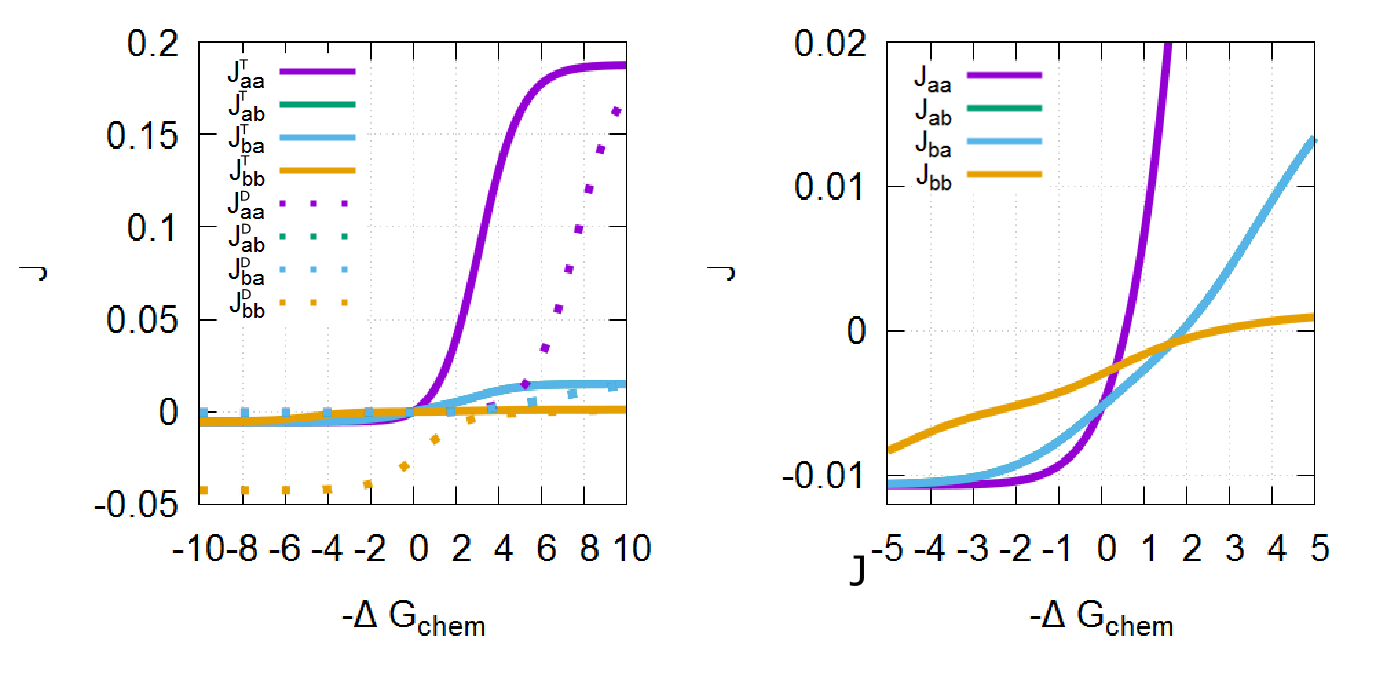}
\caption{(a) The four dimer production fluxes from a template (solid lines) and a destroyer (dashed lines). (b) The net flux of dimer creation through the template and destroyer for each of the four dimer types. Here \(\Delta G^\plimsoll_{\rm fuel}=5\) and \(\Delta G_{\rm diff}=2\). The monomer and polymer baths, \(p({\bf s})\) and \(t({\bf s})\), are unbiased and \(M_{\rm tot}=P_{\rm tot}=1\). Alongside the fact that \(|{\bf s}|=2\) this mean that \(\Delta G^\plimsoll_{\rm dim}=\Delta G^{|{\bf s}|}_{\rm dim}=\Delta G_{\rm chem}\) and \(\Delta G^\plimsoll_{\rm fuel}=\Delta G^{|{\bf s}|}_{\rm fuel}\). Because the monomer and polymer baths are unbiased, individually all the fluxes pass through zero at the same point, \(\Delta G_{\rm dim}=0\) for the template and \(\Delta G_{\rm dim}=5\) for the destroyer. However, when summing the fluxes, \(J(\textbf{s})=J^{T}(\textbf{s})+J^{D}(\textbf{s})\), there is a region (shaded grey) in which \(J([1,1])>0\) and \(J([2,2])<J([1,2])=J([2,1])<0\), where the system net produces the most accurate dimer while destroying all other dimers.}
\label{Independent}
\end{figure*}

\section{A system in which the template and destroyer act independently between fixed monomer and oligomer baths.}
First, let us consider a setting in which all baths of monomers and oligomers are held constant. Here the individual templates and destroyers are behaving independently. However, in combination, we observe interesting net effects. We calculate the flux \(J({\bf s})\) for the steady-state of the template/destroyer interacting with the monomers and oligomers by analysing a single template/destroyer as a Markov process as outlined in Ch. 2.  Fig. \ref{Independent}a shows that the flux through the destroyer is offset from the that of the template by 5, with the template and destroyer fluxes becoming positive for all four dimer types at \(\Delta G_{\rm dim}=0\) and \(\Delta G_{\rm dim}=5\), respectively. In Fig. \ref{Independent}b we plot the net production rate of each of the four types of dimers by a system with \(p({\bf s})\) and \(t({\bf s})\) unbiased \(\Delta G_{\rm diff}=2\) and \(\Delta G_{\rm fuel}=5\).  Here, the total creation fluxes for each of the four types of dimer do not all become positive at the same \(\Delta G_{\rm dim}\) unlike the equivalent system with no fuel. Instead, there is a region where there is net production of 1,1 dimers and net destruction of 1,2 , 2,1 and 2,2 dimers. This system has the potential to be able to heavily bias the creation of the desired polymer even without changing the monomer and polymer baths.

We can understand this result through the constraints given by equation \ref{gsdot}. In Ch. 2 there was no fuel, meaning that no matter how many templates or destroyers were involved in the process
\begin{align}
    \Delta \dot{G}_{\bf s}=J({\bf s})\left(\Delta G_{\rm chem} + \log{\frac{p({\bf s})}{t({\bf s})}}\right)\leq 0.
    \label{gsdot2}
\end{align}
If \(t({\bf s})\) and \(p({\bf s})\) are unbiased and \(M_{\rm tot}=S_{\rm tot}=1\) then \(J({\bf s})\Delta G_{\rm chem}\leq0\) and \(J({\bf s})\) has the same sign for all \({\bf s}\). In the case where there is fuel, however, then even in the case where \(t({\bf s})\) and \(p({\bf s})\) are unbiased then
\begin{align}
    \Delta \dot{G}_{\bf s}=J({\bf s})\Delta G_{\rm chem}-J^{D}({\bf s})\Delta G_{\rm fuel}\leq 0,
    \label{gsdot3}
\end{align}
and \(J({\bf s})\) can have different signs for different \({\bf s}\). Thus the addition of the destroyer allows the system to only create the targeted \(1,1\) dimer.

\section{Template/destroyer system as a three-way information engine}

\begin{figure}
    \centering
    \includegraphics[scale=0.2]{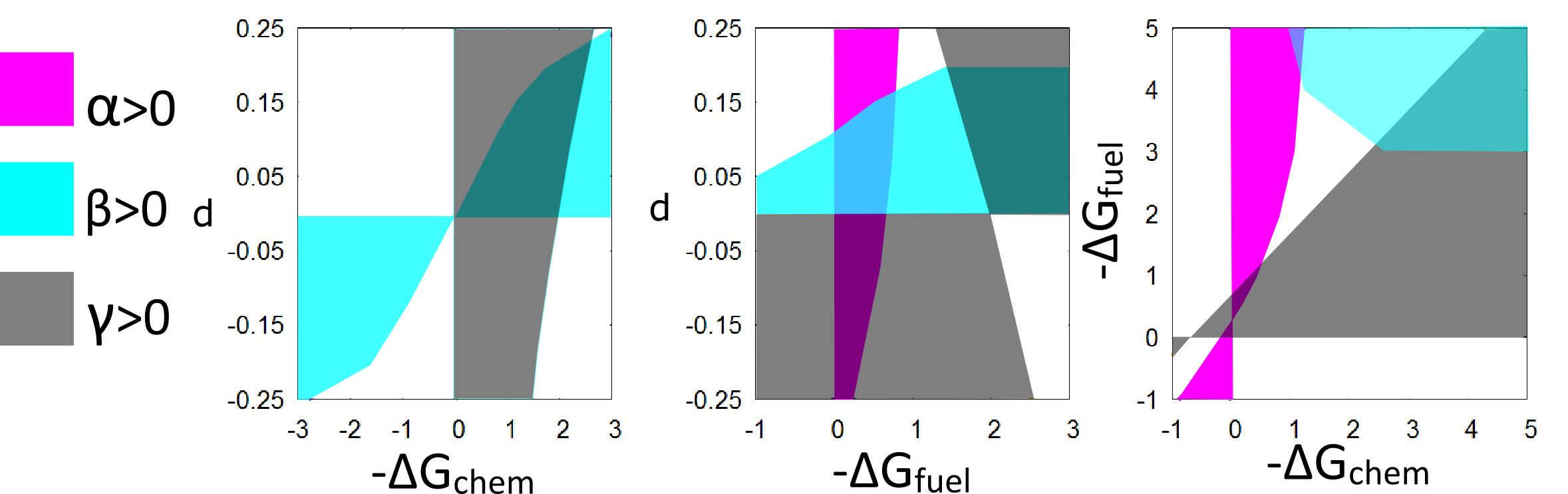}
    \caption{The system is capable of acting as a three-way engine interconverting information, the oligomerisation free-energy \(\Delta G^{|{\bf s}|}_{\rm dim}\) and the cycle free-energy  \(\Delta G^{|{\bf s}|}_{\rm fuel}\). Here the monomer bath is unbiased and \(M_{\rm tot}=1\), and \(\Delta G_{\rm diff}=5\) for both the template and destroyer. In (a) we fix \(\Delta G_{\rm fuel}=2\) and  vary \(\Delta G_{\rm dim}\) and \(d\) (refer to main text) which varies \(\Delta G_{\rm inf}\). Here \(\alpha<0\) throughout, but \(\beta\) and \(\gamma\) become positive at various points. In (b) we fix \(\Delta G_{\rm dim}=2\) and  vary \(\Delta G_{\rm fuel}\) and \(d\). Here we see that at any given point, no more than two of \(\alpha\), \(\beta\) and \(\gamma\) are positive. In (c) we fix \(d=0.249\) and vary \(\Delta G_{\rm dim}\) and \(\Delta G_{\rm fuel}\). Again, we see that at any given point, no more than two of \(\alpha\), \(\beta\) and \(\gamma\) are positive. 
}
    \label{Fig3}
\end{figure}

As in Ch. 2 section 6, it is possible for each individual template and destroyer to act independently as an engine that exchanges chemical and information free-energy. Constraints \ref{INFCHEMT} and \ref{INFCHEMD} which constrain the individual behaviours of the template and the destroyer neatly describe the separate behaviours of the two catalysts. However, if we consider the system of template and destroyer as a whole then we can see that this system can act as a three-way engine. Consider equation \ref{entire},
\begin{align}
    \Delta \dot{G}=J_{\rm tot}\Delta G_{\rm chem}+J_{\rm tot}\Delta G_{\rm inf}-J^{D}_{\rm tot}\Delta G^{|{\bf s}|}_{\rm fuel}\leq0.
\end{align}
Here we have three terms \(\alpha=J_{\rm tot}\Delta G_{\rm chem}\), \(\beta=J_{\rm tot}\Delta G_{\rm inf}\) and \(\gamma=-J_{\rm tot}^{D}\Delta G^{|{\bf s}|}_{\rm fuel}\). If the system is in a non-equilibrium steady-state then \(\Delta \dot{G}<0\), and at least one of \(\alpha\), \(\beta\) and \(\gamma\) must be negative. Here positive information free-energy change is when the system has a oligomer creation flux that has a high free-energy with respect to the baths (has a distribution like that of the polymer baths and unlike that of the monomer baths). Positive chemical free-energy change is due a to a large positive free-energy of oligomerisation and  oligomer baths having large concentration \([{\bf s}]\) compared to monomer concentrations, making it harder to create oligomers. Finally, positive free-energy change of fuel is when the turnover of a fuel molecule absorbs rather than releases free-energy; ie fuel is created rather than consumed. If \(\alpha<0\) and \(\beta,\gamma>0\), chemical information is being converted into both high free-energy fuel molecules and a high information oligomer creation flux. If \(\beta<0\) and \(\alpha,\gamma>0\) then information free-energy is converted into chemical free-energy and high energy fuel molecules. Finally, if \(\gamma<0\) and \(\alpha,\beta>0\) then the system spends high energy fuel molecules to both drive against the chemical free-energy and the produce a high information free-energy oligomer output flux.

In Fig. \ref{Fig3}, we show regions for each of these cases for a system where the monomer bath is unbiased and \(M_{\rm tot}=1\), and \(\Delta G_{\rm diff}=5\) for both the template and destroyer. In the first plot, we fix \(\Delta G_{\rm fuel}=2\) and  vary \(\Delta G_{\rm dim}\) and a quantity which varies \(\Delta G_{\rm inf}\) which we call \(d\). Here \(d\) sets the skewedness of the polymer bath distribution \(p(1,1)=0.25+d\), \(p(1,2)=0.25\), \(p(2,1)=0.25\), \(p(2,2)=0.25-d\).  In this case we see that \(\alpha<0\) throughout, but that both \(\beta\) and \(\gamma\) become positive at various points. In the second plot we fix \(\Delta G_{\rm dim}=2\) and  vary \(\Delta G_{\rm fuel}\) and \(d\). In this case we see that at any given point, no more than two of \(\alpha\), \(\beta\) and \(\gamma\) are positive. In the third plot we fix \(d=0.249\) and vary \(\Delta G_{\rm dim}\) and \(\Delta G_{\rm fuel}\). Again, we see that at any given point, no more than two of \(\alpha\), \(\beta\) and \(\gamma\) are positive.

We observe all seven possible regimes of behaviour, in which either one, two or three of \(\alpha\), \(\beta\) and \(\gamma\) are negative. In ch. 2, there were only three regimes of behaviour; chemical free-energy being transformed to informational free-energy, informational free-energy being transformed to chemical free-energy, and the ``dud" case in which both chemical free-energy and informational free-energy were being dissipated. The addition of a second destroyer pathway and that destroyer-fuel coupling adds four extra regimes.

\section{General discussion of a system in which the concentration oligomers is not externally controlled}

In order to address Bennett's conjecture, which we predict is incorrect, we consider a system in which the monomer concentrations are held constant, and the template and destroyer determine the oligomer concentration relative to the monomer concentration. In the introduction of this chapter, we discussed how it is necessary oligomer degradation can prevent oligomer populations from growing indefinitely when the system is out of equilibrium. If the degradation process is coupled to fuel turnover, then the system can reach a non-equilibrium steady-state, with the oligomer baths set by the template and destroyer. This minimal model of a system in which templates are used to set output populations is analogous to the biological system for creating proteins in which proteins are created via specific interactions with RNA templates and then degraded by the cellular environment. In this system we have the monomer distribution \(m({\bf s})\) unbiased and \(m_{\rm tot}=1\). The steady-state occurs when \(J_{\rm tot}=0\). We return to our third fundamental constraint on our system, equation \ref{entire}, which applies to the entire system. In the case where \(J_{\rm tot}=0\), then equation \ref{entire} can be rewritten as
\begin{align}
    \Delta \dot{G} = -J^{D}_{\rm tot}\Delta G^{|{\bf s}|}_{\rm fuel}\leq0.
    \label{constrain}
\end{align}
The more negative \(\Delta \dot{G}\), the more free-energy is dissipated into the environment as entropy generation. If this system is to reach steady state without equilibrating, it is vital that the system cycles. Equation \ref{constrain} tells us it is necessary to drive the system round a cycle using fuel, and that fuel is entirely dissipated as entropy generation, however, no other constraints necessarily follow. The system needs to generate housekeeping entropy in order to maintain the non-equilibrium flux\cite{speck2005integral,chetrite2019martingale}. However, the fuel does not scale with the information being created/destroyer by the cycle. The sequence specificity of copying, either through the net flux of creating sequences \(q({\bf s})\) or the product pool \(p({\bf s})\), doesn't constrain \(\Delta G^{|{\bf s}|}_{\rm fuel}\) based on equation \ref{constrain} alone. Using the equation \ref{gsdot}, which is the constraint for an individual sequence gives the same result; \(\Delta G^{|{\bf s}|}_{\rm fuel}\) is not dependent on the individual sequence flux of creation or the individual sequence concentration in the baths.

We now consider the template and destroyer individually, to see if we can find a dependence on sequence specificity. Recall our constraint for the template alone, equation \ref{template}, written in the form
\begin{equation}
 \Delta \dot{G^{T}}=J^{T}_{\rm{tot}}\Delta G_{\rm{\rm chem}} + J^{T}_{\rm{tot}}\Delta G^{T}_{\rm{\rm inf}}\leq0.
\end{equation}
Again, the more negative \(\Delta \dot{G^{T}}\), the more free-energy is dissipated into the environment as entropy. The second term \(\Delta G^{T}_{\rm{\rm inf}}\) will increase if the oligomer production flux \(q({\bf s})\) becomes more like the product pool distribution \(p({\bf s})\) than the randomly assembled monomer pool distribution \(t({\bf s})\), meaning that if \(p({\bf s})\) and \(q({\bf s})\) (the creation flux distribution) are sharply peaked at the same sequence then for a fixed \(\Delta G_{\rm chem}\) this results in less entropy generation.
 
However, while \(q({\bf s})\) can become arbitrarily sharp without necessarily increasing the heat dissipated in the baths, the fact that \(\Delta G^{T}_{\rm{\rm inf}}\) will increase if \(q({\bf s})\) becomes more like the polymer pool distribution \(p({\bf s})\) than the randomly assembled monomer pool distribution \(t({\bf s})\) means that a more negative \(\Delta G_{\rm chem}\) is required to keep \(\Delta \dot{G^{T}}<0\) so that the reaction can progress. Thus for a given \(\Delta G_{\rm dim}\) and \(m_{\rm tot}\), as accuracy increases, \(S_{tot}\) must decrease. This could be considered a ``price" of accuracy, but it does not represent entropy generation.
 
It is clear the template alone does not imply a dependence on sequence specificity, so we now consider the destroyer term. Now recall equation \ref{destroyer} for the destroyer, rewritten in the form
\begin{equation}
\Delta \dot{G^{D}}=J^{D}_{\rm{tot}}\Delta G_{\rm{\rm chem}} - J^{D}_{\rm{tot}}\Delta G^{|{\bf s}|}_{\rm{fuel}} + J^{D}_{\rm{tot}}\Delta G^{D}_{\rm{\rm inf}}\leq0.
\end{equation}
In the case of a single template and destroyer in a steady-state with \(J_{\rm tot}=0\), \(J^{D}_{\rm tot}=-J^{T}_{\rm tot}\) and \(\Delta G^{T}_{\rm inf}=-\Delta G^{D}_{\rm inf}\). Therefore
\begin{equation}
\Delta \dot{G^{D}}= -\Delta \dot{G^{T}} - J^{D}_{\rm{tot}}\Delta G^{|{\bf s}|}_{\rm fuel} \leq0.
\end{equation}
Again, for the reaction to progress, both \(\Delta \dot{G^{T}}\leq0\) and \(\Delta \dot{G^{D}}\leq0\). In order for the system to exist in a non-equilibrium steady-state, \(\Delta G^{|{\bf s}|}_{\rm fuel}\leq0\) but there is no bound on \(\Delta G^{|{\bf s}|}_{\rm fuel}\) at a non-zero quantity which scales with the information content.

Fuel consumption is required to drive the system in a cycle and that fuel is dissipated as generated entropy, sometimes called ``housekeeping entropy"\cite{speck2005integral,chetrite2019martingale}. A system that cycles is in a non-equilibrium steady-state. If a system is not cycling, then the steady-state is the equilibrium state. In a system which has reached a self consistent steady state with the oligomer baths, (as opposed to one in which there is a fixed, out of equilibrium, distribution of oligomers already in the output baths), then the equilibrium distribution of oligomers \(p({\bf s})\) is determined by the free-energy of the sequences, irrespective of the templates. There is no accuracy in equilibrium for a system in which the template and destroyer set the oligomer population \(p({\bf s})\), as in this system there can be no bias due to an out of equilibrium distribution of oligomers in large, fixed polymer baths.

Bennett\cite{Bennett1982} states that if you produce and destroy an accurate copy by separate pathways, then the entropy generated due to dissipated free-energy must be greater than the entropy reduction due to the distribution created. We define the entropy change \(\Delta S=S_{\rm fin}-S_{\rm eqb}\), as the difference between the Shannon entropy of a random distribution of outputs \(S_{\rm eqb}=\ln{2^{|{\bf s}|}}\) and the Shannon entropy of the created distribution \(S_{\rm fin}=-\sum_{s}p({\bf s})\ln{p({\bf s})}\). Bennett's suggested constraint suggests that \(\Delta G^{|{\bf s}|}_{\rm fuel}>\Delta S\) which is not something we observe in our derived thermodynamic constraints. It is not clear from our results that the entropy generated per copy sets any kind of bound on the accuracy of either \(p({\bf s})\) or \(q({\bf s})\). 

Therefore, we make the hypothesis that for a system with an unbiased pool of monomers, it is possible to maintain a steady-state with a sharply peaked \(p({\bf s})\), while dissipating free-energy in order to generate entropy that is less than \(\Delta S\) per cycle. We will explore this possibility in candidate systems. 

\section{Two explicit models for a template/destroyer system which maintains a steady-state of oligomers where \(|{\bf s}|\geq 2\)}

\begin{figure}
    \centering
    \includegraphics[scale=0.5]{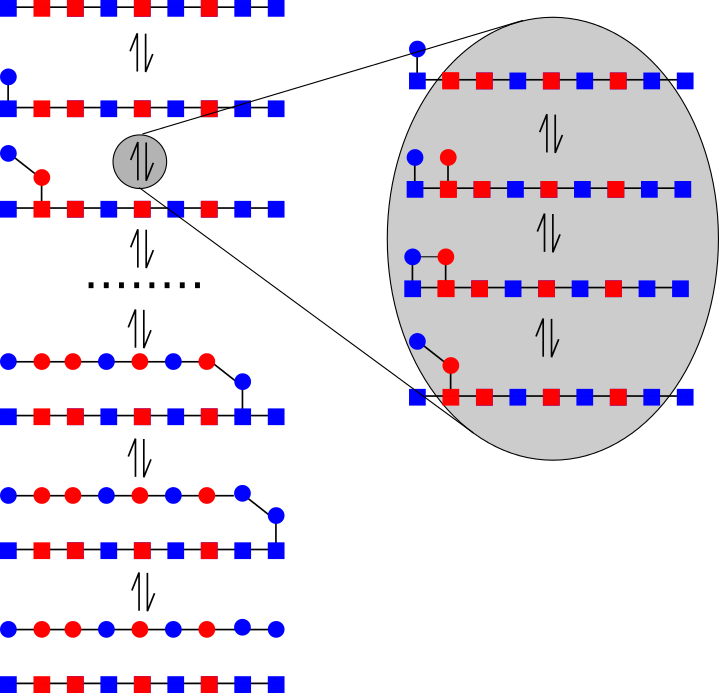}
    \caption{Left, a coarse-grained model in which the polymer grows sequentially, left to right with the tail separating as it grows. Right, an explicit set of sub-steps is required to calculate the time taken for an interaction between template/destroyer and a growing oligomer to resolve. One simple configuration of sub-steps is defined here and used in this work. In these steps, a monomer comes out of solution and binds to the template, the monomer is then polymerised into the growing chain and then the previous final monomer detaches from the chain.}
    \label{generic2}
\end{figure}

We consider a system in which there is one template and one destroyer which can each create and destroy oligomers of length \(|s|\). The system has a fixed, unbiased distribution of monomers and the oligomer baths are set by the template and destroyer systems. We consider templates of length \(|s|\geq2\). For a given accuracy per monomer \(\epsilon\), the entropy change of generating a non-equilibrium polymer distribution \(\Delta S=S_{\rm fin}-S_{\rm eqb}\) grows with length. With \(\Delta S\) growing with length, at longer lengths even a relatively large  \(\Delta G^{|{\bf s}|}_{\rm fuel}\) (allowing the cycle to be well established) could be beaten by \(\Delta S\). We can find a signature of this effect as follows; if we can find a region in which the same \(\Delta G^{|{\bf s}|}_{\rm fuel}\) leads to the same error \(\epsilon\) for progressively longer polymers, then eventually, providing the scaling holds as \(|{\bf s}|\rightarrow\infty\), we can predict \(\Delta G^{|{\bf s}|}_{\rm fuel}<\Delta S\).

We extend our Markovian model outlined in ch. 2 section 8, based on the models of ch. 1 with the addition of an initiation and termination step. We consider two catalysts, one template and one destroyer coupled to fuel. 

For reasons discussed below, we are required to explicitly consider a specific ordering of sub-steps which have been left deliberately undefined in the models of ch. 1 and 2. As illustrated in fig. \ref{generic2}, until now we have been considering a coarse-grained model in which a single monomer is added or removed, encompassing many individual chemical sub-steps. These sub-steps can be loosely described as follows; first, a monomer comes out of solution and binds to the template, the monomer is polymerised into the growing chain and then the previous final monomer detaches from the chain. Therefore, the model requires the following parameters:
\begin{itemize}
    \item The rate at which a correct or incorrect monomer comes out of solution and binds to the template; \(k^{1}_{\rm on}[1]\) or \(k^{2}_{\rm on}[2]\). Here \(k^{1}_{\rm on}\) is the rate at which a monomer of type \(1\) binds to the template and \(k^{2}_{\rm on}\) is the equivalent for monomer type \(2\).
    \item The rate at which a correct or incorrect monomer which has just come out of solution falls off the template; \(k^{1}_{\rm on}e^{-\Delta G^{\plimsoll}_{r}}\) or \(k^{2}_{\rm on}e^{-\Delta G^{\plimsoll}_{w}}\). 
    \item The rate at which a polymer which has come out of solution is polymerised into the growing chain; \(k_{\rm cat}\).
    \item The rate at which the tip monomer is depolymerised from the growing chain; \(k_{\rm cat}\).
    \item The rate at which the penultimate correct or incorrect monomer separates from the template \(k_{\rm on}e^{-\Delta G^\plimsoll_{r}+\Delta G^\plimsoll_{\rm dim}}\) or \(k_{\rm on}e^{-\Delta G^\plimsoll_{w}+\Delta G^\plimsoll_{\rm dim}}\).
    \item The rate at which the penultimate correct or incorrect monomer rebinds to the template \(k_{\rm on}\).
    \item The rate at which the final correct or incorrect monomer separates from the template \(k_{\rm on}e^{-\Delta G^{\plimsoll}_{r}}\) or \(k_{\rm on}e^{-\Delta G^{\plimsoll}_{w}}\).
\end{itemize}

We use a Monte Carlo simulation to generate the probability of various outcomes given a first step of a system using these rates. We can also generate the average time taken for a system to reach any output state given an input state. Using a dimerisation system as an example, the probability of outputting a dimer of type \(1,1\), given a monomer of type \(1\) has attached to the template is denoted \(P^{T}(1,1|1)\), equally the probability of outputting a dimer type \(1,1\) given a dimer of type \(1,2\) has attached to the destroyer is \(P^{D}(1,1|1,2)\), the probability of destroying the dimer given a dimer of type \(1,1\) has attached is \(P^{D}(\emptyset|1,1)\) etc. In ch. 2 it was necessary for us to solve a set of simultaneous equations based on these probabilities in order to solve for the fluxes of creation \(q({\bf s})\). For dimers, these equations are of the form;

\small
\begin{equation}
    \left(P^{T}(1,1|1)[1]k^{1}_{\rm on} + \sum_{s}P^{T}(1,1|s)[s]k_{\rm on}    -[1,1]k_{\rm on}\left(P^{T}(\emptyset|1,1)+\sum_{s}P^{T}(s|1,1) \right)\right)P(T={\rm \emptyset})=\\ \nonumber
\end{equation}
\begin{equation}
     -\left(P^{D}(1,1|1)[1]k^{1}_{\rm on} + \sum_{s}P^{D}(1,1|s)[s]k_{\rm on}    -[1,1]k_{\rm on}\left(P^{D}(\emptyset|1,1)+\sum_{s}P^{D}(s|1,1) \right)\right)P(D={\rm \emptyset}),
     \label{sim1}
\end{equation}
\begin{equation}
    \left(P^{T}(1,2|1)[1]k^{1}_{\rm on} + \sum_{s}P^{T}(1,2|s)[s]k_{\rm on}    -[1,2]k_{\rm on}\left(P^{T}(\emptyset|1,2)+\sum_{s}P^{T}(s|1,2) \right)\right)P(T={\rm \emptyset})=\\ \nonumber
\end{equation}
\begin{equation}
     -\left(P^{D}(1,2|1)[1]k^{1}_{\rm on} + \sum_{s}P^{D}(1,2|s)[s]k_{\rm on}    -[1,2]k_{\rm on}\left(P^{D}(\emptyset|1,2)+\sum_{s}P^{D}(s|1,2) \right)\right)P(D={\rm \emptyset}),\label{sim2}
\end{equation}
\begin{equation}
   \left(P^{T}(2,1|2)[2]k^{2}_{\rm on} + \sum_{s}P^{T}(2,1|s)[s]k_{\rm on}    -[2,1]k_{\rm on}\left(P^{T}(\emptyset|2,1)+\sum_{s}P^{T}(s|2,1) \right)\right)P(T={\rm \emptyset})=\\ \nonumber
\end{equation}
\begin{equation}
     -\left(P^{D}(2,1|2)[2]k^{2}_{\rm on} + \sum_{s}P^{D}(2,1|s)[s]k_{\rm on}    -[2,1]k_{\rm on}\left(P^{D}(\emptyset|2,1)+\sum_{s}P^{D}(s|2,1) \right)\right)P(D={\rm \emptyset}),\label{sim3}
\end{equation}
\begin{equation}
  \left(P^{T}(2,2|2)[2]k^{2}_{\rm on} + \sum_{s}P^{T}(2,2|s)[s]k_{\rm on}    -[2,2]k_{\rm on}\left(P^{T}(\emptyset|2,2)+\sum_{s}P^{T}(s|2,2) \right)\right)P(T={\rm \emptyset})=\\ \nonumber
\end{equation}
\begin{equation}
     -\left(P^{D}(2,2|2)[2]k^{2}_{\rm on} + \sum_{s}P^{D}(2,2|s)[s]k_{\rm on}    -[2,2]k_{\rm on}\left(P^{D}(\emptyset|2,2)+\sum_{s}P^{D}(s|2,2) \right)\right)P(D={\rm \emptyset}).\label{sim4}
\end{equation}
\normalsize

The left-hand side of each equation is the net flux of that dimer type via the template, with \(P(T={\rm \emptyset})\) the probability that the template is empty. The first term on the LHS is the flux of the target dimer being created due to a monomer initiation. The second term on the LHS is the flux of dimers being transformed to our target dimer. The third term on the LHS is the flux of our target dimer being destroyed given a dimer initiation, and the final term is the flux of our target dimer being transformed to other dimers. The right-hand side of each equation is the equivalent quantity for the destroyer with \(P(D={\rm \emptyset})\) being the probability that the destroyer is empty.

There is additional complexity in this model compared to ch. 2. \(P(T={\rm \emptyset})\) and \(P(D={\rm \emptyset})\) are both dependent on \([{\bf s}]\), which was fixed in ch. 2, but here is our unknown. These two terms are time-dependent; it becomes essential to be able to estimate the time in which it takes our reactions on the template and destroyer to resolve. This is why breaking down our model into its subsets is essential, it allows us to assume exponential waiting times for each step and reliably calculate the time taken. Therefore, we calculate time-dependent quantities; \(\tau^{T}({\emptyset}|1)\) is the time taken for a system to return to the empty template state, given a \(1\) monomer has attached to the template etc. To calculate the probability that the template and destroyer are empty \(P(T={\rm \emptyset})\) and \(P(D={\rm \emptyset})\) we start by defining a normalisation constant \(\rho_{tot}=[1]k^{1}_{\rm on}+[2]k^{2}_{\rm on}+[1,1]k_{\rm on}+[1,2]k_{\rm on}+[2,1]k_{\rm on}+[2,2]k_{\rm on}\). We define the average time taken for the system to resolve back to the empty state, given an object has bound to the template or destroyer as;
\small
\begin{align}
    \tau^{T}_{\rm res} = \frac{[1]k^{1}_{\rm on}\tau^{T}({\emptyset}|1)+[2]k^{2}_{\rm on}\tau^{T}({\emptyset}|2)+\sum_{\bf s}[{\bf s}]\tau^{T}({\emptyset}|{\bf s})}{\rho_{tot}},\label{Tres}
\end{align}
\begin{align}
     \tau^{D}_{\rm res} = \frac{[1]k^{1}_{\rm on}\tau^{D}({\emptyset}|1)+[2]k^{2}_{\rm on}\tau^{D}({\emptyset}|2)+\sum_{\bf s}[{\bf s}]\tau^{D}({\emptyset}|{\bf s})}{\rho_{tot}}.
\end{align}
\normalsize
The average time taken for the system to bind to the template or destroyer from the empty state is then;
\small
\begin{align}
     \tau^{T}_{\rm bind}=\frac{1}{\rho_{tot}},\label{Tbind}
\end{align}
\begin{align}
      \tau^{D}_{\rm bind}=\frac{1}{\rho_{tot}}.
\end{align}
\normalsize
The probability of the template being empty is then;
\small
\begin{align}
      P(T={\rm \emptyset})=\frac{\tau^{T}_{\rm bind}}{\tau^{T}_{\rm bind}+ \tau^{T}_{\rm res}}\label{empty1},
\end{align}
\begin{align}
       P(D={\rm \emptyset})=\frac{\tau^{D}_{\rm bind}}{\tau^{D}_{\rm bind}+ \tau^{D}_{\rm res}}\label{empty2}.
\end{align}
\normalsize

Solving both simultaneous equations \ref{sim1}-\ref{sim4} and the two empty template equations \ref{empty1} and \ref{empty2} to find \([{\bf s}]\) is not analytically tractable for dimers, let alone longer oligomers. Therefore we iterate the system, initialising all \([{\bf s}]=e^{\Delta G^\plimsoll_{\rm dim}}\), then calculating a value of \(P(T={\rm \emptyset})\) and \(P(D={\rm \emptyset})\), and inserting these values into the simultaneous equations \ref{sim1}-\ref{sim4} to get \([{\bf s}]\) and repeating. The system converges quickly on a solution.

\begin{figure}
    \centering
    \includegraphics[scale=0.25]{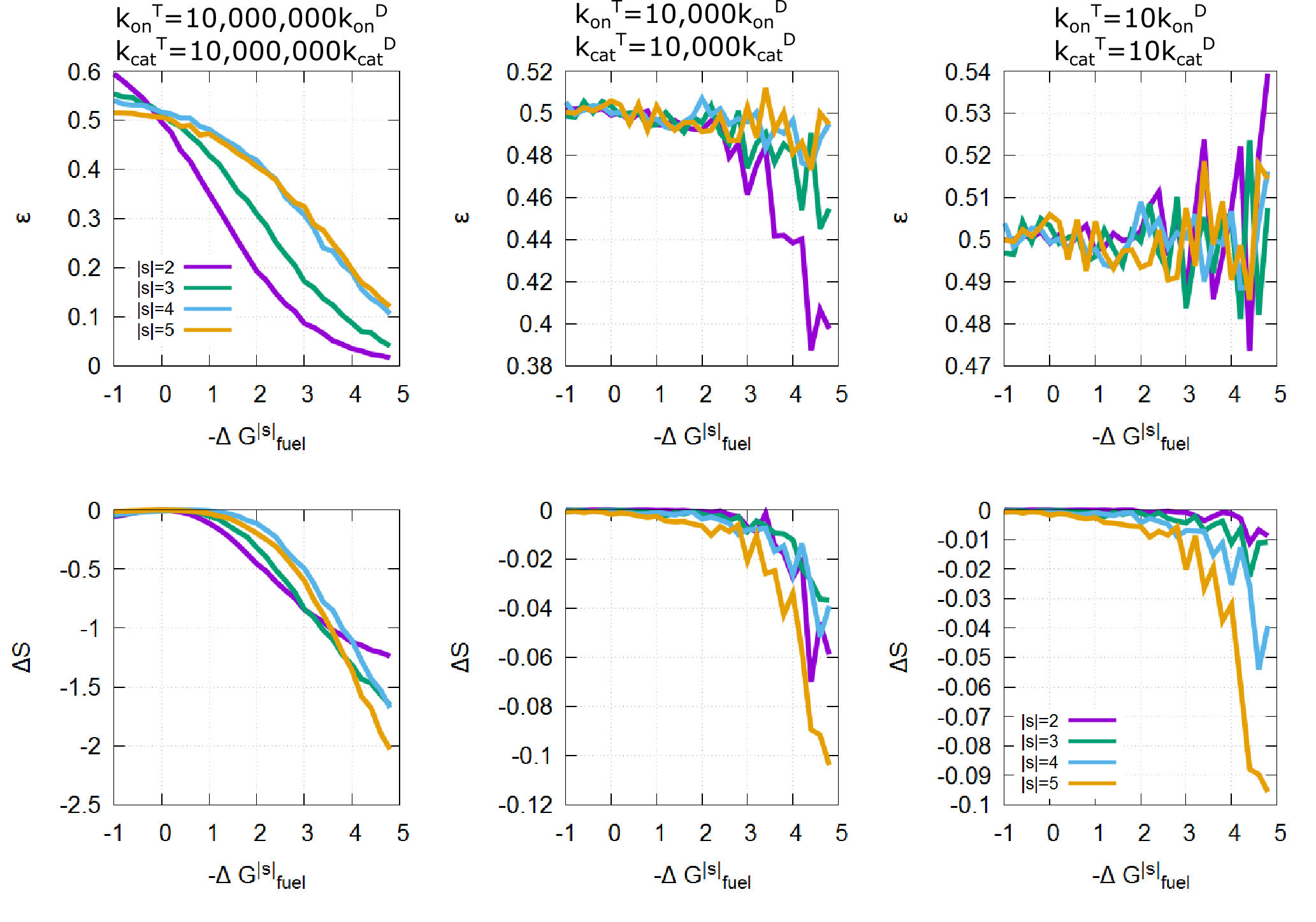}
    \caption{We plot the error and the entropy change for systems of length 2, 3, 4 and 5, for three different ratios of \(k^{T}_{\rm on}:k^{D}_{\rm on}\). Here there is off-rate discrimination alone via the template, \(\Delta G^\plimsoll_{r}=8\), \(\Delta G^\plimsoll_{w}=-8\), \(\Delta G_{\rm diff}=16\) and \(k^{1}_{\rm on}=k^{2}_{\rm on}=k_{on}\) for the template. The destroyer has no specificity \(\Delta G^\plimsoll_{r}=\Delta G^\plimsoll_{w}=\Delta G_{\rm diff}=0\). \(\Delta G_{\rm dim}=0\) throughout. The system is highly stochastic unless the destroyer is slowed considerably compared to the template. While the system fails to beat \(\Delta S<\Delta G^{|{\bf s}|}_{\rm fuel}\), we do observe that when \(k_{\rm on}^T=10,000,000k_{\rm on}^T\) that the error for \(|{\bf s}|=4\) and \(|{\bf s}|=5\) is the same; a signature that could imply the conjectured constraint would be beaten at longer lengths.}
    \label{OffRate}
\end{figure}

We set up a system in which the the template has high specificity through off-rate discrimination, \(\Delta G^\plimsoll_{r}=8\), \(\Delta G^\plimsoll_{w}=-8\), \(\Delta G_{\rm diff}=16\) and \(k^{1}_{\rm on}=k^{2}_{\rm on}=k_{on}\). The destroyer has no specificity \(\Delta G^\plimsoll_{r}=\Delta G^\plimsoll_{w}=\Delta G_{\rm diff}=0\). We keep \(\Delta G_{\rm dim}=0\) throughout, and vary \(\Delta G^{|{\bf s}|}_{\rm fuel}\). Because increasing \(\Delta G_{\rm diff}\) increases the number of steps taken, it increases the average time taken by the template to many orders of magnitude of that of the destroyer. Therefore, we increase \(k^{T}_{\rm on}\) and \(k^{T}_{\rm cat}\) for the template relative to \(k^{D}_{\rm on}\) and \(k^{D}_{\rm cat}\) for the destroyer.

In fig. \ref{OffRate} we plot the error and the entropy change for systems of length 2, 3, 4 and 5, for three different ratios of \(k^{T}_{\rm on}:k^{D}_{\rm on}\). Here the error is the probability that a given site in a chain of any length contains a \(2\) monomer; \(\epsilon=\langle\frac{\sum_{i}\delta(s_{i}=2)}{|{\bf s}|}\rangle\). The entropy change \(\Delta S=S_{\rm fin}-S_{\rm eqb}\), is the difference between the Shannon entropy of a random distribution of outputs \(S_{\rm eqb}=\ln{2^{|{\bf s}|}}\) and the Shannon entropy of the created distribution \(S_{\rm fin}=-\sum_{s}p({\bf s})\ln{p({\bf s})}\). 

In general we observe that \(\epsilon=0.5\) and \(\Delta S=0\) when \(\Delta G^{|{\bf s}|}_{\rm fuel}=0\) and that both the error and the entropy change decrease as \(\Delta G^{|{\bf s}|}_{\rm fuel}=0\) becomes more negative. In the case where \(k^{T}_{\rm on}=10k^{D}_{\rm on}\), the destroyer dominates the reaction, and both \(\Delta S\) and \(\epsilon\) become highly stochastic around \(0\) and \(0.5\) respectively. As the speed of the destroyer in decreased, the errors achieved become markedly lower. In order to disprove Bennett's conjecture, we need to find a region in which \(\Delta S<\Delta G^{|{\bf s}|}_{\rm fuel}\). It is clear that even for \(k^{T}_{\rm on}=10,000,000k^{D}_{\rm on}\) that the system doesn't achieve this. What is interesting is that for lengths \(|s|=4\) and \(|s|=5\) the system has the same error per site for the same \(\Delta G^{|{\bf s}|}_{\rm fuel}\), but \(|s|=5\) has a larger \(\Delta S\). If this trend were to continue to longer lengths, this would eventually lead to the conjectured constraint \(\Delta S<\Delta G^{|{\bf s}|}_{\rm fuel}\) being beaten, as for a given error per monomer \(\Delta S\) grows with length.

Therefore, we continue our search for a system in which \(\Delta S<\Delta G^{|{\bf s}|}_{\rm fuel}\). Following from ch. 1\cite{Poulton} we set up a system which contains both off-rate and on-rate discrimination, in analogy to the ``combined regime" of ch.1  section 1a. In ch.1 the combined system performs better than systems with on or off-rate discrimination in isolation. Therefore, we hypothesise that it might achieve better accuracy in this system too. To accomplish this we set \(k^{1}_{\rm on}=e^{\Delta G^\plimsoll_{r}}\) and \(k^{1}_{\rm on}=e^{\Delta G^\plimsoll_{w}}\).

\begin{figure}
    \centering
    \includegraphics[scale=0.25]{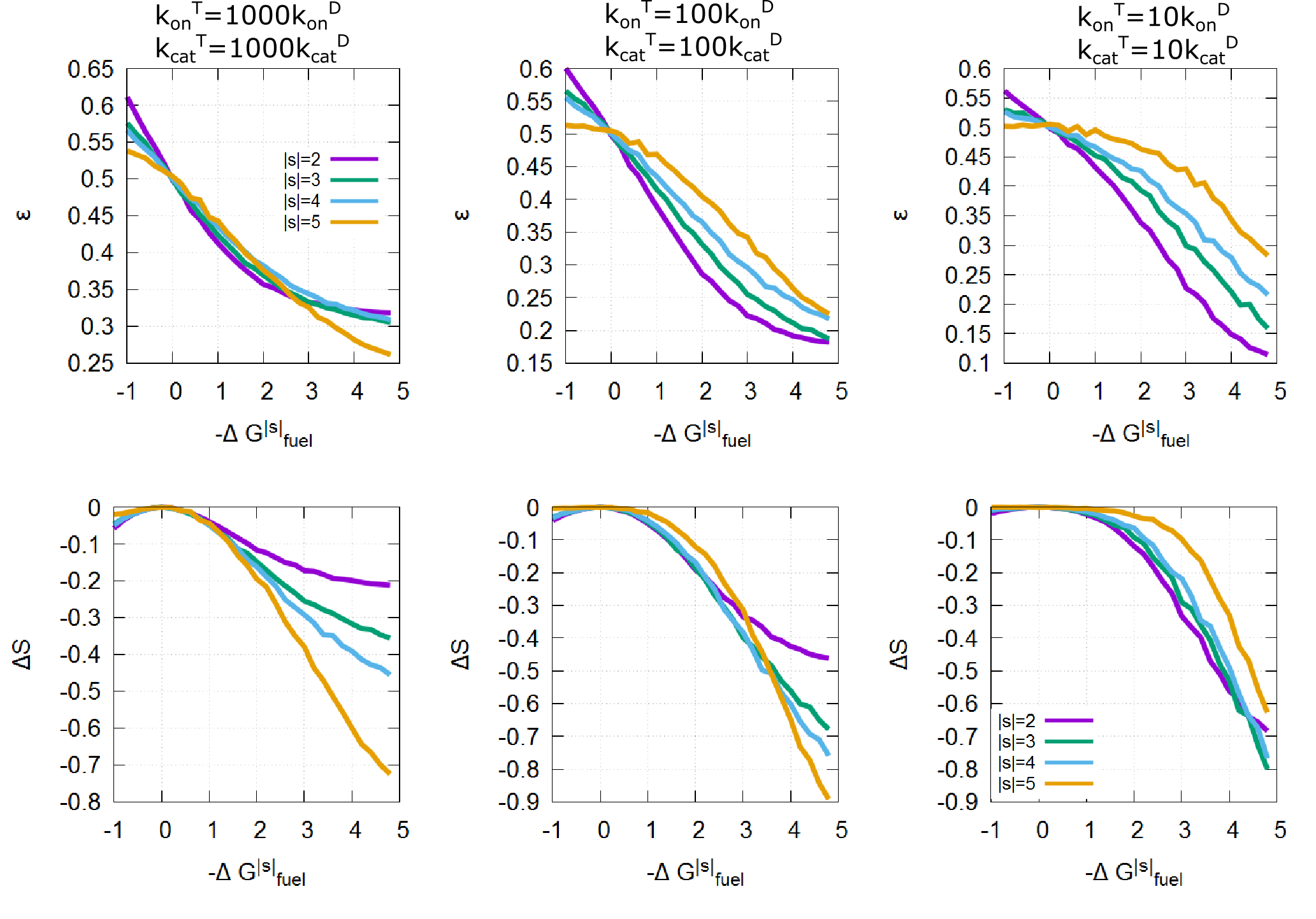}
    \caption{ We plot the error and the entropy change for systems of length 2, 3, 4 and 5, for three different ratios of \(k^{T}_{\rm on}:k^{D}_{\rm on}\). The system contains a single template and destroyer. The template has moderate specificity using on- and off-rate discrimination, \(\Delta G^\plimsoll_{r}=3\), \(\Delta G^\plimsoll_{w}=-3\), \(\Delta G_{\rm diff}=6\), \(k^{1}_{\rm on}=e^{\Delta G^\plimsoll_{r}}\) and \(k^{1}_{\rm on}=e^{\Delta G^\plimsoll_{w}}\). The destroyer has no specificity \(\Delta G^\plimsoll_{r}=\Delta G^\plimsoll_{w}=\Delta G_{\rm diff}=0\). \(\Delta G_{\rm dim}=0\) throughout.While the system fails to beat \(\Delta S<\Delta G^{|{\bf s}|}_{\rm fuel}\), we do observe that in all three ratios of \(k_{\rm on}^T:k_{\rm on}^T\) that the error improves as \(|{\bf s}|\) increase. This is a signature that could imply the conjecture constraint would be beaten at longer lengths.}
    \label{OnRate}
\end{figure}

We set up a system in which the the template has moderate specificity, \(\Delta G^\plimsoll_{r}=3\), \(\Delta G^\plimsoll_{w}=-3\), \(\Delta G_{\rm diff}=6\), and the destroyer has no specificity \(\Delta G^\plimsoll_{r}=\Delta G^\plimsoll_{w}=\Delta G_{\rm diff}=0\). We keep \(\Delta G_{\rm dim}=0\) throughout, and vary \(\Delta G^{|{\bf s}|}_{\rm fuel}\). Because increasing \(\Delta G_{\rm diff}\) increases the number of steps taken, it increases the average time taken by the template to many orders of magnitude of that of the destroyer. Therefore, we decrease \(k^{T}_{\rm on}\) and \(k^{T}_{\rm cat}\) for the template relative to \(k^{D}_{\rm on}\) and \(k^{D}_{\rm cat}\) for the destroyer. In fig. \ref{OnRate} we plot the error and the entropy change for systems of length 2, 3, 4 and 5, for three different ratios of \(k^{T}_{\rm on}:k^{D}_{\rm on}\).

In figure \ref{OnRate}, we observe that, even with both on-rate and off-rate discrimination, that the conjectured constraint proposed by Bennett are not even reached, let alone broken. There are certain regions in which the longer length oligomers have a lower error than shorter length oligomers for the same \(\Delta G^{|{\bf s}|}_{\rm fuel}\), and the function \(\Delta S\) gets progressively steeper for larger \(|s|\). If this trend continues as \(|{\bf s}|\rightarrow\infty\) then this would lead to an eventual breaking of Bennett's constraint. Thus it seems that exploring extending the length further might be the key to breaking Bennett's proposed constraint.

Here we have studied two systems, one with just off-rate discrimination, and one with both on- and off-rate discrimination. In the two specific systems we have studied, we see no violation of our generalisation of Bennett's presumption that if a system uses a cycle to generate an out of equilibrium distribution, the irreversible entropy generation must exceed the entropy reduction of the system due to the creation of the distribution ie \(\Delta G^{|{\bf s}|}_{\rm fuel}>\Delta S\). Ref\cite{Ouldridge}  demonstrates that there is a minimum free-energy cost per unit information required to {\it create} a non-equilibrium distribution. However, it is clear that in our system we are not {\it creating} but merely {\it maintaining} non-equilibrium steady-state. We hypothesise that there is no minimum cost per unit of information entropy to {\it maintain} a non-equilibrium steady-state. This hypothesis has not yet been proven using the systems above, but we will continue to search for systems which have sharply peaked \(p({\bf s})\), but have a low irreversible entropy generation that violates Bennett's presumption. It seems natural to start by extending our current models to significantly longer lengths.

\section{Conclusion}

In this chapter, we've studied a system with multiple catalysts and differential driving. Here, the differential driving is implemented by coupling the reverse oligomerisation process to the consumption of a fuel molecule in ``destroyer" molecules, while leaving ``template" molecules unchanged from those considered in ch. 2. The introduction of differential driving introduces new physics. Equations \ref{INFCHEMT} and \ref{INFCHEMD} describe constraints on the template and destroyer individually. Equation \ref{INFCHEMT} is identical to the result in ch. 2, but even considering a destroyer in isolation in \ref{INFCHEMD} demonstrates the effects of coupling to fuel. Fuel can be depleted or charged depending on the chemical and informational energy contributions to the system; we can now use fuel to compensate for an increase in both the information free-energy \(\Delta G^D_{\rm inf}\) and the chemical free-energy \(\Delta G^D_{\rm chem}\) . 

Equations \ref{entire} and \ref{gsdot} are constraints on a system of one template and one destroyer. It is clear that considering the two in combination has some notable consequences. Firstly, for a system with fixed, unbiased monomer and polymer baths a region exists in which the combined system can net create the most accurate dimer, while all others are net destroyed. We can also set up the system as an information engine with added complexity above that discussed in ch. 2. There are now seven regimes in which one, two or three out of the informational free-energy, chemical free-energy or energy stored as fuel can be depleted, increasing the other energies.

Considering the system in its entirety also allows us to explore a system in which a free-running system with fixed unbiased monomer pools sets its oligomer pools in a self-consistent way. The existence of a separate destruction pathway, coupled to fuel, allows a system to reach a non-equilibrium steady-state. In the absence of this pathway, the only way the system could reach steady-state is by equilibrating the system completely. We have questioned a generalisation of Bennett's conjecture that a system in which a unit of information is created and destroyed must dissipate an amount of free-energy as irreversible entropy generation which scales with the amount of information. In our formulation this constraint can be restated thus; in a system in which driven cycle causes a non-equilibrium distribution of products the free-energy dissipation which leads to irreversible entropy generation per cycle, must be greater than the entropy change due to the formation of the non-equilibrium distribution \(\Delta G^{|{\bf s}|}_{\rm fuel}>\Delta S\). 

Our constraints on the template and destroyer individually (equations \ref{INFCHEMT} and \ref{INFCHEMD}) and, on a system of one template and one destroyer (equations \ref{entire} and \ref{gsdot}) do not suggest a non zero bound on \(\Delta G^{|{\bf s}|}_{\rm fuel}\). While it is essential that \(\Delta G^{|{\bf s}|}_{\rm fuel}<0\) for a non-equilibrium steady-state to be maintained, there is no clear requirement that it should scale with the information content of the oligomer baths.

In chapter 1, we demonstrate a minimum thermodynamic cost for a non-equilibrium, accurate polymer to systematically grow on a template. Both chapters 2 and ref. \cite{Ouldridge} demonstrate that there is a thermodynamic cost for creating a non-equilibrium distribution which grows with the entropy change due to the creation of the distribution. This is because the free-energy cost of creating a free-energy distribution is stored in the distribution. Using idealised time-varying driving, it is possible to create a non-equilibrium distribution reversibly, because you are transducing energy between two forms; chemical and informational. If we manage to find a system which beats Bennett's conjectured constraint, then this would tell us that something fundamentally different is happening in the system which reaches a self consistent steady state considered in this paper. We predict that this is because this system is not {\em creating} a non-equilibrium state, but instead {\it maintaining} one. During {\em maintenance} of the steady-state the free-energy cost is dissipated as entropy generation in the environment, rather than stored in the distribution. Our results suggest that the free-energy cost to maintain a non-equilibrium steady-state can be arbitrarily close to zero.

Therefore, we began the search for a system which beats Bennett's conjectured constraint for short templates. While our results for a simple ``off rate discrimination only" and ``combined on- and off-rate discrimination" systems have failed to get close to Bennett's constraint we have seen one important signature. For the same error per monomer, \(\Delta S\) scales with length, if we can achieve the same or better error for the same \(\Delta G^{|{\bf s}|}_{\rm fuel}\) for progressively longer lengths, then at long lengths then \(\Delta G^{|{\bf s}|}_{\rm fuel}>\Delta S\). We therefore will continue our search at longer lengths.

Our method quickly becomes analytically intractable at long lengths, so our next step should be to do a full simulation; use a Markov state simulation to simulate a template and a destroyer in solution with full binding and unbinding. This would give us a non-exhaustive sample of oligomers in the baths. From a non-exhaustive sample, it's straightforward to calculate error rate; error can be calculated for every single site in the copy and so we don't need to worry about correlations {\it within} the copy sequence. Entropy, however, is more difficult. While in the infinite length limit correlations within the copy sequence were limited to length two \cite{Poulton}, we are uncertain how the initiation and termination steps might affect this. It might be implausible to calculate the full entropy rate of the final system, but we can use the single site entropy \(H<|{\bf s}|H_{ss}\) as an upper bound on the copy entropy giving a lower bound on \(\Delta S\). Since we are looking for \(\Delta S>\Delta G^\plimsoll_{\rm fuel}\), this would make Bennett's conjectured constraint harder to beat. Therefore, using the single site entropy would not invalidate our results if we did find a regime where the constraint is crossed; in fact, it would make our argument stronger.

\newpage
\setcounter{equation}{0}
\setcounter{figure}{0}
\setcounter{table}{0}
\setcounter{section}{0}

\section*{Conclusion}

\setcounter{equation}{0}
\setcounter{figure}{0}
\setcounter{table}{0}
\setcounter{section}{0}

Up until now, attempts to describe the templated copying process have contained a serious omission. The failure to explicitly consider separation of copy and template fundamentally changes the thermodynamics of a copying process. Historical attempts to describe a copying process which fail to explicitly consider separation\cite{Bennett,Cady,Andrieux,Sartori1,Sartori2,Gaspard,esposito2010,EHRENBERG1980333,Johansson} have led to serious misunderstandings\cite{pigolotti2016protocols,Sartori1,Sartori2}. If separation is omitted, the free energy released by the formation of strong copy-template bonds for accurate matches can compensate for the entropic cost of correlating copy and template. When separation is considered these strong copy template bonds must be broken.

In chapter 1\cite{Poulton}, we consider a set of biologically inspired models of templated copying on an infinitely long template. These models describe a system in which reactions are chemically driven, do not rely on time varying conditions and are the first of their kind to explicitly consider separation of copy and template. We find that separation forces a copying system to spend free-energy correlating copy and template. In order to systematically make progress along the template, while creating a completely accurate copy, the system must be driven by a free energy of no less than \(\ln{2}\) per monomer\cite{Ouldridge}. The biologically inspired systems we consider, in which the copy and template separate in a continuous fashion from behind the growing tip of the template, creates within chain correlations above the correlations between copy and template. These correlations require an extra input of free energy above that required to correlate copy and template.

In chapter 2, we consider templates of finite length, therefore necessarily including the final step in which a polymer detaches from the template and mixes with the surrounding monomer and polymer baths. This again changes the fundamental thermodynamics of the system. For an infinite length template, in order to systematically progress along the template and generate an accurate copy, the system had to spend free energy. More free energy needs to be spent to generate an accurate copy than an unbiased copy, \(\ln{2}\) per monomer. However the final process of detachment can provide a theoretically unlimited adjustment to the overall free energy cost of the process, meaning that in some circumstances, creating an accurate copy can be thermodynamically favourable compared to a random copy. Thus, although extending the copy polymer costs free energy, if a system can overcome this free energy barrier temporarily, it will end up in a lower free energy state. Thus the creation of finite length copy relies on exploiting fluctuations; if the system can briefly gain enough energy from thermal fluctuations to overcome the free energy cost of extending the polymer, then when that polymer detaches from the template into the baths, energy will be released and the system will have reached a lower free energy state. This is much like a system crossing an energy barrier to move between two meta stable states. The more accurate the copy polymer, the higher the energy barrier; making the previous absolute thermodynamic limit into a kinetic limit. Equally, the longer the polymer created, the higher the energy barrier. Thus as the length of the template increases, we observe a kinetic transition back to the thermodynamic limit found in chapter one.

In chapter 2 we further observe that a finite length templated copying system can act as a chemical/information engine, moving free energy between the out-of-equilibrium monomer and polymer baths. By splitting the overall free energy of creating a copy into two terms - a chemical and an informational term - we observe that a system must convert chemical free energy into informational free energy if it is creating a copy distribution like the polymer baths and unlike the monomer baths. It can equally turn informational free energy back into chemical free energy, pumping against the chemical cost of creating a copy if the system is creating a copy distribution like the monomer baths and unlike the polymer baths. Thus the selectivity of the template catalyst can affect the workings of the system; for a given biased polymer distribution, the selectivity affects whether the system converts from informational to chemical free energy or vice versa.

In chapter 3 we take on the challenge of describing a more complex biological environment. It is clear that a biological copying system such as mRNA translation can be approximated by a non-equilibrium steady state, proteins are continually produced by the ribosome in a highly selective way while being constantly degraded by the cellular environment. Previous work\cite{OuldridgePRX} has shown that a second reverse pathway can increase accuracy in the overall system and we observe that this is the case when we introduce a destroyer pathway in our system. 

We also challenge Bennett's prediction that the creation and destruction of a bit of information will cause \(\ln{4}\) of useful free-energy to be dissipated into the environment as entropy\cite{Bennett1982}. We propose that while it is essential that a non-zero amount of entropy is generated, that the amount does not scale with the information content of the object being created and destroyed. We demonstrate this analytically, and while we do not find a system which breaks Bennett's prediction, we find signatures of a system which could break it for polymers of long length.

The major limitation of this work is that the models considered have been highly simplified. While they are capable of providing interesting physical insights, they require more complexity in order to capture the behaviour of biological systems. Over the course of my PhD I have therefore also been working with Master's and PhD students, supervising their projects that attempt to move towards a more biologically realistic model of templated copying. I will introduce this work here, as a summary of the next steps available for this project.

This work can be split into two main strands. The first is kinetic proofreading. The process of transcription has an error rate per base of \(10^{-4}\), well below that reached by even the most accurate of our models\cite{Alberts2002}. Kinetic proofreading is a process by which cells error check and so it is important to include in a model of copying. The second is turn order in the copying process. Throughout my work I have enforced a strict turn order, artificially excluding many theoretically plausible reactions in order to simplify the system. For example I have not asked the question ``under what circumstances does a copy reliably detach from the template"; I have merely assumed that separation occurs as part of the turn order. In order to fully understand the copying process, it is vital to explore a system which allows alternate reactions, including allowing the copy to remain stuck to the template. Further work includes finding parameters, in the absence of enforced separation, which allow the process of templated copying, including separation, to occur successfully.

\section{Kinetic Proofreading}

In his seminal paper on kinetic proofreading\cite{Hopfield}, Hopfield proposed a theoretical method for error correction which was accepted by biologists as accurately describing real biological systems\cite{McKeithan5042,blanchard2004trna}. In order to understand kinetic proofreading, we should return to the underlying reason for accuracy generation in all templated copying systems.

\begin{figure}
    \centering
    \includegraphics[scale=0.5]{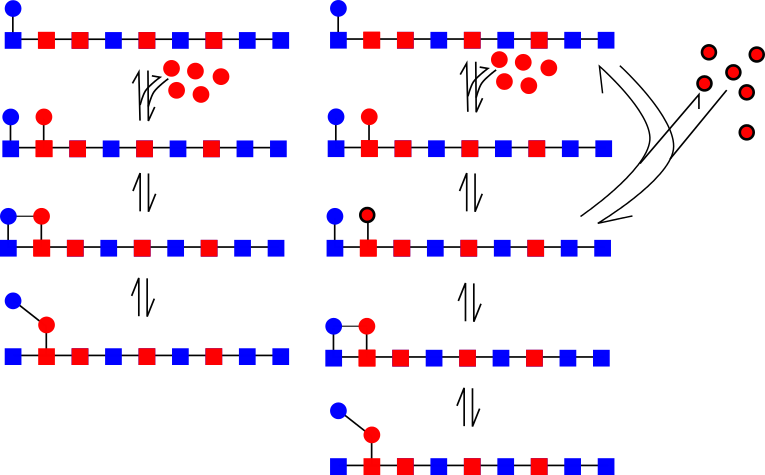}
    \caption{On the left we see a copying process without proofreading. A monomer comes out of solution and binds to the template, is polymerised into the growing chain and the previous final monomer detaches. Accuracy is generated because correct monomers are more likely to stay bound to the template for long enough to be polymerised into the copy polymer. On the right we see a copying process with proofreading. A monomer comes out of solution, is activated, is polymerised into the growing chain and the previous final monomer detaches. Monomers now have two opportunities to fall off, before and after activation. As long as the activation step is driven; either by the process of activation releasing free-energy into the surroundings, or by a higher concentration of inactive than active monomers. The system is now capable of cycling, meaning the system is no longer tightly coupled. This means that the amount of free-energy it takes to generate the polymer no longer exactly predicts the state of the polymer, unlike a system without futile cycles.}
    \label{proofreading}
\end{figure}

As illustrated in fig. \ref{proofreading}, if we break down the process of extending a monomer by a single step, then the first step in adding a monomer to a growing chain is for a monomer to come out of solution and bind to a template. It then has to remain on the template for long enough to be polymerised into the growing chain. The likelihood of it remaining bound to the template depends on the strength of a copy template bond. If the monomer bound to the template is a match, forming the bond releases a free energy of \(\Delta G^{\plimsoll}_{r}\) and is more stable than a mismatch which releases a free energy of \(\Delta G^{\plimsoll}_{w}<\Delta G^{\plimsoll}_{r}\). The rate at which matches leave the template is proportional to \(e^{-\Delta G^{\plimsoll}_{r}}\) and the rate at which mismatches leave the template is proportional to \(e^{-\Delta G^{\plimsoll}_{w}}\). Thus the matching monomer is more likely to remain on the template for long enough to be polymerised into the chain. This is the essence of the method of accuracy generation in copying. It should be noted that, due to the fact that all copy/template bonds are transitory, this method only works if the system is driven out-of-equilibrium. Otherwise this fundamentally kinetic effect would be averaged out over the equilibrium time period.

Kinetic proofreading works in general by adding further discard pathways. In Hopfield's proofreading model\cite{Hopfield}, this means adding a second step. Before the monomer is polymerised into the chain, it first has to undergo a transformation, sometimes referred to as an activation. This allows it a second opportunity to fall off before polymerisation, once before the transformation and once afterwards. In fig. \ref{proofreading}, a proofreading cycle would be the addition of a red inactivated monomer, the transformation of the monomer to an active monomer, and then the release of the activated monomer back into solution. In order for this process to be effective, it is essential that the system is driven round the proofreading cycle. This could be either because the process of activation releases free-energy, because the reaction is coupled to fuel, or because there is a chemical gradient; a larger concentration of inactive than active monomers. As long as the system is driven round the proofreading cycle, we can get the discrimination effect twice.

Kinetic proofreading has, until now, only been studied for templated self-assembly systems, in which separation was not explicitly modelled\cite{Sartori1,pigolotti2016protocols}. This allowed a permanent thermodynamic bias towards accuracy at equilibrium, even in the infinite length limit, because of permanent stabilising bonds between copy and template. This means that in these systems, the out-of-equilibrium driving used in kinetic proofreading was used to improve an already existing equilibrium bias.

Adrian Beersing-Vasquez, a masters student in academic year 2018-2019, built and analysed the first model that explicitly considers both proofreading cycles and separation of copy and template. His work highlighted an important property of systems that contain proofreading cycles.

\begin{figure}
    \centering
    \includegraphics[scale=0.25]{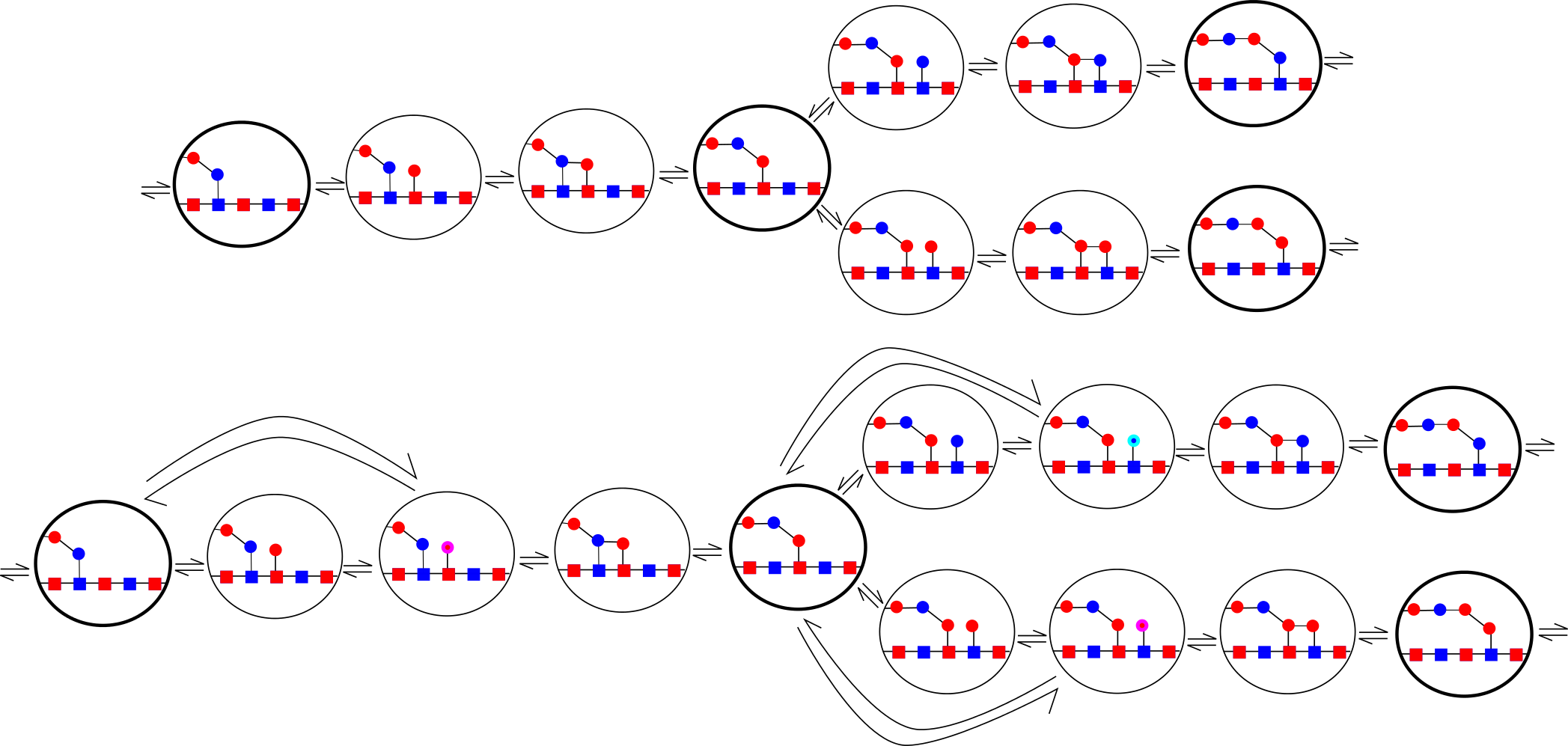}
    \caption{Drawing out the full state tree we see that in the non-proofreading case there are two paths forward: adding a correct or incorrect match, and only one paths backwards: removing the final monomer. Thus the polymer can grow from purely entropic driving.  Adding kinetic proofreading adds extra paths backwards, reducing the capacity for entropic driving.}
    \label{entropic}
\end{figure}

It is possible, under certain conditions, for copying systems in the absence of proofreading cycles to grow entropically\cite{Esposito}. Consider the system in fig. \ref{entropic}, entropic growth is a result of the shape of the network. Here it is possible for a system to step forward in more than one way (adding a match or a mismatch), but there is only one way to step backwards (here removing the final monomer in the chain). If all three pathways had equal probabilities then the system would step forward twice as often as it steps backwards. One has to increase the relative probability of the reverse step to twice that of each of the individual forward transitions before the system would become stationary. This corresponds to the system being able to grow against a maximum free energy bias of \(\ln{2}\)\cite{Poulton}.

However, the selectivity of the system affects entropic driving. As accuracy increases, the second route forward (ie adding a mismatch) is cut off slowly and driving must increase in order for the polymer to grow. This is another intuitive explanation for the need for a system to be driven above the equilibrium required to grow an unbiased polymer, in order to grow an accurate polymer in a copying system that includes separation.

A proofreading cycle reduces the effectiveness of entropic driving. Given the activation step provides a second opportunity for a monomer to fall off the template, it effectively acts as an additional backward pathway. Not only is it an additional backward pathway, it is also a backward pathway with the potential to be a very low free-energy state relative to the input pool of inactivated monomer. This means the system has a strong drive to disassemble the polymer via the proofreading pathway. However accurate the polymer produced, the kinetic proofreading pathway requires the system to be driven harder in the direction of growth than would be required in its absence. This is above the free energy required to drive the proofreading cycle. This result provides a new understanding as to the requirements of a system to dissipate free energy in order to create a templated copy. The next step in this project is to extend this research to more generic proofreading systems, such as the proofreading ladders of Murugan's work\cite{murugan2012speed, murugan2014discriminatory}.

\section{Turn order in copying systems}
\begin{figure}
    \centering
    \includegraphics[scale=0.5]{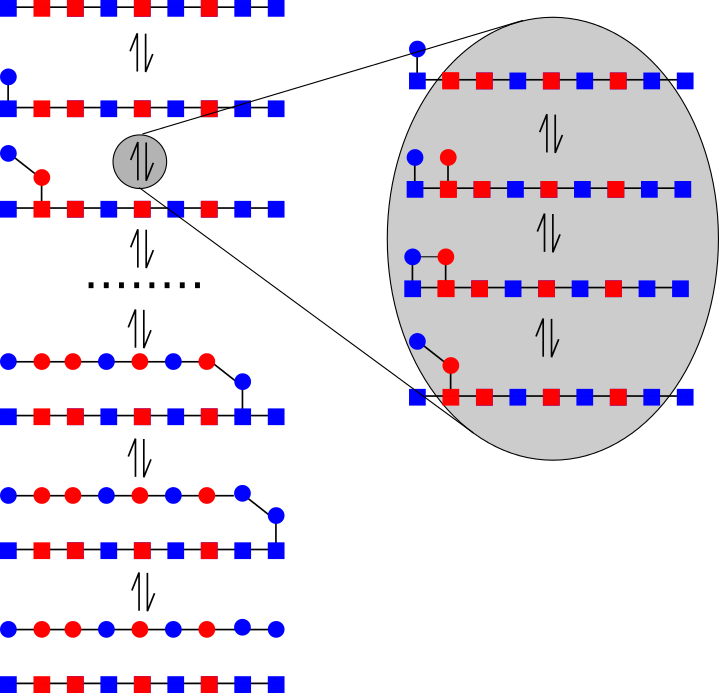}
    \caption{The models used throughout this thesis have enforced well defined turn order. An empty template exists in a bath of monomers. A monomer comes out of solution and attaches at the leftmost site on the template. Monomers come out of solution, join the leftmost available site and are polymerised into the chain. The previous final monomer must separate from the template before the next monomer can come out of solution. The polymer grows to full length and can only fall off into solution from the final position on the template.}
    \label{generic3}
\end{figure}

Throughout this work, we have enforced a very specific turn order in our models of the process of templated copying. As in fig. \ref{generic3}a, we start with an empty template. A monomer can come out of solution and bind at the leftmost site of the template only. The polymer then grows left to right, with all sites except for the site next to the current tip blocked. The polymer then detaches from the final site. Fig, \ref{generic3}b shows how the process of growth enforces separation, once the monomer comes out of solution, it is polymerised into the chain, and then the (new) penultimate monomer in the chain detaches from the template. Only then can the next monomer come out of solution.

\begin{figure}
    \centering
    \includegraphics[scale=0.4]{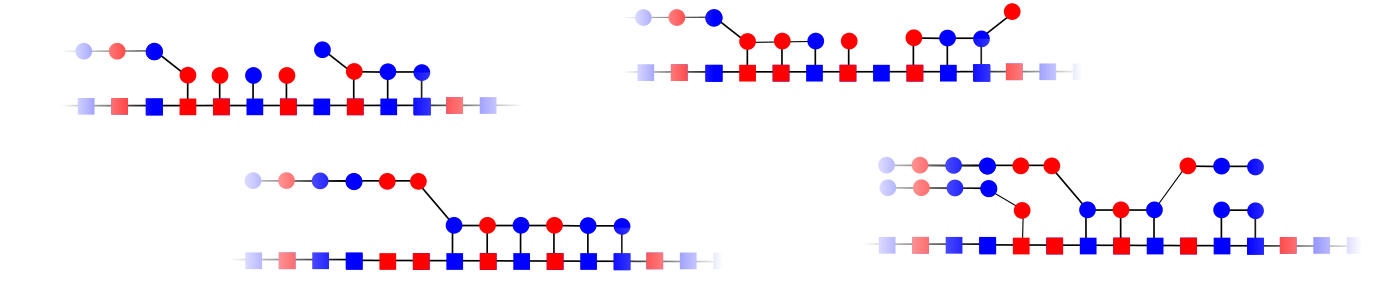}
    \caption{A selection of plausible copy/template configurations if strict turn order is relaxed. Products can remain bound to the template (product inhibition) or small fragmentary polymers can be formed. Avoiding these requires careful engineering of the copy/template interactions.}
    \label{chimera}
\end{figure}

This turn order is very artificial. In theory there is nothing to stop a monomer from binding at any point on the template, or for the tail to remain stuck down as the polymer grows, or for a polymer to break off the template early, or any number of other possible outcomes. Plausible intermediate states are shown in fig. \ref{chimera}, demonstrating that these reactions are undesirable; leading to copy fragments or product inhibition. Our models are sufficient to investigate the {\it consequences} of the separation of copy and template but are unable to deduce the \textit{mechanism} by which separation of copy and template occurs. It is clear that we cannot capture all the subtleties of preventing these undesirable outcomes with the two or three parameters we use in our systems.

The work of investigating the underlying mechanisms of templated copying was begun by Akash Bhattacharjee; a masters student 2017-2018. His work lifted the restriction that the penultimate copy/template bond must separate before the next monomer could come out of solution, and also allowed copy templates to separate early from the template causing fragments. This allows both product inhibition - the copy remaining stuck to the template - and early detachment - the formation of copy fragments. He found that with the parameters used in chapter 1 of this text, there is no region of parameter space in which the polymer grows fully on the template and then detaches. 

He proposed several additional parameters that allow the system to be varied in ways which may prevent product inhibition or early detachment. The first was having a parameter to describe an ``anti-cooperativity" effect, weakening copy template bonds which have another bond to their right. This solved the problem of product inhibition, allowing the tail of the copy polymer to separate from the template, but failed to solve the issue of early detachment. He then added a kinetic parameter which slows down reactions behind the leading tip, effectively creating a region in which the tail monomers are able to come off the template, but at a slower rate than the growth of the polymer. With these two parameters, it was possible for a system to grow on the template to full length, and then detach.

\begin{figure}
    \centering
    \includegraphics[scale=0.2]{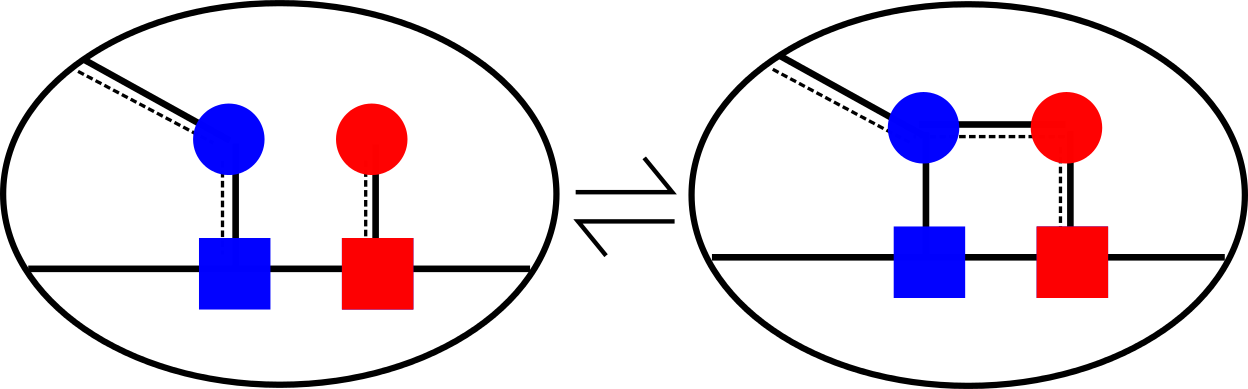}
    \caption{A proposed solution to the problem of product inhibition. If the process of forming a backbone bond between consecutive copy monomer weakens the copy template bond at the penultimate site, then it helps force the copy polymer off the template behind the tip, while keeping one strong copy/template bond at the tip.}
    \label{twobond}
\end{figure}

Inspired by this, Jordan Juritz, a PhD student from 2018-present proposed a synthetically realisable mechanism which captures the effects of these two parameters. Illustrated by fig. \ref{twobond} he proposed that when a monomer attached to the template is polymerised into the growing chain, that this should weaken the copy template bond of the previous monomer. This has proved to be a surprisingly resilient way of encouraging reliable growth and separation. I will return to the ways this can be synthetically realised in a section on strand displacement reactions below.

Jordan's work has considered a whole range of leak reactions, including allowing monomers to come out of solution and bind to any point on the template. He found that a significant difficulty is actually producing polymers of consistent length. He has proposed several solutions to this problem which will be discussed in an upcoming paper. Excitingly, many of them reflect solutions used by biological machinery; specialised ``start" and ``stop" codons and restricting access to the template to a small ``bubble" around the tip of the growing copy polymer. This seems to suggest that even these highly simplified models capture the essence of biological copying. 

\section{Synthetic copying systems}

A long term aim of the group is to create synthetic copying systems which do not use complex biomachinery such as enzymes, and instead are chemically driven just by the thermodynamic properties of the components of the system. Due to its high level of programmability, DNA is an ideal material for building synthetic copying systems. Inspired by the success of toehold mediated strand displacement\cite{yurke2000dna, zhang2009control} as a way of programming reactions between strands of DNA, my colleague Javier Cabello Garcia developed a method of using strand displacement reactions to realise a reaction capable of performing templated copying, using short strands of DNA as monomers\cite{cabello2020handhold}.

\begin{figure}
    \centering
    \includegraphics[scale=0.25]{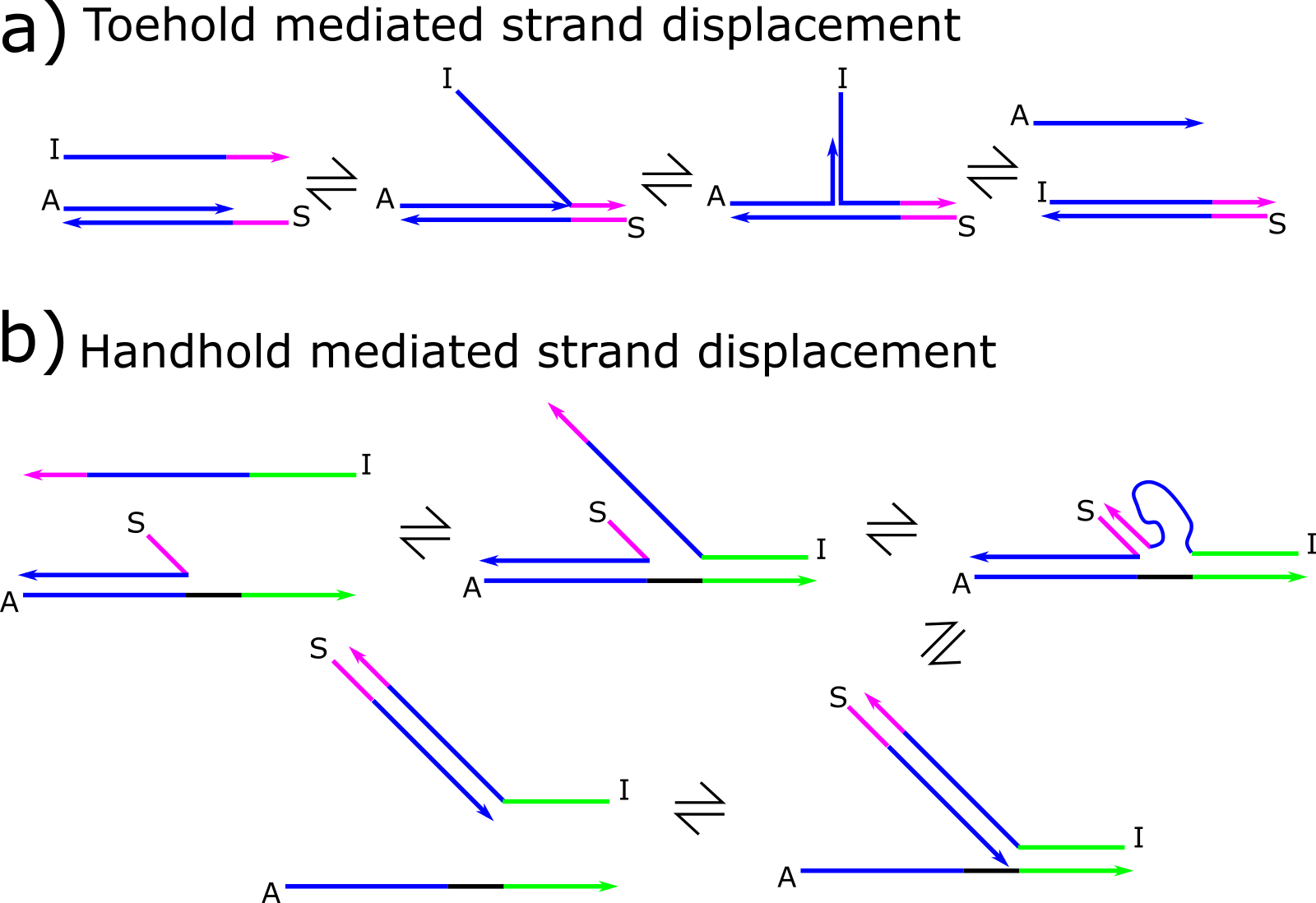}
    \caption{a) Toehold mediated strand displacement. A substrate strand \(S\) contains a sequence \(\beta\) (blue) is bound to a complementary strand \(A\) containing \(\beta^{*}\) (blue). A toehold sequence \(\gamma\) (pink) is left uncovered by \(A\). An invader strand \(I\) containing both \(\beta^{*}\) and \(\gamma^{*}\) attaches to the toehold and then slowly displaces the initial strand \(A\). This process releases free energy due to the bonds between \(I\) and \(S\) in the toehold section. b) Handhold mediated strand displacement. This reaction allows the invader strand \(I\) to interact with both the substrate strand \(S\) and the initially bound strand \(A\). The substrate strand \(S\) contains both the sequence \(\beta\) (blue) and the toehold \(\gamma\) (pink). The initially bound strand contains \(\beta^{*}\) and also a handhold sequence \(\omega\) (green). The invader strand initially binds to the handhold on \(A\) via the sequence \(\omega^{*}\) which brings the sequence \(\beta^{*}\) into proximity to the toehold on the substrate. The strand displacement reaction occurs and the handhold detaches. If we label \(A\) the template, \(S\) the final monomer attached to the template and \(I\) the new monomer, we can see that this reaction performs templated copying.}
    \label{stranddisplacement}
\end{figure}

Toehold mediated strand displacement as shown in fig. \ref{stranddisplacement} is a process by which a DNA strand can be encouraged to displace and replace another strand which is bound to a third strand. The substrate strand \(S\) has a section complementary to both of the other two strands \(\beta\). A strand \(A\) contains the complementary sequence \(\beta^{*}\) starts bound to the target strand. A small toehold on the substrate strand \(\gamma\)is left uncovered. An invader strand \(I\) is comprised of both \(\beta^{*}\) and also a section complementary to the toehold \(\gamma^{*}\). It attaches first to the toehold \(\gamma\) and then due to its sequestering near the sequence \(\beta\), invades and displaces strand \(A\). While in theory displacement can happen in both directions, the lack of a toehold for strand \(A\) makes the reverse reaction significantly less likely. Not only that but the process is thermodynamically downhill because of the extra free energy released by the formation of the base pair bonds between the toehold and the template. This is a highly versatile reaction, but in its current form, is insufficient for templated copying.

Templated copying requires that an incoming monomer be able to interact with two different strands; a transient link with the template, and a permanent link with the neighbouring monomer. As shown in fig. \ref{stranddisplacement}, Javier's handhold mediated strand displacement reaction\cite{cabello2020handhold} solves this problem. Now the substrate strand \(S\) is the tip monomer in a growing polymer chain, which contains both \(\beta\) and \(\gamma\). The template contains strand \(A\) which is initially bonded to the tip monomer via it's complementary section \(\beta^{*}\). However now, on top of the toehold on the previous monomer \(S\), Javier proposed a handhold \(\omega\) on the template strand \(A\). The invader strand has all three domains \(\beta^{*}\), \(\gamma^{*}\) and \(\omega^{*}\). It attaches first to the handhold, which brings it into close proximity with the toehold. This then allows the toehold to attach, and the invading monomer \(I\) to displace the previous monomer \(S\) from the template \(A\) and bind to it in its place. Handhold mediated strand displacement is therefore a very effective method of generating a persistent copy.

Handhold and toehold mediated strand-displacement has the advantage of being highly tunable. It should therefore be possible to design experiments in order to directly observe some of the more surprising results outlined in this thesis. For sufficiently short polymers it should be possible to bias the system towards creating accurate copies even when the free energy released by extending the polymer less than \(\ln{2}\) per monomer (or equivalent for more monomer types), just by manipulating the monomer and polymer baths. Equally, it should be possible to maintain a non-equilibrium bath of polymers at a housekeeping entropy cost which does not scale with the information content of the bath. It should however be clarified that while these results are helpful to our greater understanding of the thermodynamics of copying, they do not speak to extant biological systems. Biological reactions are coupled to the hydrolysis of a fuel called ATP, which releases \(12k_{B}T\) of energy. Thus many of the subtle effects outlined in this thesis would be masked by the huge influx of energy, much of which is lost as heat. A greater understanding of these effects is however useful in the design of synthetic copying systems, which may not necessarily be coupled to ATP.

\section{Conclusion}

The accurate, templated copying of long polymers, despite being the central process of the central dogma of molecular biology, has hitherto proved impossible to replicate in the absence of complex biological machinery. Theoretical studies of the process have completely failed to consider separation of copy and template. This separation is not only essential to the usability of the products; a protein is no use to the cell if it remains bound to RNA, but it also fundamentally changes the underlying thermodynamics of the system. This thesis is the next part of the process of fully understanding templated copying, which culminates in the building of a fully autonomous, chemically driven, templated copying process. In this section I have outlined some of the work of my colleagues who incrementally bridge the gap between my models, and a full copying system, build using strands of DNA.
 
\bibliography{aipsamp}

\newpage
\section*{Acknowledgements}
I would like to thank my supervisor Thomas Ouldridge for both his academic support and inspiration, and also for his pastoral support. He went above and beyond to support me to complete this thesis.\\

I would like to thank my colleagues and students Jordan Juritz, Akash Bhattacharjee and Adrian Beersing-Vasquez for their interesting conversations which shaped many of the ideas in this thesis.\\

I would like to thank my proofreaders, Kit Roskelly, Hattie Gemmill, Ben Mackay, David Proctor and Jordan Juritz (again).\\

Finally I would like to thank my parents: Jackie and Derek Poulton, my aunt: Debbie Mudie, and my friends: Sarah-Jane Roberts, Ruth Narramore, Jay Holder, and Gabi Falquero, who along with countless others have supported me enormously over these four years.

\end{document}